\documentclass[journal, 12pt, draftcls, onecolumn]{IEEEtran}

\usepackage[noadjust]{cite}

\ifCLASSINFOpdf
  \usepackage[pdftex]{graphicx}
  
\else
  \usepackage[dvips]{graphicx}
\fi
\usepackage{microtype}
\DisableLigatures[f]{encoding=T1}
\usepackage[T1]{fontenc}
\usepackage{footmisc}
\usepackage[cmex10]{amsmath}
\usepackage{array}
\usepackage{mdwmath}
\usepackage{mdwtab}
\usepackage{amssymb}
\usepackage{bm}
\usepackage{mathrsfs}
\usepackage{filecontents}
\usepackage{stfloats}
\usepackage{color}
\definecolor{ao}{rgb}{0.0, 0.5, 0.0}
\definecolor{copper}{rgb}{0.72, 0.45, 0.2}

\ifCLASSOPTIONcompsoc
	\usepackage[caption=false,font=normalsize,labelfont=sf,textfont=sf]{subfig}
\else
	 \usepackage[caption=false,font=footnotesize]{subfig}
\fi

\usepackage{stfloats}
\usepackage{url}
\newtheorem{Theorem}{Theorem}
\newtheorem{Corollary}{Corollary}
\newtheorem{Lemma}{Lemma}
\newtheorem{Example}{Example}
\newcommand{\goodchi}{\protect\raisebox{2pt}{$\chi$}}
\newtheorem{Remark}{Remark}

\newtheorem{Definition}{Definition}

\newtheorem{Proposition}{Proposition}

\usepackage[ruled,vlined]{algorithm2e}
\usepackage{algcompatible}

\SetKwProg{Fn}{Function}{}{}
\begin{document}
\title{Structural Characteristics of Two-Sender Index Coding}
\author{\IEEEauthorblockN{Chandra Thapa$^{1}$, Lawrence Ong$^{2}$, Sarah J.~Johnson$^{2}$, and Min Li$^{3}$\\
\small{$^{1}$CSIRO Data61, Australia; 
		$^{2}$The University of Newcastle, Australia;
		$^{3}$Zhejiang University, China}}
\thanks{This work is supported by the Australian Research Council under Grant FT140100219 and Grant DP150100903.} 
	}
\markboth{Draft}%
{Shell \MakeLowercase{\textit{et al.}}: Bare Demo of IEEEtran.cls for Journals}
\maketitle 

\begin{abstract}
	This paper studies index coding with two senders. In this setup, source messages are distributed among the senders possibly with common messages. In addition, there are multiple receivers, with each receiver having some messages a priori, known as side-information, and requesting one unique message such that each message is requested by only one receiver. Index coding in this setup is called two-sender unicast index coding (TSUIC). The main goal is to find the shortest aggregate normalized codelength, which is expressed as the optimal broadcast rate. In this work, firstly, for a given TSUIC problem, we form three independent sub-problems each consisting of the only subset of the messages, based on whether the messages are available only in one of the senders or in both senders. Then we express the optimal broadcast rate of the TSUIC problem as a function of the optimal broadcast rates of those independent sub-problems. In this way, we discover the structural characteristics of TSUIC. For the proofs of our results, we utilize confusion graphs and coding techniques used in single-sender index coding. To adapt the confusion graph technique in TSUIC, we introduce a new graph-coloring approach that is different from the normal graph coloring, which we call two-sender graph coloring, and propose a way of grouping the vertices to analyze the number of colors used. We further determine a class of TSUIC instances where a certain type of side-information can be removed without affecting their optimal broadcast rates. Finally, we generalize the results of a class of TSUIC problems to multiple senders.
\end{abstract}


\IEEEpeerreviewmaketitle


\section{Introduction}
Consider a communication scenario over a noiseless channel where a sender is required to broadcast messages to multiple receivers, each caching some messages requested by other receivers a priori. The messages cached at each receiver is known as its \emph{side-information}. In this scenario, if the sender is informed about the side-information available at all receivers, then it can leverage that information whilst encoding to reduce the required number of broadcast transmissions, in comparison with a naive approach of transmitting all requested messages uncoded and separately. Such an encoding process is called index coding, and the resulting sequence of coded messages is known as an index code. Moreover, each receiver upon receiving the index code will be able to decode its required message by utilizing its side-information. The main aim of index coding is to find the optimal (shortest) codelength and the corresponding coding scheme.
Index coding was introduced by Birk and Kol~\cite{ISCODfirst,ISCODtran}, and further studied in subsequent works~\cite{maisbound,broadcastrate,netcoderelation,interferencealignment,composite,structuralproperties1,criticalgraphs,localgarphcoloring,linearoptimaljournal,interlinkedcycle, interlinkedcyclecorrection}.

Most existing works on index coding deal only with a single sender, capturing scenarios with centralized transmissions. However, many communication scenarios such as the following have messages distributed among multiple senders:
\begin{itemize} 
	\item Macro-cell networks with caching helpers \cite{femtocaching}~\textemdash~cellular networks deploying dedicated nodes, called helpers, with large storage capacity instead of femto-cell access points to reduce backhaul loads,
	\item cooperative data exchange \cite{coopdata1}~\textemdash~peer-to-peer networks with data exchange within a group of closely-located wireless nodes, and
	\item distributed storage~\textemdash~storage networks where data are distributed over multiple storage devices/locations.
\end{itemize}
In addition, each sender can be constrained to know only a subset of the total messages due to reasons such as limited storage, or error whilst receiving some messages over noisy channels, or server failure to deliver all messages. In this case, distributed transmissions are required, where multiple senders broadcast messages to the receivers. One metric to maximize the transmission efficiency in this scenario is to minimize the aggregate number of transmissions from all senders in such a way that all receivers' demands can be fulfilled. 
As this problem is more general than an index-coding problem with a single sender and is of practical interest (e.g., reducing delay in content delivery, and energy efficient broadcasting), it is a useful research avenue to study index-coding problems with multiple senders, known as multi-sender index-coding problems. 

\subsection{Prior works}
The multi-sender index-coding problem was first studied by Ong et al.~\cite{lawrencemultisender}. They considered the problems where multiple senders are connected to receivers via noiseless broadcast links (orthogonal to each other) with flexible capacities. In their setup, each sender knows only a subset of the messages; each receiver knows only one message requested by some receiver a priori, but may request multiple messages; also, one message is known to only one receiver. For this setup, they aimed to characterize the optimal aggregate codelengths, also known as the optimal broadcast rates, of the problems. This problem formulation model is called \emph{broadcast-rate formulation} of the problems. In their work, they devised lower and upper bounds on the optimal broadcast rate by implementing a graph-theoretic approach. The results were established using \emph{information-flow graphs}, which represent receivers' request, and \emph{message graphs}, which represent senders' message setting. Furthermore, they showed problem instances for which the upper and lower bounds coincide. A class of such instances is where no two senders have messages in common. 

In another work, Thapa et al. \cite{ourpaper3} considered a model similar to Ong et al.~\cite{lawrencemultisender} but with the \emph{unicast} message setting, meaning each message is requested by only one receiver, each receiver requests only one message, and each receiver knows a subset of messages requested by other receivers a priori. Based on graph-theoretic approaches, they established upper bounds on the optimal broadcast rate. In particular, they focused on the two-sender case, called two-sender unicast index coding (TSUIC). They extended existing single-sender index-coding schemes, namely the cycle-cover scheme~\cite{chaudhary, neely}, the clique-cover scheme~\cite{ISCODfirst, ISCODtran} and the local-chromatic scheme~\cite{localgarphcoloring} to the corresponding schemes in TSUIC. 

Sadeghi et al.~\cite{parastoomultisender1} considered multi-sender index-coding problems where the senders are connected to receivers via noiseless broadcast links of arbitrary but fixed capacities. They aimed to characterize the closure of the set of all achievable rate\footnote{The rate of a message is the number of message bits per encoded/broadcast bits.} tuples of messages, known as the capacity region. They devised inner bounds on the capacity region using random-coding approaches (which requires infinitely long messages), and outer bounds using Shannon-type inequalities. In particular, the first general inner bound was attained by a partitioned distributed-composite-coding scheme, built on the single-sender composite-coding scheme (an existing single-sender scheme that is based on a random-coding approach \cite{composite}). This scheme was further enhanced to a fractional distributed-composite-coding scheme by Liu et al.~\cite{parastoomultisender2}. Preliminary and improved polymatroidal outer bounds were also developed in the work by Sadeghi et al.~\cite{parastoomultisender1} and Liu et al.~\cite{parastoomultisender2}, respectively. As a result, the capacity region was established for all problem instances up to 3 receivers, and the sum capacity is established for all instances with 4 receivers and with unit link capacity from each sender. Independent of and in parallel with the work by Liu et al.~\cite{parastoomultisender2}, Li et al.~\cite{minli, mintranit} introduced new techniques of joint link-and-sender partitioning and cooperative compression of composite messages and developed a multi-sender cooperative composite-coding scheme. 
In a recent work by Li et al.~\cite{minminrank}, a new rank-minimization framework for multiple-sender index coding with the unicast message setting, i.e., MSUIC, was developed on the classic single-sender minrank concept. The framework enabled the authors to establish the optimal broadcast rate for all critical MSUIC instances up to four receivers. In addition, they presented a heuristic algorithm to study MSUIC instances with more receivers. 

Wan et al.~\cite{wan} introduced \emph{decentralized data shuffling} problems in which the receivers/workers can communicate with one another via a shared link. The decentralized data shuffling phase with uncoded storage (which stores a subset of bits of the data set) is equivalent to a multi-sender index coding problem. For this problem, they proposed converse and achievable bounds that are to within a factor of 3/2 of one another. Moreover, the proposed schemes were shown to be optimal for some classes of the problem. Recently, Porter et al.~\cite{embeddedindexcoding} introduced a special case of multi-sender index coding, called \emph{embedded index coding} (EIC), in which each node acts as both sender and receiver. With the help of several results, they showed the relationship between single-sender index coding and EICs. Furthermore, they developed heuristics to solve EIC problems efficiently.

\subsection{Our work and contributions}
Different approaches have been attempted to solve the multi-sender index-coding problems. However, the problems are more difficult and computationally complex than their single-sender counterparts, and we know very little about the characteristics of the problems. This paper studies the broadcast-rate formulation of TSUIC problems by implementing a graph-theoretic approach. More precisely, in the same spirit of studying structural properties of index-coding capacity in the single-sender case by Arbabjolfaei et al.~\cite{structuralproperties1}, we examine the structural characteristics of TSUIC problems. This kind of study embraces the "divide-and-conquer" approach and provides us an insight into the problems where we can solve a larger problem by solving its smaller sub-problems. Note that in the work by Arbabjolfaei et al.~\cite{structuralproperties1,Fthesis}, the capacity region of a given single-sender index-coding problem is shown to be a simple function of the capacity regions of its independent sub-problems by generalizing the notion of lexicographic graph product. In the TSUIC setup, due to the distributed message setting among senders, we cannot directly implement the notion of graph products and the existing approaches of single-sender index coding. In this work, we consider interactions between three independent sub-problems at a time in the TSUIC setup. By applying the notion of \emph{confusion graphs} in index coding~\cite{fractionalchromatic} along with the introduction of a two-sender graph coloring and a code-forming technique, we bound the optimal broadcast rate in both asymptotic and non-asymptotic regimes, and show it to be tight for some classes of TSUIC instances. Moreover, even for the single-sender cases, the non-asymptotic cases (especially index coding in non-linear finite fields) are less explored. For an index coding instance in the unicast message setting and the non-asymptotic regime in the message size, our techniques in this paper can be used to upper bound the optimal broadcast rate of this instance by a function of the optimal broadcast rates of its sub-instances in single-sender unicast index coding.

The contributions of this paper are summarized as follows:
\begin{enumerate}
	\item \textbf{Proposing a new coloring concept for confusion graphs in TSUIC, called two-sender graph coloring (Definition~\ref{def:2scoloring}, Section~\ref{sec:C})}: For SSUIC, the chromatic number of its confusion graph gives the optimal broadcast rate and the corresponding index code (for a specific message size). However, for TSUIC, as the two senders (encoders) contain some messages in common, the standard method of graph coloring of the confusion graph may not lead us to an index code. In this regard, we need a different kind of coloring function in TSUIC, and thus, in this paper, we propose a novel coloring technique to color the confusion graphs in TSUIC, and its optimization gives the optimal broadcast rate and optimal index code. 
	
	\item \textbf{Presenting a way of grouping the vertices of confusion graphs in TSUIC (Appendix~\ref{append2a})}: By exploiting the symmetry of the confusion graph, we propose a way of grouping its vertices for analysis purposes mainly in its two-sender graph coloring. In particular, this grouping helps us to analyze the number of colors used in two-sender graph coloring of a confusion graph.
	
	\item \textbf{Deriving the optimal broadcast rates of TSUIC problems as a function of the optimal broadcast rates of its sub-problems (Theorem~\ref{theorem:acyclic}--\ref{theorem:caseIIE1})}: We divide a TSUIC problem into three independent sub-problems based on the requested messages by receivers, specifically whether the messages are present in only one of the senders or in both senders. Now in TSUIC, considering the interactions (defined by side-information available at the receivers) between these three independent sub-problems, we derive the optimal broadcast rate (in both asymptotic and non-asymptotic regimes in the message size) of the problem as a function of the optimal broadcast rates of its sub-problems. Moreover, we bound the optimal broadcast rate, and show that the bounds are tight for several classes of TSUIC instances (sometimes with conditions). Furthermore, we find a class of TSUIC instances where a TSUIC scheme can achieve the same optimal broadcast rate as the same instances when the two senders form a single sender having all messages.
	
	\item \textbf{Characterizing a class of TSUIC instances where a certain type of side-information is not critical (Corollary~\ref{cor:1})}: For a class of TSUIC instances, we prove that certain interactions between the three independent sub-problems can be removed without affecting the optimal broadcast rate (in the asymptotic regime). This means that those interactions are not \emph{critical}.
	
	\item \textbf{Generalizing the results of some classes of TSUIC problems to multiple senders (Section~\ref{generalsetup})}: For some classes of TSUIC problems, we generalize the two-sender graph coloring of confusion graphs and the proposed grouping of their vertices. Then we compute the optimal broadcast rates of those problems as a function of the optimal broadcast rates of their sub-problems.
\end{enumerate}

After posting the first draft of this paper~\cite{ourtwosenderpaper} on Arxiv, this work had led to the following works in TSUIC. Arunachala et al.~\cite{arunahala1,arunahala2} claimed that they derived the optimal linear broadcast rates of classes of TSUIC problems as a function of their sub-problems by analyzing special matrices and linear code constructions. In another work by Arunachala et al.~\cite{arunahala3}, the optimal asymptotic broadcast rates (asymptotic in the message size) of TSUIC problems were derived as a function of their sub-problems with fully-participated interactions. They affirmed that for some classes of TSUIC problems, the upper bounds of the optimal broadcast rates presented in our paper are tight. For their results, they used a similar graph-based technique as presented in our first draft (\cite{ourtwosenderpaper}). In this paper, we consider general broadcast rates (which includes both linear and non-linear broadcast rates) for both asymptotic and non-asymptotic regimes in the message size.

\section{Problem Definitions and Graphical Representation} \label{background}
\subsection{Problem setup}
\begin{figure} [t]
	\centering
	\includegraphics[width=0.75\linewidth]{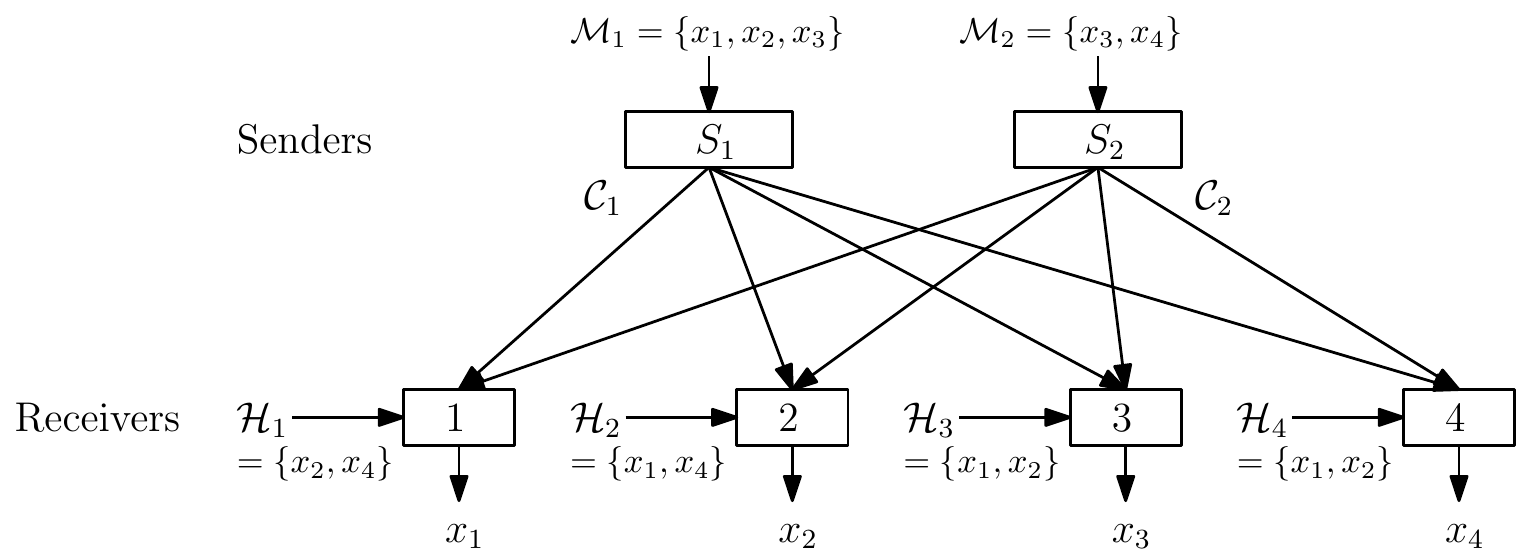}
	\caption{An example of a TSUIC problem with four receivers: The total message set $  \mathcal{M}=\mathcal{M}_1\cup \mathcal{M}_2 $ is distributed among two senders $ S_1 $ and $ S_2 $. Each sender is connected to all receivers via a noiseless broadcast channel. Each receiver, $ r\in \{1,2,3,4\} $, having some side-information represented by a set $  \mathcal{H}_r $, requests a unique message $ x_r $. We assume that each sender $ S_s $, $ s\in \{1,2\} $, is broadcasting a sub-codeword $\mathcal{C}_s $ of length $|\mathcal{C}_s |$, and they cooperate with each other to reduce their aggregate transmissions. Precisely, we aim to find a two-sender index code with the minimum sum of lengths $|\mathcal{C}_1|+|\mathcal{C}_2| $.}
	\label{fig:systemmodel} \vspace{-2ex}
\end{figure}

In this paper, we consider unicast index coding. There are $ N $ independent messages $  \mathcal{M} = \{x_1,x_2,\dotsc,x_N\} $, where $ x_i \in \{0,1\}^t $ for all $ i\in\{1,2\dotsc, N\} $ and some integer $ t\geq 1 $, i.e., each message consists of $t$ binary bits. There are $ N $ receivers $ \{1,2,\dotsc,N\} $, where each receiver $ r\in \{1,2,\dotsc,N\}$ requests a message $ x_r $, and has an ordered\footnote{\label{f1}The elements are ordered in increasing indices.} set $ \mathcal{H}_r \subseteq \mathcal{M}\setminus \{x_r\} $ of messages as its side-information a priori. This paper deals with the following two types of unicast index coding (UIC) based on the number of senders: (i) Single-sender unicast index coding (SSUIC)~\textemdash~it has only one sender, denoted $ S $, having all $ N $ messages $\mathcal{M} $, and (ii) two-sender unicast index coding (TSUIC)~\textemdash~it has two senders, denoted by $ S_1 $ and $ S_2 $, having (ordered) message sets $\mathcal{M}_1 \subseteq \mathcal{M}$ and $\mathcal{M}_2 \subseteq \mathcal{M}$, respectively, such that $\mathcal{M}_1\cup \mathcal{M}_2=\mathcal{M} $ (i.e., each message is available at some sender(s)). In other words, the total messages are distributed over the two senders in TSUIC. Figure~\ref{fig:systemmodel} illustrates an example of TSUIC problems with four receivers.

Given an index-coding problem, a two-sender index code is defined as follows:
\begin{Definition} [Two-sender index code]
	A two-sender index code ($ \{\mathscr{F}_s\},\{\mathscr{G}_r\} $), for $ s\in\{1,2\} $, $ r\in \{1,2,\dotsc,N\} $, is defined by
	\begin{enumerate}
		\item [(i)] an encoding function for each sender $ S_s $, $ \mathscr{F}_{s}: \{0,1\}^{|\mathcal{M}_{s}|\times t}\rightarrow \{0,1\}^{p_{s}}$ such that $\mathcal{C}_s= \mathscr{F}_{s}(\mathcal{M}_{s})$, and   
		\item [(ii)] a decoding function for every receiver $ r $, $ \mathscr{G}_r: \{0,1\}^{(\Sigma_{s=1}^2 p_{s} + |\mathcal{H}_r| \times t)}\rightarrow \{0,1\}^t $ such that $ x_r = \mathscr{G}_r(\mathcal{C}_1,\mathcal{C}_2,\mathcal{H}_r) $.  
	\end{enumerate}
\end{Definition}
This means each sender $ S_s $ encodes its known messages to a $ p_{s}$-bit sub-codeword, for some non-negative integer $ p_s $. We assume that each receiver $ r $ receives sub-codewords from both senders without any noise, and decodes $x_r $ from the received sub-codewords and $ \mathcal{H}_r $. The sub-codewords $ (\mathcal{C}_1,\mathcal{C}_2) $ form an index code in TSUIC.

Now we define the aggregate normalized codelength, which measures the performance of a code $(\mathcal{C}_1, \mathcal{C}_2) $, in the following. 
\begin{Definition}[Broadcast rate or aggregate normalized codelength]
	The broadcast rate of an index code (with a single sender or two senders) is the total number of transmitted bits (if two senders, then it is a sum of transmitted bits by both senders) per received message bit. In \textnormal{TSUIC}, it is denoted by $  \ell_{\mathsf{TSUIC}}\triangleq \frac{(p_{1}+p_{2})}{t}$ for an index code ($ \{\mathscr{F}_s\},\{\mathscr{G}_r\} $). The broadcast rate is also referred to as the aggregate normalized codelength of the index code. We say that $ \ell$ is \emph{achievable} for a \textnormal{UIC} problem if there exists an index code of normalized length $ \ell$.  
\end{Definition}

For the rest of the paper, we refer to normalized codelength simply as codelength.

\begin{Definition} [Optimal broadcast rate] \label{def:boradcastrate}
	The optimal broadcast rate for a given index-coding problem with $ t $-bit messages is $ \beta_t\triangleq  \underset{\mathcal{E}}{\min}\ \ell $, where $ \mathcal{E}=\mathscr{F} $ for \textnormal{SSUIC} and $ \mathcal{E}=\{ \mathscr{F}_s \}$ for \textnormal{TSUIC}. The optimal broadcast rate over all $ t $ is defined as $ \beta \triangleq  \underset{t}{\mathrm{inf}}\ \beta_{t}=\underset{t\rightarrow \infty}{\lim}\ \beta_{t}$. The limit exists and is equal to the infimum due to the subadditivity of $t\beta_{t}=p_1+p_2$ and Fekete's lemma~\cite{fekete}.  
\end{Definition}
\begin{Remark}
	With the (optimal) broadcast rate as a performance metric, we can treat \textnormal{SSUIC} as a special case of \textnormal{TSUIC} when $\mathcal{M}_1=\mathcal{M}$ or $ \mathcal{M}_2=\mathcal{M}$. Furthermore, for this case, the sender with $ \mathcal{M} $ alone will be responsible for fulfilling the demands made by all receivers.
\end{Remark}

\subsection{Representation of the receivers' side-information and the senders' message setting in \textnormal{TSUIC} problems}

An index-coding problem can be modeled by graphs, which are defined as follows: 
\begin{Definition} [Directed graphs and undirected graphs]
	A directed graph is an ordered pair $D =(V(D),A(D))$, where $ V(D)$ is a set of vertices, and $ A(D) $, usually called an arc set, is a set of ordered pairs of vertices. An undirected graph is an ordered pair $G =(V(G),E(G))$, where $ V(G) $ is a set of vertices, and $ E(G) $, usually called an edge set, is a set of unordered pairs of vertices. 
\end{Definition}
From now on in this paper, we call directed graphs simply digraphs, and undirected graphs simply graphs.

The receivers' message setting of a UIC problem is represented by a side-information digraph $D =(V(D),A(D))$, where $ V(D)=\{1,2,\dotsc,N\} $ represents the $ N$ receivers, and the arc set $  A (D) $ represents the side-information available at each receiver. More precisely, an arc $ (i,j) \in A (D) $ exists from vertex $ i $ to vertex $ j$ if and only if receiver~$ i $ has message $ x_j $ (the message requested by receiver $ j $) in its side-information. So, in a side-information digraph, $\mathcal{H}_i \triangleq \{x_j: j\in N_D^+(i)\} $, where $ N_D^+ (i) $ is the out-neighborhood of a vertex $ i $ in $ D $. In this paper, for convenience, a receiver $ i $ is also referred to as a vertex $ i $, and vice versa. We also use the compact form of representation of an instance of UIC problems as used by Arbabjolfaei et al.~\cite{composite}, where a sequence $ (i|N_D^+ (i)) $, for all $ i\in V(D) $, represents a UIC problem.

In TSUIC, $ S_1 $ (sender one) encodes the messages in $ \mathcal{M}_1 $, and $ S_2 $ (sender two) encodes the messages in $ \mathcal{M}_2 $. In general, each sender has \emph{private messages} and \emph{common messages} defined as follows: Let $ \mathcal{P}_1 \triangleq \mathcal{M}_1\setminus\mathcal{M}_2$ and $ \mathcal{P}_2 \triangleq \mathcal{M}_2\setminus\mathcal{M}_1$ be the set of private messages at senders $ S_1 $ and $ S_2 $, respectively, and $ \mathcal{P}_3 \triangleq \mathcal{M}_1\cap \mathcal{M}_2$ be the set of common messages at both senders. Now for a given side-information digraph $ D $, without loss of generality, we define the following sub-digraphs induced by the following vertex subsets that partition $V(D) $: For $ i\in \{1,2,3\} $, let $ D_i $ be the sub-digraph of $ D $ induced by vertices $ \{j:x_j\in \mathcal{P}_i \} $. We refer to $ D_1 $, $ D_2 $ and $ D_3 $ as per this definition throughout this paper unless stated otherwise. From the definition, it is clear that $ D_1 $, $ D_2 $ and $ D_3 $ are the three sub-digraphs of $ D $ such that $ V(D)=V(D_1)\cup V(D_2)\cup V(D_3) $ and $V(D_i)\cap V(D_k)=\emptyset $ for any $ i\neq k $, $ i,k\in \{1,2,3\} $. In TSUIC, the senders are limited to transmit only their messages, and this limitation is defined formally as a constraint due to the two senders as follows:
\begin{Definition} [Constraint due to the two senders] \label{def:senderconstraint}
	The constraint due to the two senders is the following: Whilst encoding, any two private messages $ x_i\in \mathcal{P}_1 $ and $ x_j\in \mathcal{P}_2 $ should not be encoded together (with or without other messages) to construct one coded symbol, or alternatively any two-sender index code can be written as $ (\mathcal{C}_1,\mathcal{C}_2) $ such that $\mathcal{C}_1= \mathscr{F}_{1}(\mathcal{M}\setminus \mathcal{P}_2 )$ and $\mathcal{C}_2= \mathscr{F}_{2}(\mathcal{M}\setminus \mathcal{P}_1)$. 
\end{Definition}

In TSUIC, to reflect the senders' message setting, we introduce an undirected graph, denoted by $ G_o=(V(G_o),E(G_o))$, that is constructed in the following way:
(i) $ V(G_o)=V(D)$, and (ii) for all $ i,j\in V(G_o) $, an undirected arc, i.e., an edge $ (i,j)\in E(G_o)$ exists if and only if $x_i\in \mathcal{P}_1 $ and $ x_j\in \mathcal{P}_2$, or vice versa. This means, there is an edge connecting two vertices in $ G_o $ if and only if no sender has both the corresponding messages. We call the graph $ G_o $ the \emph{sender-constraint} graph. 

As a TSUIC problem is described by $ D $ and $ G_o $, it is represented by $ (D,G_o) $ in this paper. For a given $ (D,G_o) $, let $ \ell(D,G_o) $ denote the index codelength, $ \beta_t(D,G_o)$ and $\beta(D,G_o)$ represent the optimal broadcast rate for a fixed $ t $, and over all $ t $, respectively. $ \ell(D) $, $ \beta_t(D)$ and $\beta(D)$ are the respective terms used for single-sender problems. 
\section{A new way of classifying TSUIC problems and the main results} \label{sec:interaction}
In a TSUIC problem, if there is no common message, i.e., $ \mathcal{P}_3=\emptyset$, then in our earlier work, we have proved that the problem is equivalent to two separate SSUIC problems (\cite[Theorem~1]{ourpaper3}). However, if $ \mathcal{P}_3\neq \emptyset$, then the problem is less well understood. We propose to tackle this problem by dividing it into three sub-problems based on the type of messages at the senders (whether they are common or private), and then study the interactions among these sub-problems due to the side-information present at the receivers. In this way, we can devise the structural characteristics of TSUIC problems. For a given problem $ D $, three sub-problems based on the type of messages are $ D_1 $, $ D_2 $ and $ D_3 $. The side-information present at receivers of one sub-problem about messages requested by receivers of other sub-problems are formally referred to as an interaction between those sub-problems, defined in the following. We will see that this allows us to derive $ \beta_{t}(D,G_o) $ in terms of the single-sender characterizations $ \{\beta_{t}(D_i): i\in \{1,2,3\}\} $ for a number of TSUIC instances.

\begin{figure}[t!]
	\centering
	\subfloat[]{\includegraphics[width=6.2cm, keepaspectratio]{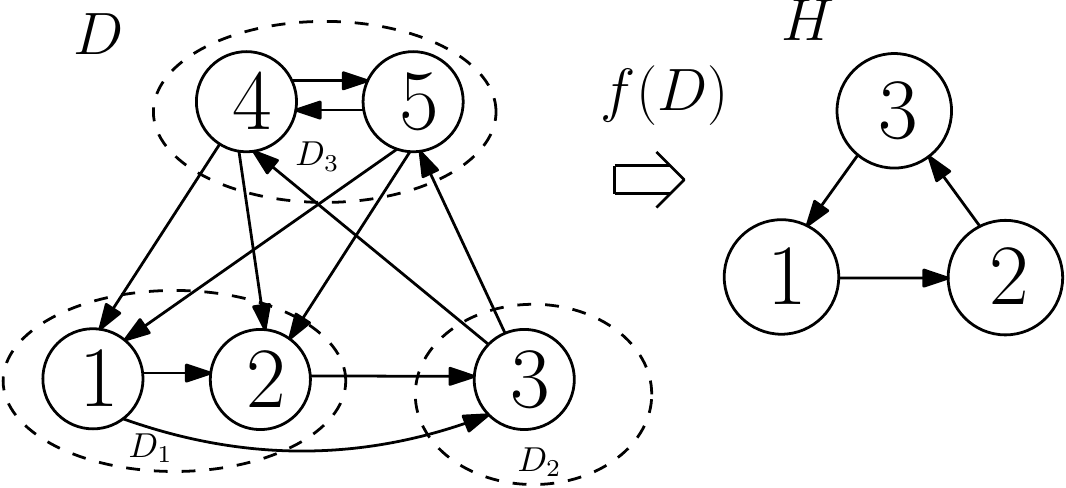}%
		\label{fig:fullyparticipated}}				
	\hfil
	\subfloat[]{\includegraphics[width=6.3cm, keepaspectratio]{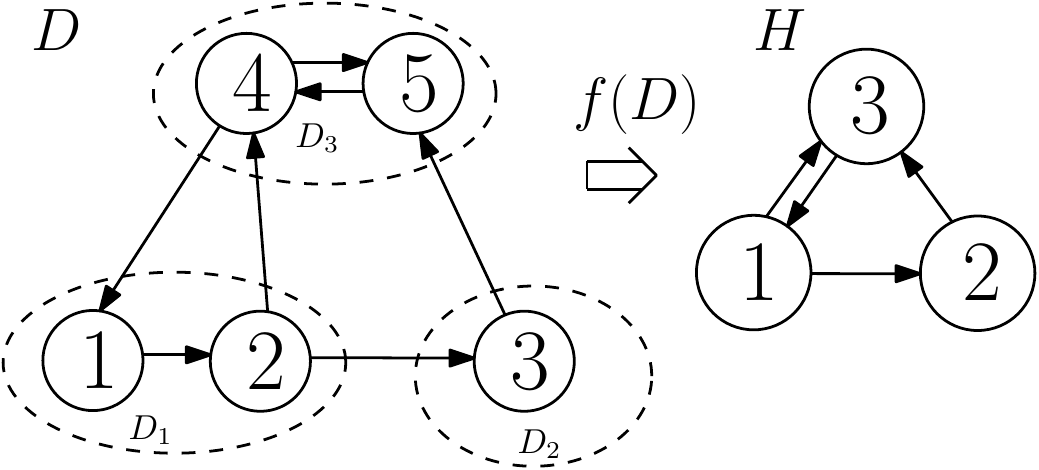}%
		\label{fig:partiallyparticipated}}
	\caption{ (a) An example of the fully participated interaction between $ D_1 $, $ D_2 $ and $ D_3 $ of a side-information digraph $ D$ along with the digraph $ H $ (having three vertices) obtained by mapping $ D $ by the function $ f $, and (b) an example of the partially participated interaction between $ D_1 $, $ D_2 $ and $ D_3 $ of another side-information digraph $D$ along with the digraph $ H $ obtained by mapping $ D $ by the function $ f $.}
	\label{hgraph1}	 \vspace{-2ex}
\end{figure}

\subsubsection{Interactions between $ D_1 $, $ D_2 $ and $ D_3 $}

Arcs between $ V(D_1) $, $ V(D_2) $ and $ V(D_3) $, each originating from some vertices of $ V(D_i) $, $ i\in \{1,2,3\} $, and terminating at some vertices of $ V(D)\setminus V(D_i) $ in $ D $ are called an interaction between $ D_1 $, $ D_2 $ and $ D_3 $. It is called a \emph{fully-participated} interaction between $ D_1 $, $ D_2 $ and $ D_3 $ if and only if we have the following:
If there exists an arc from a vertex of $ D_i $ to a vertex of $ D_j$ for any $ i,j\in \{1,2,3\}, i\neq j $, then $ V(D_j)\subseteq N^+_{D}(r)$ for every $ r\in V(D_i) $. In other words, all the vertices of the sub-digraph $ D_i $ interact in the same way to all the vertices of the sub-digraph $ D_j $. For an example of a fully-participated interaction see Figure~\ref{fig:fullyparticipated}.
If an interaction between the sub-digraphs is not a fully-participated interaction, then it is called a \emph{partially-participated} interaction among the sub-digraphs of the digraph. For example of a partially-participated interaction see Figure~\ref{fig:partiallyparticipated}. 
For the sub-digraphs of $ D $, if some vertices in $ V(D_i) $ have out-going arcs to some vertices in $ V(D_j) $, $ i\neq j,\ i,j\in \{1,2,3\} $, then it is denoted as $ D_i \rightarrow D_j $. If we write $ D_i \rightleftarrows D_j $, then it means $ D_i\rightarrow D_j $ and $ D_j\rightarrow D_i$. These representations are used for the indication of interaction, which does not explicitly specify the type of interactions.

\subsubsection{A compact representation of interactions}

For simplicity, an interaction between the sub-digraphs $ D_1 $, $ D_2 $ and $ D_3 $ of $ D $ can be viewed as an interaction between three vertices, where each vertex represents one of the sub-digraphs. In this regard, we define a function that maps a digraph $ D $ (with its sub-digraphs $D_1 $, $ D_2 $ and $ D_3 $) to a digraph having three vertices, denoted $H $, in the following:
$ f:D\rightarrow H $ such that (i) all the vertices in $ V(D_i) $ are mapped to a single vertex $ i $ of $ H$, so $ V(H)=\{1,2,3\} $, and (ii) $ (i,j)\in A(H) $ if and only if there exist an arc $ (u,v)\in A(D) $ for some $ u\in V(D_i) $ and some $ v\in V(D_j) $. For example see Figure~2. By referring to the definition of the fully or partially participated interaction, one can find that for a given $ D_1 $, $ D_2 $ and $ D_3 $ of $ D $, we can retrieve $ D $ by observing $ f(D) $ if $ D $ has a fully-participated interaction among the sub-digraphs, but this is not true if $ D $ has a partially-participated interaction among the sub-digraphs. Observe that for any $ D $, $ D_i\rightarrow D_j $, if and only if $ i\rightarrow j $ in $f(D)$ (i.e., $ H $).
\begin{figure}[t!]
	\centering
	\includegraphics[width=0.7\linewidth]{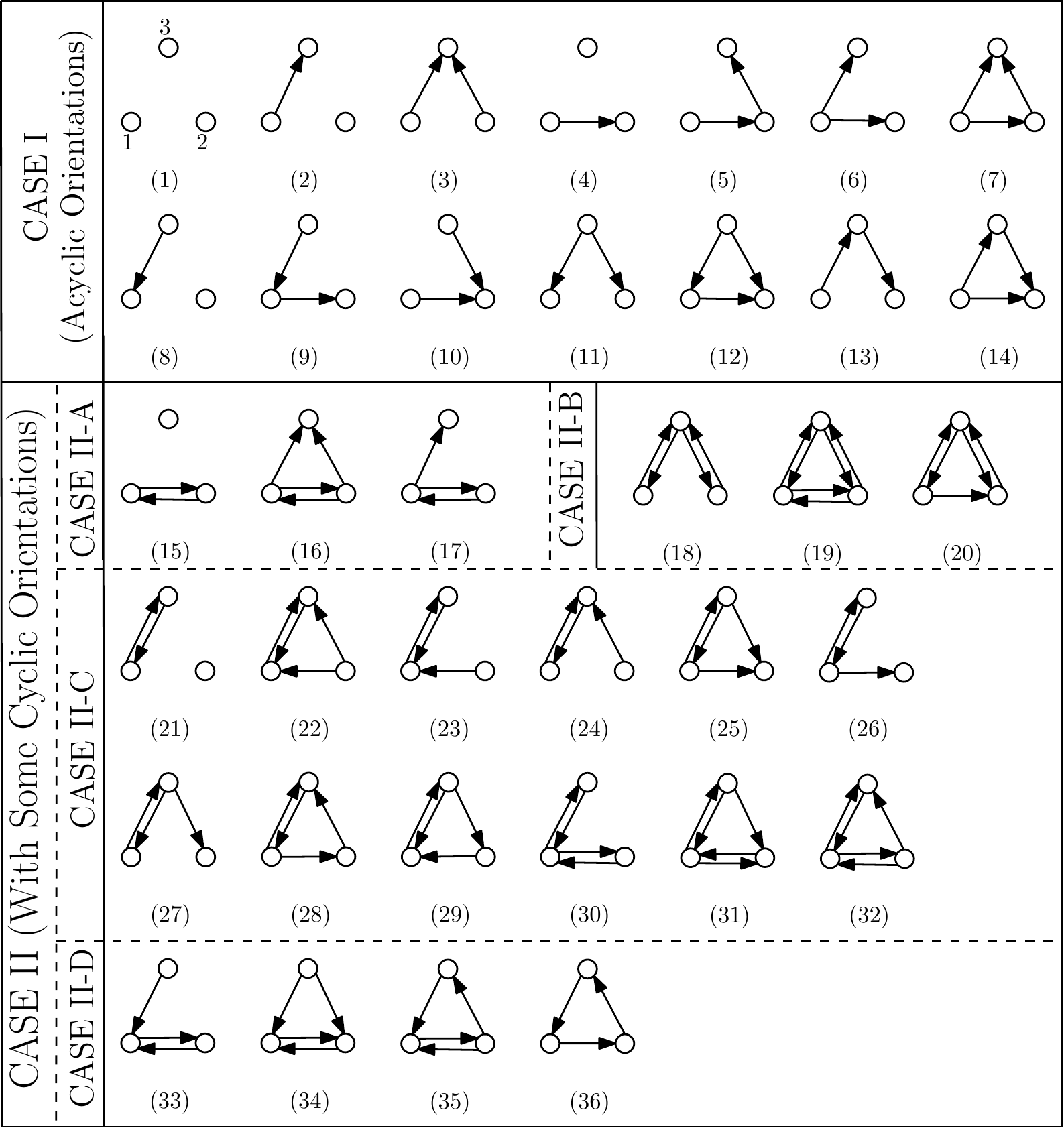}
	\caption{All unique interactions among the vertices of $ H $. The digraph of a number $ i $ is labeled by $ \mathrm{H}_i$, $ i \in \{1,2,\dotsc,36\} $. For example, the digraph of the number $ 30 $ is labeled $ \mathrm{H}_{30}$.}
	\label{fig:allgraphscasepart1} \vspace{-2ex}
\end{figure}

\subsubsection{A classification of the interactions}

Considering the digraph $ H $, we get a total of 64 possible cases of the orientation of arcs among its vertices. As the vertices $ 1 $ and $ 2 $ of $ H $ can be swapped because we can interchange $ D_1 $ and $ D_2 $ (by swapping the labels of the senders), we get 36 unique cases (out of 64 cases) of interactions between the vertices of $ H $. 
Now depending upon the type of orientation of arcs among the vertices of $ H $, we classify all unique cases into two categories: (i) CASE~I --- Acyclic orientation (14 cases in total), and (ii) CASE~II --- with some cyclic orientation (22 cases). CASE~II is further classified into smaller sub-cases II-A, II-B, II-C, and II-D. Refer to Figure~\ref{fig:allgraphscasepart1} for details, where each digraph of $ H$ is labeled $ \mathrm{H}_i $ for $ i\in \{1,2,\dotsc,36\} $. Note that an interaction between $ D_1 $, $ D_2 $ and $ D_3 $ of $ D $ defines arcs between them (not within the sub-digraph), and the cases of interactions (acyclic or cyclic) are defined with respect to the orientation of the arcs between the sub-digraphs. In this paper, a fully-participated interaction and a partially-participated interaction between $ D_1 $, $ D_2 $ and $ D_3 $ of $ D $ are called a \emph{cyclic-fully-participated} interaction and a \emph{cyclic-partially-participated} interaction between the sub-digraphs, respectively, if and only if $ f(D) $ has some cycles (for example, see CASE~II in Figure~\ref{fig:allgraphscasepart1}).

\subsubsection{Main results} 
For SSUIC, Arbabjolfaei and Kim \cite[Prop.~1]{structuralproperties1} argued that using structural properties can reduce the number of problems that need to be studied. This paper investigates the structural characteristics of TSUIC problems for the same purpose by studying the interactions among $ D_1 $, $ D_2 $ and $ D_3 $ of $ D $. Moreover, structural properties can be used to determine the criticality/non-criticality of arcs in TSUIC as in its SSUIC counterpart \cite{criticalgraphs,criticalarcs}. An arc is said to be critical if removing the arc strictly increases the optimal broadcast rate. 

This paper analyzes all cases of fully-participated and some cases of partially-participated interactions between $ D_1 $, $ D_2 $ and $ D_3 $ of $ D $, and establishes their optimal broadcast rates ($ \beta(D,G_o) $ and $ \beta_t(D,G_o) $) as a function of the optimal broadcast rates of $ D_1 $, $ D_2 $ and $ D_3 $ for TSUIC. For fully-participated interactions, the results are summarized in Table~\ref{table:1}. Furthermore, similar results are presented for $ D $ whose $ f(D) $ is of CASE~I and Case~II-A, and it has partially-participated interactions between the sub-digraphs (refer to Theorem~\ref{theorem:acyclic} and Theorem~\ref{theorem:caseIIA}).
The results are established by utilizing existing SSUIC's results and our proposed coloring of confusion graphs for TSUIC, which we discuss in the subsequent sections.

\begin{table}[t]
	\centering
	\caption{Summary of the results for any $ D $ with fully-participated interactions between $D_1 $, $ D_2 $ and $ D_3 $ in TSUIC.}
	\includegraphics[width=1\linewidth,keepaspectratio]{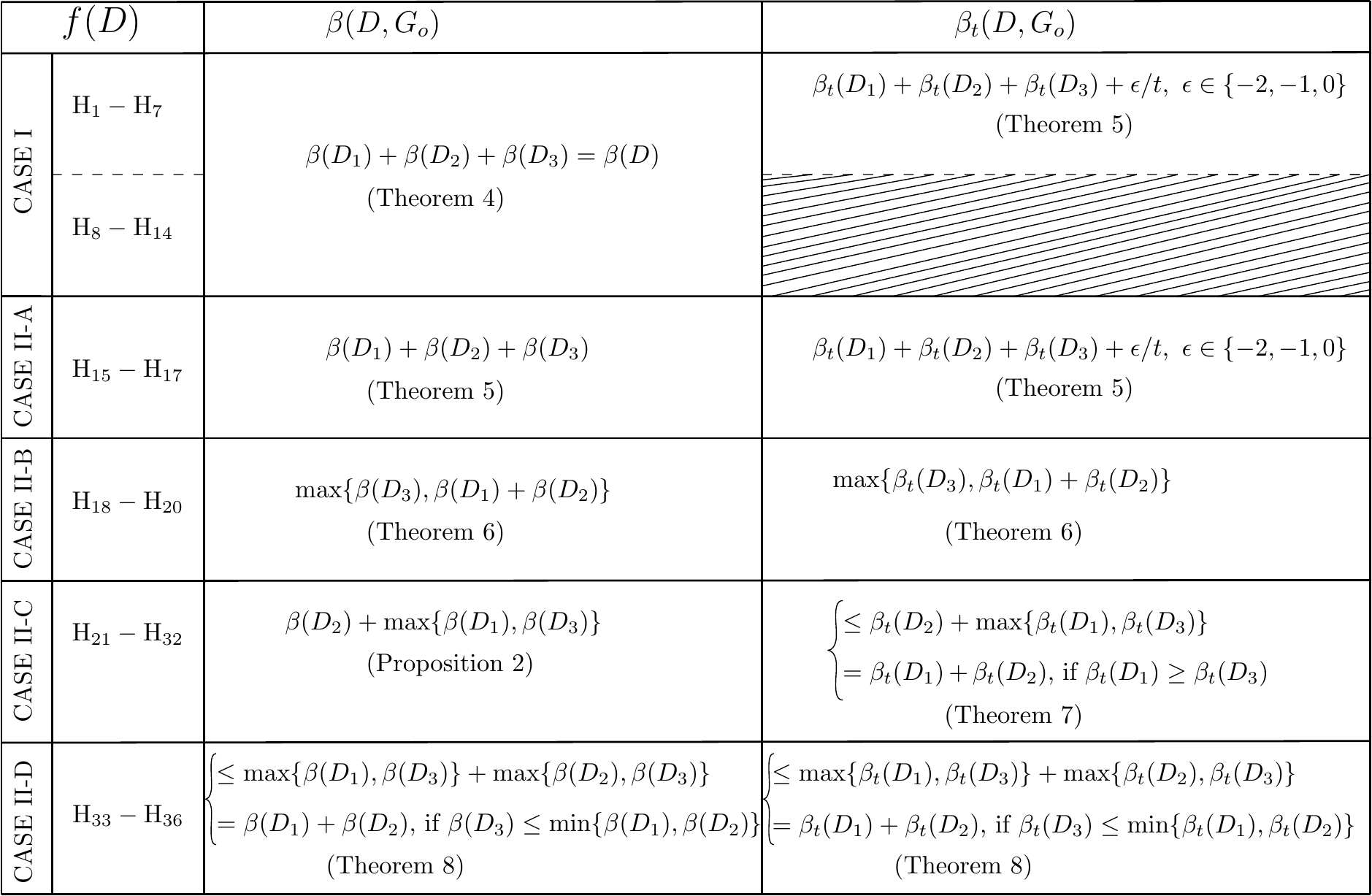}
	\label{table:1} \vspace{-2ex}
\end{table}

\section{Confusion graphs and their coloring} \label{confusiongraphcoloring}
\subsection{Confusion graphs}
For an index-coding problem modeled by a side-information digraph $ D $ with $ N $ vertices, two realizations of $ N $ messages, say, $ u^N=(u_1, u_2, \dotsc, u_N)$ and $ v^N=(v_1, v_2, \dotsc, v_N)$, are said to be confusable at a vertex (receiver) $ r\in \{1,2,\dotsc,N\} $, if $u_r\neq v_r $ and $ u_{i}=v_{i} $ for all $ i\in N^+_{D}(r) $, where, by definition, $u_{j}, v_{j} \in \{0,1\}^{t}$ for all $ j\in \{1,2,\dotsc,N\} $. We say that two tuples are confusable if they are confusable at some receiver $ r $. Clearly, in an index coding, we cannot encode message tuples that are confusable to the same codeword; otherwise one of the receivers may not always decode its requested message successfully. The confusability among all possible $ N $-tuples of messages (each message having $ t $ bits) for an index-coding problem is represented by a graph called a \emph{confusion graph}, defined as follows:		
\begin{Definition}[Confusion graph]
	The confusion graph, denoted $ \Gamma_t(D)=(V(\Gamma_t(D)),E(\Gamma_t(D))) $, of a side-information digraph $ D $ with $ N $ vertices and $ t $-bit messages is an undirected graph with the following:
	\begin{enumerate}
		\item [(i)] $V(\Gamma_t(D))=\{u^N: u^N\in \{0,1\}^{t\times N}\}$, and
		\item [(ii)] $ E(\Gamma_t(D))=\{(u^N,v^N):u^N,v^N \in V(\Gamma_t(D)),\ \text{and}\ u^N\ \text{and}\ v^N\ \text{are confusable}  \}$.
	\end{enumerate}
\end{Definition}

\subsection{A review of confusion graph coloring for  \textnormal{SSUIC}}
Before proposing a notion of coloring for TSUIC, we first recall the standard definition of the graph coloring in the following:
\begin{Definition} [Graph coloring and Chromatic number]
	A proper graph coloring of a graph $ G $ is an onto function $ J:V(G)\rightarrow\mathcal{J} $, where $ \mathcal{J} $ is a set of colors, in such a way that if $ i $ and $ j $ are adjacent vertices of $ G $, then $ J(i)\neq J(j) $. The minimum number of colors over all possible proper coloring of a graph $ G $ is called the chromatic number of $ G $, and it is denoted by $ \goodchi(G) $.
\end{Definition}

Consider coloring a confusion graph $\Gamma_t(D)$ with a set of colors $ \mathcal{J}$. Now we get a family of \emph{sets of independent vertices} where all vertices belonging to one set are assigned with the same color in the graph coloring. Here a set of independent vertices refers to a vertex set where any pair of vertices are not connected by an edge in $\Gamma_t(D)$, and we call such a set an \emph{independent vertex set}. The tuples representing vertices within an independent vertex set are not confusable, and hence they can be coded into the same codeword. Assigning each independent vertex set (whose vertices are all colored by a unique color) a unique codeword provides us a valid index code having $|\mathcal{J}|$ codewords. Thus there exists a bijective mapping $ I: \mathcal{J}\rightarrow \mathcal{C}$, where $ \mathcal{C} $ is an index code (or a set of codewords that satisfies the demands made by all receivers). We know that $ \goodchi(\Gamma_t(D))=\underset{J}{\min}\ |\mathcal{J}| $. In SSUIC, it is shown that the optimal broadcast rate of an index-coding problem $ D $ with $ t $-bit messages can be obtained by using confusion graphs. This is stated in the following theorem.

\begin{Theorem} (Alon et al. \cite[Th.~1.1]{fractionalchromatic}, Arbabjolfaei and Kim \cite[Prop.~1]{structuralproperties1})   \label{theorem:1}
	The optimal broadcast rate for a \textnormal{SSUIC} problem with $ t $-bit messages is 
	\begin{equation} \label{eq:betat}
	\beta_t(D)= \frac{\lceil \log_2 \goodchi(\Gamma_t(D)) \rceil}{t}. 
	\end{equation}
\end{Theorem}

The notion of confusion graphs has been considered in the index coding literature, and it has been shown to be an effective tool for proving important results, for example, Bar-Yossef et al.~\cite{maisbound}, Alon et al.~\cite{fractionalchromatic}, and Arbabjolfaei et al.~\cite{structuralproperties1} in their respective works, referred to the confusion graph for the proof of results related to the odd hole and the odd anti-hole~\cite{maisbound}, the gap between $ \beta $ and $ \beta_{t=1}$ of hypergraphs~\cite{fractionalchromatic}, and the structural properties of the index-coding problems~\cite{structuralproperties1}, respectively.

\subsection{Proposed confusion graph coloring for  \textnormal{TSUIC}} \label{sec:C}
The confusion graph, which is only a function of the side-information graph, does not depend on the number of senders. Its coloring function described above for SSUIC may not lead to an index code for TSUIC because of the constraint due to the two senders. In this work, we propose a way of coloring the confusion graphs in TSUIC, which we call two-sender graph coloring.
Before presenting a formal definition, we first define some notations that will be used in the remainder of this paper, unless stated otherwise.
\begin{enumerate} 
	\item Without loss of generality, we assume $ x_1,x_2,\dotsc,x_{n_1} $ to be the messages requested by vertices in $ V(D_1)$, $ x_{n_1+1},x_{n_1+2},\dotsc,x_{n_1+n_2}$ the messages requested by vertices in $ V(D_2) $, and $ x_{n_1+n_2+1},x_{n_1+n_2+2},\dotsc,x_{n_1+n_2+n_3}$ the messages requested by vertices in $ V(D_3) $ with $N=n_1+n_2+n_3$.
	\item Indices $ i,i_1,i_2\in \{1,2,\dotsc,2^{t n_1}\} $, $ j,j_1,j_2\in \{1,2,\dotsc,2^{t n_2}\} $ and $ k, k_1,k_2\in  \{1,2,\dotsc,2^{t n_3}\}$ are used in the representation of possible realizations of words of $ tn_1 $, $ tn_2 $ and $ tn_3 $ bits, respectively. For convenience, we use three indices (e.g., $ i,i_1,i_2 $) for the same set of numbers, where the first index (e.g., $ i $) is used for a general case, and the remaining two indices (e.g., $ i_1 $ and $ i_2 $) are used to indicate any two words within the group of words.		\item We group the bits associated with the messages requested by vertices of $ D_{i'} $, $ i'\in \{1,2,3\} $. Within each group, each realization of the bits, i.e., each member in $ \{0,1\}^{t n_{i'}} $ is represented by a unique label $\textbf{b}_{D_{i'}}^{j'}$, $ j'\in \{ 1,2,\dotsc, 2^{t n_{i'}}\} $. The Figure~\ref{table:blockmap} in Appendix~\ref{append1} outlines each tuple $\textbf{b}_{D_{i'}}^{j'}$ for $ t=1 $. Each message tuple $ (x_1,\dotsc,x_N) $ realization can then be uniquely written as $ (\textbf{b}_{D_{1}}^{i},\textbf{b}_{D_{2}}^{j},\textbf{b}_{D_{3}}^{k}) $ for some $ i,j,k $.		
\end{enumerate}

\begin{Definition} [Two-sender graph coloring of $ \Gamma_t(D) $] \label{def:2scoloring}
	Let two onto functions $ J_1:\{0,1\}^{tn_1} \times \{0,1\}^{tn_3} \rightarrow \mathcal{J}_1$, and $J_2:\{0,1\}^{tn_2}\times \{0,1\}^{tn_3} \rightarrow \mathcal{J}_2$ be the coloring functions carried out by senders $ S_1 $ and $ S_2 $, respectively. A proper two-sender graph coloring of $ \Gamma_t(D) $ is an onto function $ J_o: \{0,1\}^{tn_1}\times \{0,1\}^{tn_2}\times \{0,1\}^{tn_3}\rightarrow \mathcal{J}_1\times \mathcal{J}_2 $ where $ J_o((\textbf{b}_{D_{1}}^{i},\textbf{b}_{D_{2}}^{j},\textbf{b}_{D_{3}}^{k}))=(J_1(\textbf{b}_{D_{1}}^{i},\textbf{b}_{D_{3}}^{k}), J_2(\textbf{b}_{D_{2}}^{j},\textbf{b}_{D_{3}}^{k})) $ such that if $ (\textbf{b}_{D_{1}}^{i_1},\textbf{b}_{D_{2}}^{j_1},\textbf{b}_{D_{3}}^{k_1}) $ and $ (\textbf{b}_{D_{1}}^{i_2},\textbf{b}_{D_{2}}^{j_2},\textbf{b}_{D_{3}}^{k_2}) $ are adjacent vertices of $ \Gamma_t(D) $, then $ J_o((\textbf{b}_{D_{1}}^{i_1},\textbf{b}_{D_{2}}^{j_1},\textbf{b}_{D_{3}}^{k_1}))\\ \neq J_o((\textbf{b}_{D_{1}}^{i_2},\textbf{b}_{D_{2}}^{j_2},\textbf{b}_{D_{3}}^{k_2}) ) $.
\end{Definition}

\begin{Remark}
	The two-sender graph coloring is not a $ b $-fold coloring that assigns a set of $ b $ colors to each vertex such that the color sets corresponding to two adjacent vertices do not share any color (refer to the definition of the fractional graph coloring \cite{structuralproperties1}). In our definition, the color sets can share colors, as long as the color vectors (i.e., ordered pairs) are different. 
\end{Remark}

\subsection{A few lemmas for the \textnormal{TSUIC} confusion graph coloring}
In the form of lemmas, we discuss two-sender graph coloring of $ \Gamma_t(D) $ in detail. Before this, we first assume the following: For any indices $ i',j'$, assume that $ c_{i'}$ and $ c_{j'} $ are any two distinct colors if $ i'\neq j'$, and let $ (c_{i'},c_{j'}) $ be an ordered pair of colors. Any two ordered pairs of colors, $ (c_{i'_1},c_{j'_1}) $ and $ (c_{i'_2},c_{j'_2}) $, are said to be different (or not equal) if and only if $i'_1\neq i'_2$ or $ j'_1\neq j'_2$ or both. If a color $ c_{i'} $ is associated to a sender $ S_s $, $ s\in \{1,2\} $, then we denote it by $ c_{i'}^s$.

In TSUIC, the two senders encode separately, so in the aforementioned definition, we need to assign an ordered pair of colors for each vertex, where the first color is associated with $ S_1 $ and the second color with $ S_2 $. Now we have the following lemmas. 

\begin{Lemma} \label{lemmaA}
	For any two distinct vertices $ u^N,v^N \in V(\Gamma_t(D)) $ that are labeled by $(\textbf{b}_{D_{1}}^{i_1},\textbf{b}_{D_{2}}^{j},\textbf{b}_{D_{3}}^{k})$ and $(\textbf{b}_{D_{1}}^{i_2},\textbf{b}_{D_{2}}^{j},\textbf{b}_{D_{3}}^{k})$, respectively, if $ (u^N,v^N)\in E(\Gamma_t(D)) $, then we must have $J_o(u^N)=(c^1_{i'_1},c^2_{j'_1})$ and $J_o(v^N)=(c^1_{i'_2},c^2_{j'_2})$ such that $ c^1_{i'_1}\neq c^1_{i'_2}$ and $ c^2_{j'_1}=c^2_{j'_2}$ for some indices $ i'_1,i'_2,j'_1,j'_2 $. 
\end{Lemma} 
\begin{IEEEproof}
	Since $ (u^N,v^N)\in E(\Gamma_t(D)) $, $ u^N$ and $v^N $ are confusable. Moreover, these two tuples are confusable only at some vertex in $ V(D_1) $. This is because the labels $(\textbf{b}_{D_{1}}^{i_1},\textbf{b}_{D_{2}}^{j},\textbf{b}_{D_{3}}^{k})$ and $(\textbf{b}_{D_{1}}^{i_2},\textbf{b}_{D_{2}}^{j},\textbf{b}_{D_{3}}^{k})$ of $ u^N$ and $v^N $, respectively, are different only in $ \textbf{b}_{D_{1}}^{i}$ sub-label (which is representing $ tn_1 $-bit tuples of the messages requested by vertices in $ V(D_1) $). Now for the sender $ S_2 $, which does not contain any message in $ \mathcal{P}_1 $ (messages requested by receivers in $ V(D_1) $), the coloring function $ J_2(\textbf{b}_{D_{2}}^{j},\textbf{b}_{D_{3}}^{k}) $ provides the same color to both vertices. Thus $ c^2_{j'_1}=c^2_{j'_2}$. On the other hand, for the sender $ S_1 $, which contains all messages in $ \mathcal{P}_1 $, it is necessary to have $ J_1(\textbf{b}_{D_{1}}^{i_1},\textbf{b}_{D_{3}}^{k})\neq J_1(\textbf{b}_{D_{1}}^{i_2},\textbf{b}_{D_{3}}^{k})$ because these two tuples $ (\textbf{b}_{D_{1}}^{i_1},\textbf{b}_{D_{3}}^{k})$ and $(\textbf{b}_{D_{1}}^{i_2},\textbf{b}_{D_{3}}^{k}) $ are confusable given that $(\textbf{b}_{D_{1}}^{i_1},\textbf{b}_{D_{2}}^{j},\textbf{b}_{D_{3}}^{k})$ and $(\textbf{b}_{D_{1}}^{i_2},\textbf{b}_{D_{2}}^{j},\textbf{b}_{D_{3}}^{k})$ are confusable at some receiver in $ V(D_1) $. Thus $ c^1_{i'_1}\neq c^1_{i'_2}$. 
\end{IEEEproof}
In a similar reasoning as in the above proof (of Lemma~\ref{lemmaA}), one can prove the following lemma:
\begin{Lemma} \label{lemmaB}
	For any two distinct vertices, $ u^N,v^N \in V(\Gamma_t(D)) $ such that they are labeled by $(\textbf{b}_{D_{1}}^{i},\textbf{b}_{D_{2}}^{j_1},\textbf{b}_{D_{3}}^{k})$ and $(\textbf{b}_{D_{1}}^{i},\textbf{b}_{D_{2}}^{j_2},\textbf{b}_{D_{3}}^{k})$, respectively, if $ (u^N,v^N)\in E(\Gamma_t(D)) $, then we must have $J_o(u^N)=(c^1_{i'_1},c^2_{j'_1})$ and $J_o(v^N)=(c^1_{i'_2},c^2_{j'_2})$ such that $ c^1_{i'_1}=c^1_{i'_2}$ and $ c^2_{j'_1}\neq c^2_{j'_2}$ for some indices $ i'_1,i'_2,j'_1,j'_2 $.
\end{Lemma} 
If $ u^N$ and $v^N $ are confusable at some vertices in $ V(D_1) $ and in $ V(D_2) $, then referring to Lemma~\ref{lemmaA} and~\ref{lemmaB}, we get the following:
\begin{Lemma} \label{lemmaC}
	For any two distinct vertices, $ u^N, v^N \in V(\Gamma_t(D)) $ such that they are labeled by $(\textbf{b}_{D_{1}}^{i_1},\textbf{b}_{D_{2}}^{j_1},\textbf{b}_{D_{3}}^{k})$ and $(\textbf{b}_{D_{1}}^{i_2},\textbf{b}_{D_{2}}^{j_2},\textbf{b}_{D_{3}}^{k})$, respectively, if $ (u^N,v^N)\in E(\Gamma_t(D)) $ due to confusion at some vertices in $ V(D_1) $ and in $ V(D_2) $, then we must have $J_o(u^N)=(c^1_{i'_1},c^2_{j'_1})$ and $J_o(v^N)=(c^1_{i'_2},c^2_{j'_2})$ such that $ c^1_{i'_1}\neq c^1_{i'_2}$ and $ c^2_{j'_1}\neq c^2_{j'_2}$ for some indices $ i'_1,i'_2,j'_1,j'_2 $. 
\end{Lemma} 

If $ u^N$ and $v^N $ are confusable at some vertices in $ V(D_3)$, then whilst coloring $ \Gamma_t(D) $ in two-sender graph coloring, it suffices to have a different color associated with any one of the senders because all the messages in $ \mathcal{P}_3 $ are contained by both senders $ S_1 $ and $ S_2 $. Thus we have the following lemma:
\begin{Lemma}  \label{lemmaD}
	For any two vertices, $ u^N, v^N \in V(\Gamma_t(D)) $ such that they are labeled by $(\textbf{b}_{D_{1}}^{i},\textbf{b}_{D_{2}}^{j},\textbf{b}_{D_{3}}^{k_1})$ and $(\textbf{b}_{D_{1}}^{i},\textbf{b}_{D_{2}}^{j},\textbf{b}_{D_{3}}^{k_2})$, respectively, if $ (u^N,v^N)\in E(\Gamma_t(D)) $, then we have $J_o(u^N)=(c^1_{i'_1},c^2_{j'_1})$ and $J_o(v^N)=(c^1_{i'_2},c^2_{j'_2})$ such that either $ c^1_{i'_1}\neq c^1_{i'_2}$, or $ c^2_{j'_1}\neq c^2_{j'_2}$, or both. 
\end{Lemma}
\section{The optimal broadcast rate for TSUIC}
For a TSUIC problem with $ t $-bit messages, we have the following theorem:   
\begin{Theorem} \label{theorem:2}
	The optimal broadcast rate for a \textnormal{TSUIC} problem with $ t $-bit messages is
	\begin{equation}
	\beta_t(D,G_o)= \underset{J_1,J_2}{\min}\ \frac{\lceil \log_2 |\mathcal{J}_1| \rceil+ \lceil \log_2 |\mathcal{J}_2|  \rceil }{t}.
	\end{equation} 
\end{Theorem}
\begin{IEEEproof}
	For $ s\in \{1,2\} $, consider $ J_s $, a coloring function of the sender $ S_s $, with a set of colors $\mathcal{J}_s $. A two-sender index code is obtained by $ S_1 $ mapping distinct colors in $ \mathcal{J}_1 $ to distinct sub-codewords, and $ S_2 $ mapping distinct colors in $ \mathcal{J}_2 $ to distinct sub-codewords. By definition, all confusable vertex pairs are assigned different codewords. Now for $ s\in \{1,2\} $, the sender $ S_s $ transmits $|\mathcal{J}_s| $ sub-codewords. Equivalently, $\lceil \log_2 |\mathcal{J}_s| \rceil $ bits are transmitted by $ S_s $. This is because the number of bits required to index $ |\mathcal{J}_s| $ colors are $ \lceil \log_2 |\mathcal{J}_s| \rceil $. Minimizing the sum $(\lceil \log_2 |\mathcal{J}_1| \rceil+ \lceil \log_2 |\mathcal{J}_2|  \rceil) $ over all coloring functions $ J_1 $ (of $ S_1 $) and $ J_2 $ (of $ S_2 $) per received message bits (i.e., $ t $), we get 
	\begin{equation} \label{eq:a} 
	\beta_t(D,G_o)\leq \underset{J_1,J_2}{\min}\ \frac{\lceil \log_2 |\mathcal{J}_1| \rceil+ \lceil \log_2 |\mathcal{J}_2|  \rceil }{t}. 
	\end{equation}
	From the definition of $ \beta_t(D,G_o) $, we have $ \beta_t(D,G_o)= \underset{\mathcal{E} }{\min}\ \frac{p_1+p_2}{t} $, so there exists a two-sender index code such that $ S_1 $ and $ S_2 $ transmit $ p'_1 $-bit and $ p'_2 $-bit sub-codewords, respectively, resulting in
	\begin{equation} \label{eq:b}
	\beta_t(D,G_o)= \frac{p'_1+p'_2}{t}. 
	\end{equation}
	Now for each sender $ S_s $, we know that there are at most $ 2^{p'_s}$ possible sub-codewords. Consider a bijective function that maps each sub-codeword to a color. A valid code must translate to a valid two-sender graph coloring. So, there exists a valid two-sender graph coloring such that $|\mathcal{J}'_1|\leq 2^{p'_1}$ and $|\mathcal{J}'_2|\leq 2^{p'_2}$, or equivalently, $ p'_1\geq \lceil \log_2 |\mathcal{J}'_1| \rceil $ and $ p'_2\geq  \lceil \log_2 |\mathcal{J}'_2| \rceil $ as both are non-negative integers. Substituting the inequalities of $ p'_1 $ and $ p'_2 $ in \eqref{eq:b}, wet get
	\begin{equation} \label{eq:d}
	\beta_t(D,G_o)\geq \frac{\lceil \log_2 |\mathcal{J}'_1| \rceil +\lceil \log_2 |\mathcal{J}'_2| \rceil }{t}.
	\end{equation}  
	Now we prove equality in \eqref{eq:a}. This is done by contradiction. Suppose that
	\begin{equation} \label{eq:c} 
	\beta_t(D,G_o)< \underset{J_1,J_2}{\min}\ \frac{\lceil \log_2 |\mathcal{J}_1| \rceil+ \lceil \log_2 |\mathcal{J}_2|  \rceil }{t}. 
	\end{equation}
	From \eqref{eq:d} and \eqref{eq:c}, we get 
	\begin{equation} \label{eq:e} 
	\frac{\lceil \log_2 |\mathcal{J}'_1| \rceil +\lceil \log_2 |\mathcal{J}'_2| \rceil }{t}< \underset{J_1,J_2}{\min}\ \frac{\lceil \log_2 |\mathcal{J}_1| \rceil+ \lceil \log_2 |\mathcal{J}_2|  \rceil }{t}, 
	\end{equation}
	and this leads to a contradiction. Thus $ \beta_t(D,G_o)= \underset{J_1,J_2}{\min}\ \frac{\lceil \log_2 |\mathcal{J}_1| \rceil+ \lceil \log_2 |\mathcal{J}_2|  \rceil }{t} $.  	
\end{IEEEproof}

We illustrate two-sender graph coloring of a confusion graph in TSUIC, and a mapping function that maps colors to codewords at each sender from the following example.
\begin{Example} \label{eg1}
	\begin{figure}[t!]
		\centering
		\subfloat[]{\includegraphics[width=5.5cm, keepaspectratio]{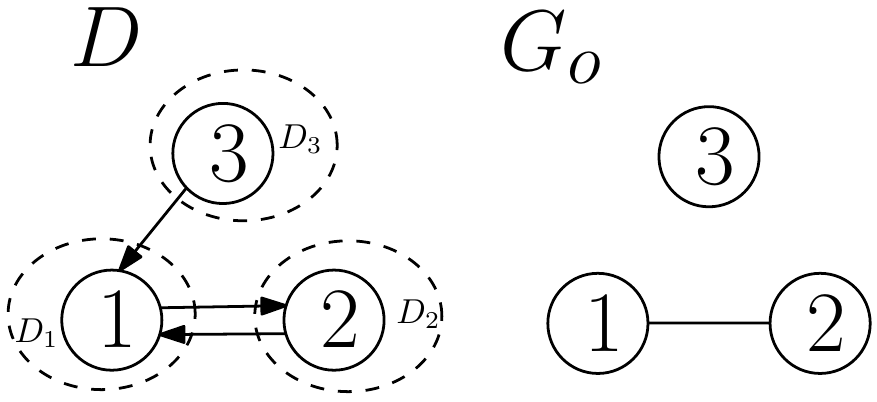}%
			\label{fig:example1}}				
		\hfil
		\subfloat[]{\includegraphics[width=7cm, keepaspectratio]{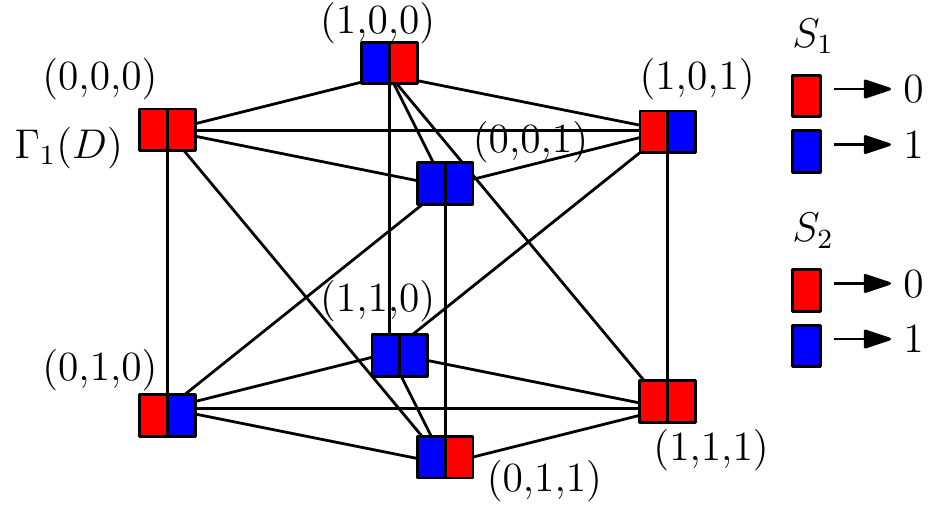}%
			\label{fig:example2}}
		\caption{ (a) A TSUIC problem $ (D,G_o) $, and (b) the confusion graph $ \Gamma_{1}(D) $, and its two-sender graph coloring.}
		\label{example} 
	\end{figure}
	Consider a \textnormal{TSUIC} problem $ (D,G_o) $ of the following: $ (1|2), (2|1), (3|1)$, and $ \mathcal{M}_1=\{1,3 \} $, $ \mathcal{M}_2=\{2,3\} $ with $ t=1 $. The problem is depicted in Figure~\ref{fig:example1}. We have $ V(D_1)=\{1\} $, $ V(D_2)=\{2\} $, $ V(D_3)=\{3\} $, and $ N=3 $. The confusion graph $ \Gamma_{1}(D) $ has $ 2^{N}=8 $ vertices labeled by all possible realizations of a word with three bits. In $ \Gamma_{1}(D) $, any two vertices are connected by an edge if the message tuples labeling the vertices are confused at some receiver. For example, $ (0,0,0) $ and $ (1,0,0) $ are connected by an edge because these two message tuples are confused at receiver $ 1 $. The confusion graph $ \Gamma_{1}(D) $ is depicted in Figure~\ref{fig:example2}. Now we perform two-sender graph coloring of the vertices of $ \Gamma_{1}(D) $. In two-sender graph coloring, each vertex of $ \Gamma_{1}(D) $ is assigned with an ordered pair of colors; the first color is always associated with $ S_1 $ and the second color is always associated with $ S_2 $, and we color the vertices as dictated by Lemma~\ref{lemmaA} to Lemma~\ref{lemmaD}. For example, consider $ (0,0,0) $ and $ (1,0,0) $. These two tuples are confused at receiver $ 1 $ (requesting $ x_1 $). As $ S_2 $, which does not know $ x_1 $, the tuples $ (0,0,0) $ and $ (1,0,0) $, which have the same second and third message bits, are treated as the same. Thus $ S_2 $ must assign the same color, say RED, to both the tuples. As $ S_1 $ knows $ x_1 $ and the tuples are confusable at receiver $ 1 $, it must assign two different colors, say RED and BLUE, to $ (0,0,0) $ and $ (1,0,0) $, respectively. In a similar way, we assign ordered pairs of colors to all vertices of $ \Gamma_{1}(D) $ as shown in Figure~\ref{fig:example2}. Altogether, one can get $ \mathcal{J}_1=\{\text{RED},\text{BLUE}\} $ and $ \mathcal{J}_2=\{\text{RED},\text{BLUE}\} $. Now we assume a mapping function that maps RED to 0 and BLUE to 1. So, we get $ (0,0,0)\rightarrow 00 $, $ (1,0,0)\rightarrow 10 $, $ (0,1,0)\rightarrow 01 $ and so on for the remaining tuples (vertices of $ \Gamma_t(D) $). Thus $ \{00,10,01,11\} $ are codewords of a valid two-sender index code for $(D,G_o) $, where each sender transmits a $ 1 $-bit sub-codeword for a message tuple, and the sum of bits to be transmitted by the two senders is two for each message tuple. Consequently, $ \beta_t(D,G_o)\leq 2$. Each sender has one private message, and that must be transmitted by that sender, so there must be at least one transmission by that sender. Thus $ \beta_t(D,G_o)\geq 2$. Altogether, we get $ \beta_t(D,G_o)=2$.            
\end{Example}




\subsection{Lower Bounds}
For any $ D $, we have $ \beta(D) \leq \beta_t(D) $ for all $ t $ (by definition). Since any index code for $ (D,G_o) $ is also an index code for $ D $, but the converse is not always true, so we have the following:
\begin{Lemma} [A simple lower bound] \label{simplebound}
	For any $ D $ and $ G_o $, $ \beta(D,G_o)\geq \beta(D) $. 
\end{Lemma}	

In TSUIC, each sender $ S_s $ transmits at least $ \beta(D_s)$, for $ s\in \{1,2\}$. We now provide a lower bound of the optimal broadcast rate for a TSUIC problem with $ t $-bit messages in the following lemma.
\begin{Lemma} [A lower bound] \label{lemma:1a}
	For any two-sender index-coding problem $ (D, G_o) $, $ \beta_t(D,G_o)\geq \beta_t(D_1)+\beta_t(D_2) $, and $ 	\beta(D,G_o)\geq \beta(D_1)+\beta(D_2) $.
\end{Lemma}
\begin{IEEEproof}
	For any two-sender index-coding problem $ (D, G_o) $, let $(D', G'_o)$ be its sub-problem induced by vertices $ V(D_1)\cup V(D_2) $. Observe that $ V(D') \cap V(D_3) =\emptyset$. Now we have $ \beta_t(D',G'_o)=\beta_t(D_1)+\beta_t(D_2) $ (\cite[Th.~1]{ourpaper3}). For any index-coding problem, its broadcast rate is always lower bounded by the broadcast rate of any sub-problem, so we get
	\begin{equation}\label{eq:AB}
	\beta_t(D,G_o)\geq \beta_t(D',G'_o)=\beta_t(D_1)+\beta_t(D_2).
	\end{equation} 
	We know that $ \underset{t\rightarrow \infty}{\lim}\ \beta_{t}(D_1)=\beta(D_1) $, $ \underset{t\rightarrow \infty}{\lim}\ \beta_{t}(D_2) =\beta(D_2)$ and $ \beta(D,G_o)= \underset{t\rightarrow \infty}{\lim}\  \beta_{t}(D,G_o)$ (by Definition~\ref{def:boradcastrate}). Now taking a limit $ t\rightarrow \infty $ on both sides in \eqref{eq:AB}, we get
	\begin{align}
	\beta(D,G_o)&\geq \beta(D_1)+\beta(D_2).
	\end{align}
\end{IEEEproof}

To compute the simple lower bound to the optimal broadcast rate of a given problem in TSUIC, we utilize the following SSUIC results by Arbabjolfaei and Kim \cite[Prop.~3, Th.~2, Th.~3]{structuralproperties1}.
\begin{Theorem} \label{theorem:betaD}
	In \textnormal{SSUIC}, for a side-information digraph $ D $ having two sub-digraphs $ D_a $ and $ D_b $ induced by vertices $ V(D_a) $ and $ V(D_b) $, respectively, such that $ V(D_a)\cup V(D_b)=V(D)$ and $ V(D_a)\cap V(D_b)=\emptyset $, we have
	\begin{enumerate} 
		\item [(i)] $ \beta(D)= \beta(D_a)+\beta(D_b) $ if there is (i) no interaction between $ D_a $ and $ D_b $ (i.e., no $ D_a~\rightarrow~D_b $ and $ D_b~\rightarrow~D_a $), or (ii) a one way interaction (either partially or fully participated) between $ D_a $ and $ D_b $, i.e., either $ D_a~\rightarrow~D_b$ or $ D_b~\rightarrow~D_a$, but not both and  
		\item [(ii)] $ \beta(D)= \text{max}\{ \beta(D_a),\beta(D_b)\} $ if there is a fully participated both way interaction between $ D_a $ and $ D_b $ (i.e., fully participated $ D_a\rightleftarrows D_b $). 
	\end{enumerate}
\end{Theorem}


\subsection{Optimal broadcast rates for CASE~I and CASE~II-A: The arcs between $ D_1 $, $ D_2 $ and $ D_3 $ are not critical in asymptotic regime in the message size.}
For a digraph $ D $ whose $ f(D) $ belongs to a digraph of CASE~I and CASE~II-A (see Figure~\ref{fig:allgraphscasepart1}), we have the following results.	 
\begin{Theorem}[CASE~I]\label{theorem:acyclic}
	For any $ D $ having any interaction (i.e., either fully participated or partially participated) between its sub-digraphs $ D_1 $, $ D_2 $ and $ D_3 $, if $ f(D)\in \{ \mathrm{H}_1,\mathrm{H}_2,\dotsc,\mathrm{H}_{14}\} $ (i.e., a digraph $ H $ of CASE~I in Figure~\ref{fig:allgraphscasepart1}), then $ \beta(D,G_o)=\beta(D_1)+\beta(D_2)+\beta(D_3)=\beta(D) $.
\end{Theorem}
\begin{IEEEproof}
	Referring to the definition of the mapping function $ f $ (in Section~\ref{sec:interaction}), we know that for any $ D $ if $ f(D)\in \{ \mathrm{H}_1,\mathrm{H}_2,\dotsc,\mathrm{H}_{14}\} $, then the interaction between $ D_1 $, $ D_2 $ and $ D_3 $ of $ D$ are acyclic. Thus one can arrange $ D_1 $, $ D_2 $ and $ D_3 $ in a sequence such that there is no arc between $ D_1 $, $ D_2 $ and $ D_3 $ in a backward direction. Without loss of generality, let the sequence be $ D_1 $, $ D_2 $ and $ D_3 $. Now for $ D$, referring to Theorem~\ref{theorem:betaD}, we get 
	\begin{equation}
	\beta(D)=\beta(D_1\cup D_2)+\beta(D_3) =\beta(D_1)+\beta ( D_2)+\beta(D_3).
	\label{eq:21}
	\end{equation}
	From Lemma~\ref{simplebound}, we have
	\begin{equation}
	\beta(D,G_o)\geq \beta(D).
	\label{eq:beta}
	\end{equation} 
	Now from \eqref{eq:21} and \eqref{eq:beta}, we get
	\begin{equation}
	\beta(D,G_o)\geq \beta(D_1)+\beta(D_2)+\beta(D_3).
	\label{eq:3}
	\end{equation}   
	
	In TSUIC, if we consider the sub-digraphs $ D_1 $, $ D_2 $ and $ D_3 $ separately, then their respective source constraint graphs are the sub-graphs of $ G_0 $ induced by vertices $ V(D_1) $, $ V(D_2) $ and $ V(D_3) $, denoted $ G_0^1 $, $ G_0^2 $ and $ G_0^3 $, respectively. These sub-graphs are edgeless graphs, and thus one can get $\beta(D_1,G_0^1)=\beta (D_1)$, $\beta(D_2,G_0^2)=\beta (D_2)$ and $\beta(D_3,G_0^3)=\beta (D_3)$. We know that the optimal broadcast rate of a side-information digraph is always less than or equal to the sum of the optimal broadcast rates of its sub-digraphs, so 
	\begin{align}	\label{eq:4}
	\beta(D,G_o)&\leq  \beta(D_1,G_0^1)+ \beta(D_2,G_0^2)+\beta(D_3,G_0^3)\nonumber \\
	&\leq  \beta(D_1)+\beta(D_2)+\beta(D_3).
	\end{align}
	From \eqref{eq:21}, \eqref{eq:3} and \eqref{eq:4}, we get $ \beta(D,G_o)=\beta(D_1)+\beta(D_2)+\beta(D_3)=\beta(D)$.    
\end{IEEEproof}	
\begin{figure} [t]
	\centering
	\includegraphics[height=2.5cm,keepaspectratio]{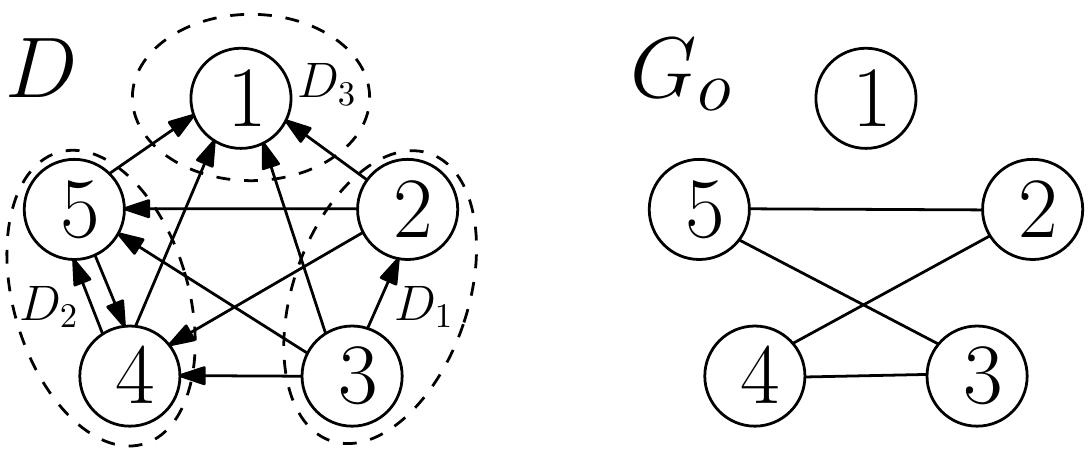}
	\caption{A given side-information digraph $ D $ such that $ f(D)=\mathrm{H}_{7} $, and a source-constraint graph $ G_o $ (for $ \mathcal{M}_1=\{1,2,3\} $ and $ \mathcal{M}_2=\{ 1,4,5\} $). Moreover, we have $ V(D_1)=\{2,3\} $, $ V(D_2)=\{4,5\} $ and $ V(D_3)=\{ 1\} $. From Theorem~\ref{theorem:acyclic}, we get $ \beta(D,G_o)=\beta(D)=\beta(D_1)+\beta(D_2)+\beta(D_3)=2+1+1=4$. It is not difficult to observe that a two-sender index code $ \{x_1,x_2,x_3,x_4\oplus x_5\} $, that is obtained by transmitting $ x_1,x_2,x_3$ from $ S_1 $, and $ x_4\oplus x_5 $ from $ S_2 $, achieves the optimal broadcast rate both in TSUIC and SSUIC.}
	\label{fig:exampletheorem1} \vspace{-2ex}
\end{figure}
\begin{Example}
	Consider a \textnormal{TSUIC} problem of the following: $ (1), (2|1,4,5), (3|1,2,4,5), (4|1,5),(5|1,4) $, and $ \mathcal{M}_1=\{1,4,5\} $, $ \mathcal{M}_2=\{1,2,3\} $. We compute its $ \beta(D,G_o) $ using Theorem~\ref{theorem:acyclic}. Refer to Figure~\ref{fig:exampletheorem1} for details. 
\end{Example}

\begin{Proposition} \label{lemma:2}
	For any $ D $ having a fully-participated interaction between its sub-digraphs $ D_1 $, $ D_2 $ and $ D_3 $, if $ f(D)\in \{\mathrm{H}_1,\mathrm{H}_{16} \} $, then $ \beta_t(D,G_o)=\beta_t(D_1)+\beta_t(D_2)+\beta_t(D_3)+\epsilon/t$ for some $\epsilon\in \{-2,-1,0\} $.  
\end{Proposition}
\begin{IEEEproof}
	Refer to  Appendix~\ref{append2}
\end{IEEEproof}

\begin{Remark}
	The proof of the Proposition~\ref{lemma:2} is based on the analysis of the confusion graph $\Gamma_t(D) $ and its coloring. This is described in section~\ref{confusiongraphcoloring}. As a confusion graph possesses some symmetry within~\textemdash~in fact, all confusion graphs are vertex-transitive\footnote{An undirected graph $ G $ is vertex-transitive if for every pair $ u,v\in V(G) $ there exists an automorphism mapping from $ u $ to $ v $. In the automorphism mapping of all vertices in $ V(G) $, the graph is mapped onto itself whilst preserving the connectivity of the vertices and edges.}~\textemdash~whilst analyzing them (especially coloring), we systematically group its vertices and then analyze the graph based on these groups (rather than individual vertices). This way, for a \textnormal{TSUIC} problem whose sub-problems interact with each other in some way, we can reduce the complexity arising during its analysis (especially finding the number of colors in a proper coloring of $ \Gamma_t(D) $) due to the number of vertices, which is exponential in $ t $ and $ N $. The proposed grouping of the vertices of the confusion graph and its characteristics are stated in Appendix~\ref{append2a}. 
\end{Remark}

\begin{Theorem} [CASE~I \& CASE~II-A]\label{theorem:caseIIA}
	For any $ D $ having any interaction (i.e., either fully participated or partially participated) between its sub-digraphs $ D_1 $, $ D_2 $ and $ D_3 $, if $ f(D)\in \{\mathrm{H}_{1},\mathrm{H}_{2},\dotsc,\mathrm{H}_{17}\}\setminus \{\mathrm{H}_{8},\mathrm{H}_{9},\dotsc,\mathrm{H}_{14}\}  $, then $ \beta_t(D,G_o)=\beta_t(D_1)+\beta_t(D_2)+\beta_t(D_3)+\epsilon/t$, where $\epsilon\in \{-2,-1,0\} $, and $ \beta(D,G_o)=\beta(D_1)+\beta(D_2)+\beta(D_3)$.      
\end{Theorem}

\begin{IEEEproof}
	For convenience, let $ D =D^{i} $ if $ f(D)=\mathrm{H}_{i} $. Now for $ D^{i} $, $ i\in \{1,16\} $, with a fully-participated interaction between the sub-digraphs, we have the following from Proposition~\ref{lemma:2}, 
	\begin{align} \label{eq:2a}
	\beta_{t}(D^{i},G_o)=\beta_{t}(D_1)+ \beta_{t}(D_2)+\beta_{t}(D_3)+\epsilon /t,
	\end{align}
	for some  $\epsilon\in \{-2,-1,0\} $.	
	For any $ D^{i}$, $ i\in \{1,2,3,\dotsc,17 \} $, having either a partially-participated or a fully-participated interaction between the sub-digraphs, we have the following observations: 
	The interactions between the sub-digraphs (i.e., 
	$ D_1 $, $ D_2$ and $D_3$) are equal to or more than that in $ D^{1}$, so
	\begin{equation} \label{eq:2c}
	\beta_{t}(D^{i},G_o)\leq \beta_{t}(D^{1},G_o),
	\end{equation} 
	and equal to or fewer than that in $ D^{14}$ (with a fully-participated interaction between the sub-digraphs), so
	\begin{equation} \label{eq:2b}
	\beta_{t}(D^i,G_o)\geq \beta_{t}(D^{14},G_o).
	\end{equation} 
	Now from \eqref{eq:2a}, \eqref{eq:2c} and \eqref{eq:2b}, we get
	\begin{equation} \label{eq:2d}
	\beta_{t}(D^i,G_o)=\beta_{t}(D,G_o)=\beta_{t}(D_1)+ \beta_{t}(D_2)+\beta_{t}(D_3)+\epsilon /t, 
	\end{equation}
	where $\epsilon\in \{-2,-1,0\} $ and $ i\in \{1,2,\dotsc,17 \} \setminus \{\mathrm{H}_{8},\mathrm{H}_{9},\dotsc,\mathrm{H}_{14}\} $. 
	
	Now taking a limit $ t\rightarrow \infty $ on both sides in \eqref{eq:2d}, we get 
	\begin{align}
	\beta(D,G_o)=\underset{t\rightarrow \infty}{\lim}\beta_{t}(D,G_o) &=\underset{t\rightarrow \infty}{\lim}(\beta_{t}(D_1)+ \beta_{t}(D_2)+\beta_{t}(D_3)+\epsilon /t) \nonumber \\
	&= \beta(D_1)+ \beta(D_2)+\beta(D_3),	
	\end{align}
	where $\underset{t\rightarrow \infty}{\lim} \frac{\epsilon}{t}=0 $.
\end{IEEEproof}


\begin{Remark}
	For any $ D $ having non-empty $ D_1 $, $ D_2 $ and $ D_3 $ with fully-participated interactions between them such that $ f(D) \in \{ \mathrm{H}_{15},\mathrm{H}_{16},\mathrm{H}_{17}\} $, in \textnormal{SSUIC}, we have $ \beta(D)=\beta(D_3)+\text{max}\{\beta(D_1),\beta(D_2)\} $ by Theorem~\ref{theorem:betaD}, however in \textnormal{TSUIC}, $ \beta(D)$, a lower bound to $ \beta(D,G_o) $, is not achievable due to Theorem~\ref{theorem:caseIIA}.
\end{Remark}
\begin{Corollary} \label{cor:1}
	For any $ D $ having any interaction (i.e., either fully participated or partially participated) between its sub-digraphs $ D_1 $, $ D_2 $ and $ D_3 $, if $ f(D)\in \{\mathrm{H}_{1},\mathrm{H}_{2},\dotsc,\mathrm{H}_{17}\} $, then the arcs (contributing to that interaction) between $ D_1 $, $ D_2 $ and $ D_3 $ of $ D $ are not critical in \textnormal{TSUIC} in the asymptotic regime in message size (considering infinitely long messages).
\end{Corollary}
\begin{IEEEproof}
	The proof follows from Theorem~\ref{theorem:acyclic} and~\ref{theorem:caseIIA}. 
\end{IEEEproof}



\subsection{Optimal broadcast rates for CASE~II-B}

\begin{Theorem} [CASE~II-B]\label{theorem:caseIIB1}
	For any $ D $ having a fully participated interaction between its sub-digraphs $ D_1 $, $ D_2 $ and $ D_3 $, and $ t $-bit messages for any $ t\geq 1 $, if $ f(D)\in \{\mathrm{H}_{18},\mathrm{H}_{19},\mathrm{H}_{20}\} $, then $\beta_t( D,G_o)=\text{max} \{\beta_t(D_3),\beta_{t}(D_1)+\beta_{t}(D_2) \} $ and $\beta( D,G_o)=\text{max} \{\beta(D_3),\beta(D_1)+\beta(D_2) \} $.
\end{Theorem}
\begin{IEEEproof}
	Refer to Appendix~\ref{append3}.
\end{IEEEproof}

\begin{Corollary}\label{prop:1}
	For any $ D $ having a fully participated interaction between its sub-digraphs $ D_1 $, $ D_2 $ and $ D_3 $ such that $ f(D)\in \{\mathrm{H}_{18},\mathrm{H}_{20}\} $, $\beta(D,G_o)=\beta(D)$. 
\end{Corollary}
\begin{IEEEproof}
	The result follows from Theorem~\ref{theorem:betaD} and~\ref{theorem:caseIIB1}.
\end{IEEEproof}

\begin{Remark}
	In \textnormal{TSUIC}, for any $ D$ having a fully participated interaction between its non-empty sub-digraphs $ D_1 $, $ D_2 $ and $ D_3 $ such that $ f(D)=\mathrm{H}_{19} $, no \textnormal{TSUIC} scheme achieves $ \beta(D)$ if $\beta(D_3)< \beta(D_1)+~\beta(D_2)$ because $ \beta(D)= \text{max}\{ \beta(D_1),\beta(D_2),\beta(D_3)\} \} $ (by Theorem~\ref{theorem:betaD}) and $ \beta(D,G_o) $ is at least $\beta(D_1)+\beta(D_2) $ (by Lemma~\ref{lemma:1a}).
\end{Remark}

\begin{figure} [t!]
	\centering
	\includegraphics[height=2.5cm,keepaspectratio]{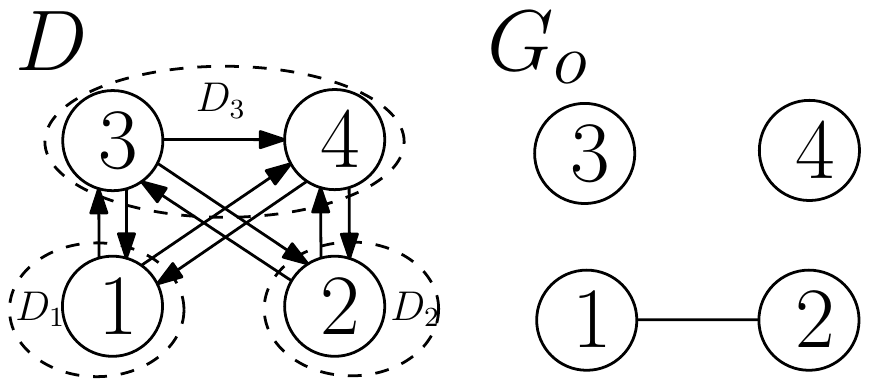}
	\caption{A given side-information digraph $ D $ such that $ f(D)=\mathrm{H}_{18} $, and a source-constraint graph $ G_o $. We have $ V(D_1)=\{1\} $, $ V(D_2)=\{2\} $ and $ V(D_3)=\{3,4\} $. From Theorem~\ref{theorem:caseIIB1}, we get $ \beta(D,G_o) =2$, and from Corollary~\ref{prop:1}, $\beta(D)=\beta(D,G_o) =2$. It is not difficult to observe that a two-sender-index code $\{x_1\oplus x_3, x_2\oplus x_4\} $, that is obtained by transmitting $ x_1\oplus x_3$ and $ x_2\oplus x_4 $ from sender 1 and sender 2, respectively, achieves its optimal broadcast rate both in TSUIC and SSUIC.}
	\label{fig:exampletheorem3} \vspace{-2ex}
\end{figure}
\begin{Example}
	Consider a \textnormal{TSUIC} problem of the following: $ (1|3,4), (2|3,4), (3|1,2,4), (4|1,2) $, and $ \mathcal{M}_1=\{1,3,4 \} $, $ \mathcal{M}_2=\{2,3,4\} $. We compute its $ \beta(D,G_o) $ and $ \beta_t(D,G_o) $ using Theorem~\ref{theorem:caseIIB1}. Refer to Figure~\ref{fig:exampletheorem3} for details. 
\end{Example}

\subsection{Optimal broadcast rates for CASE~II-C: An upper bound, and special cases where the upper bound is tight}
\begin{Theorem}[CASE~II-C] \label{theorem:caseIIC1}
	For any $ D $ having a fully participated interaction between its sub-digraphs $ D_1 $, $ D_2 $ and $ D_3 $, and $ t $-bit messages for any $ t\geq 1 $, if $ f(D)\in \{\mathrm{H}_{21},\mathrm{H}_{22},\dotsc,\mathrm{H}_{32}  \} $, then 
	\begin{enumerate} 
		\item [(i)] $ \beta_t(D,G_o) \leq \beta_t(D_2)+ \text{max}\{\beta_t(D_1),\beta_t(D_3) \} $,
		\item [(ii)] $ \beta_t(D,G_o)=\beta_t(D_1)+\beta_t(D_2)$ if $ \beta_t(D_1)\geq \beta_t(D_3) $,
		\item [(iii)] $ \beta(D,G_o) \leq \beta(D_2)+ \text{max}\{\beta(D_1),\beta(D_3) \} $, and
		\item [(iv)] $ \beta(D,G_o)=\beta(D_1)+\beta(D_2)$ if $ \beta(D_1)\geq \beta(D_3) $.
	\end{enumerate}
\end{Theorem}
\begin{IEEEproof}
	Refer to Appendix~\ref{append5}.
\end{IEEEproof}

Case $ (iii) $ in Theorem~\ref{theorem:caseIIC1} can be strengthened as follows:
\begin{Proposition}[CASE~II-C] \label{prop:extra}
	For any $ D $ having a fully participated interaction between its sub-digraphs $ D_1 $, $ D_2 $ and $ D_3 $ such that $ f(D)\in \{\mathrm{H}_{21},\mathrm{H}_{22},\dotsc,\mathrm{H}_{32}\} $, $ \beta(D,G_o)=\beta(D_2)+ \text{max}\{\beta(D_1), \beta(D_3) \} $.
\end{Proposition}
\begin{IEEEproof}
	It follows from Theorem~\ref{theorem:caseIIC1} that $ \beta(D,G_o)=\beta(D_1)+ \beta(D_2)$ if $ \beta(D_1)\geq \beta(D_3)$, and for the case when $ \beta(D_1)\leq \beta(D_3) $,  
	\begin{equation} \label{eq:check1}
	\beta(D,G_o)\leq \beta(D_2)+ \beta(D_3). 
	\end{equation}
	For $ D $ whose $ f(D)\in \{\mathrm{H}_{21},\mathrm{H}_{22},\dotsc,\mathrm{H}_{32}\} \setminus \{\mathrm{H}_{28},\mathrm{H}_{29}\} $ (all digraphs of CASE~II-C except $\mathrm{H}_{28} $ and $\mathrm{H}_{29}$), considering Theorem~\ref{theorem:betaD}, we get $ \beta(D)= \beta(D_2)+\beta(D_3)$ if $ \beta(D_1)\leq \beta(D_3) $. As $  \beta(D)\leq  \beta(D,G_o) $ (Lemma~\ref{simplebound}), so
	\begin{equation} \label{eq:check2}
	\beta(D,G_o)\geq \beta(D_2)+ \beta(D_3). 
	\end{equation}  
	The interaction among $ D_1 $, $ D_2 $ and $ D_3 $ in $ D^{28} $ (that is, $ D $ where $ f(D)=\mathrm{H}_{28}$) is less than that in $D^{32}$. Thus, 
	\begin{equation}\label{newone}
	\beta(D^{28},G_o)\geq \beta(D^{32},G_o).
	\end{equation}
	From \eqref{eq:check2} and \eqref{newone}, we get 
	\begin{equation}\label{eq:check3}
	\beta(D^{28},G_o)\geq \beta(D^{32},G_o)\geq \beta(D_2)+ \beta(D_3),
	\end{equation}
	if $ \beta(D_1)\leq \beta(D_3) $. 
	Due to the similar aforementioned reasoning, we get 
	\begin{equation} \label{eq:check4}
	\beta(D^{29},G_o)\geq \beta(D^{31},G_o)\geq \beta(D_2)+ \beta(D_3).
	\end{equation}
	From \eqref{eq:check1}, \eqref{eq:check2}, \eqref{eq:check3} and \eqref{eq:check4}, we get $ \beta(D,G_o)=\beta(D_2)+ \beta(D_3)  $ if $ \beta(D_1)\leq \beta(D_3) $ for any $ D $ with $ f(D)\in \{\mathrm{H}_{21},\mathrm{H}_{22},\dotsc,\mathrm{H}_{32}\} $. Altogether, $ \beta(D,G_o)=\beta(D_2)+ \text{max}\{\beta(D_1), \beta(D_3)\} $.   
\end{IEEEproof}
\begin{Corollary} \label{prop:3}
	For any $ D $ having a fully participated interaction between its sub-digraphs $ D_1 $, $ D_2 $ and $ D_3 $ such that $ f(D)\in \{\mathrm{H}_{21},\mathrm{H}_{22},\dotsc,\mathrm{H}_{27}\} $, $\beta(D,G_o)=\beta(D)$.
\end{Corollary}
\begin{IEEEproof}
	For the given digraph $ D $, by applying Theorem~\ref{theorem:betaD}, one can get $ \beta(D)= \beta(D_2)+\text{max} \{\beta(D_1),\beta(D_3)\}$, and this equals $\beta(D,G_o)$ by Proposition~\ref{prop:extra}.	
\end{IEEEproof}

\begin{Corollary} \label{coro:new}
	For any $ D $ such that $f(D)\in \{\mathrm{H}_{30}, \mathrm{H}_{31},\mathrm{H}_{32}  \}$,
	\begin{itemize} 
		\item [(i)] if $ \beta(D_1)\leq \beta(D_3)$, then $ \beta(D,G_o)=  \beta(D_2)+\beta(D_3)=\beta(D) $,
		\item [(ii)] if $ \beta(D_1)> \beta(D_3)$, then 
		\begin{itemize}
			\item [(a)] if $ \beta(D_1)\geq  \beta(D_2)+\beta(D_3)$, then $ \beta(D,G_o)=  \beta(D_1)+\beta(D_2)\geq \beta(D_1)=\beta(D) $, with a strict inequality if $ D_2 $ is non-empty, and
			\item [(b)] if $ \beta(D_1)\leq  \beta(D_2)+\beta(D_3)$, then $ \beta(D,G_o)=  \beta(D_1)+\beta(D_2)> \beta(D_2)+ \beta(D_3)=\beta(D) $ for a non-empty $ D_2 $.
		\end{itemize}
	\end{itemize}
\end{Corollary}
\begin{IEEEproof}
	If $ \beta(D_1)\leq \beta(D_3)$, then $ \beta(D_1)\leq \beta(D_3)+ \beta(D_2)$. Now from Proposition~\ref{prop:extra} and Theorem~\ref{theorem:betaD}, $ \beta(D,G_o)=  \beta(D_2)+\beta(D_3)=\beta(D)$. For the case $ \beta(D_1)> \beta(D_3)$, the results directly follows from Proposition~\ref{prop:extra} and Theorem~\ref{theorem:betaD}. 
\end{IEEEproof}

\begin{Remark} \label{remark:2}
	Let $ D =D^{i} $ if $ f(D)=\mathrm{H}_{i} $. Now for any $ D $ such that $ f(D)\in \{\mathrm{H}_{28}, \mathrm{H}_{29}\} $, we have the following if $ \beta(D_1)\leq \beta(D_3)$: 
	\begin{itemize}
		\item [(i)] $ \{\beta(D^{32},G_o)=	\beta(D^{32})=\beta(D_2)+\beta(D_3) \} \leq \beta(D^{28})\leq \beta(D^{28},G_o) \leq \{ \beta(D^{26})=\beta(D^{26},G_o)=\beta(D_2)+\beta(D_3) \}$ from Proposition~\ref{prop:extra}, Theorem~\ref{theorem:betaD}, Corollary~\ref{prop:3}, Corollary~\ref{coro:new}, and Lemma~\ref{simplebound}. This implies $ \beta(D^{28})= \beta(D^{28},G_o)= \beta(D_2)+\beta(D_3) $.
		\item [(ii)] $ \{\beta(D^{31},G_o)=	\beta(D^{31})=\beta(D_2)+\beta(D_3) \} \leq 	\beta(D^{29})\leq \beta(D^{29},G_o) \leq \{ \beta(D^{27})=\beta(D^{27},G_o)=\beta(D_2)+\beta(D_3) \}$ from Proposition~\ref{prop:extra}, Theorem~\ref{theorem:betaD}, Corollary~\ref{prop:3}, Corollary~\ref{coro:new}, and Lemma~\ref{simplebound}. This implies $ \beta(D^{29})= \beta(D^{29},G_o)= \beta(D_2)+\beta(D_3) $. 
	\end{itemize}
\end{Remark}
\begin{figure}[t!]
	\centering
	\includegraphics[width=5.5cm, keepaspectratio]{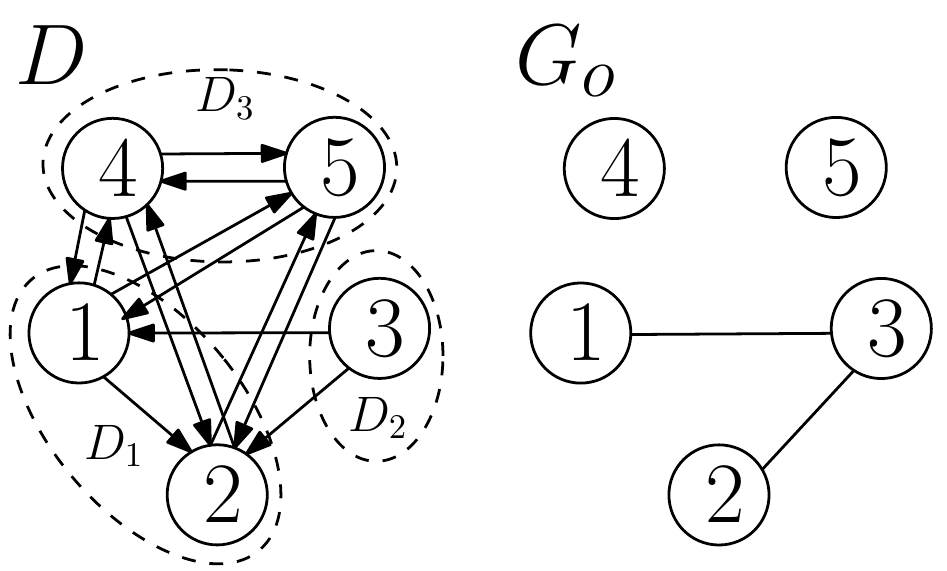}%
	\caption{ (a) A given side-information digraph $ D $ such that $f(D)=\mathrm{H}_{23}$, and a source-constraint graph $ G_o $. We have $ V(D_1)=\{1,2 \}$, $ V(D_2)=\{3 \} $, and $ V(D_3)=\{4,5 \} $. From Theorem~\ref{theorem:caseIIC1}, we get $ \beta(D,G_o)=3 $ and from Corollary~\ref{prop:3}, $ \beta(D)=\beta(D,G_o)=3$. It is not difficult to observe that a two-sender index code $\{x_1\oplus x_4\oplus x_5, x_2,  x_3\} $, where $( x_1\oplus x_4\oplus x_5, x_2 )$ and $ x_3 $ are transmitted by sender 1 and sender 2, respectively, achieves its optimal broadcast rate both in TSUIC and SSUIC.}
	\label{fig:exampletheorem4} \vspace{-1ex}
\end{figure}
\begin{Example}
	Consider a \textnormal{TSUIC} problem $ D $ of the following: $ (1|2,4,5), (2|4,5), (3|1,2), (4|1,2,5),(5|1,2,4) $, and $ \mathcal{M}_1=\{1,2,4,5 \} $, $ \mathcal{M}_2=\{3,4,5\} $. We compute its $ \beta(D,G_o) $ and $ \beta_{t}(D,G_o) $ using Theorem~\ref{theorem:caseIIC1}. Refer to Figure~\ref{fig:exampletheorem4} for details.
\end{Example}

\subsection{Optimal broadcast rates for CASE~II-D}

\begin{Theorem}[CASE~II-D] \label{theorem:caseIIE1}
	For any $ D $ having a fully participated interaction between its sub-digraphs $ D_1 $, $ D_2 $ and $ D_3 $, and $ t $-bit messages for any $ t\geq 1 $, if $ f(D)\in \{\mathrm{H}_{33},\mathrm{H}_{34},\mathrm{H}_{35},\mathrm{H}_{36}\} $, then 
	\begin{itemize} 
		\item [(i)] $\beta_t(D,G_o)\leq  \text{max} \{\beta_t(D_1),\beta_t(D_3)\}+ \text{max} \{\beta_t(D_2),\beta_t(D_3)\}$,
		\item [(ii)] $ \beta_t( D,G_o)= \beta_t(D_1)+\beta_t(D_2)$ if $ \beta_t(D_3)\leq \text{min}\{\beta_t(D_1),\beta_t(D_2) \} $,
		\item [(iii)] $\beta( D,G_o)\leq  \text{max}\{\beta(D_1),\beta(D_3)\}+ \text{max}\{\beta(D_2),\beta(D_3)\}$ , and 
		\item [(iv)] $ \beta(D,G_o)= \beta(D_1)+\beta(D_2)$ if $ \beta(D_3)\leq \text{min}\{\beta(D_1),\beta(D_2) \} $. 
	\end{itemize}
\end{Theorem}
\begin{IEEEproof}
	Refer to Appendix~\ref{append7}.	
\end{IEEEproof}

\begin{Proposition} \label{prop:extra2}
	For any $ D $ having a fully participated interaction between its sub-digraphs $ D_1 $, $ D_2 $ and $ D_3 $ such that $ f(D)\in \{\mathrm{H}_{33},\mathrm{H}_{34}\}$, $ \beta(D,G_o)=\beta(D)=\beta(D_3)+\text{max}\{ \beta(D_1),\beta(D_2)\}$ if $ \text{min}\{ \beta(D_1),\beta(D_2)\} \leq \beta(D_3)\leq \text{max}\{ \beta(D_1),\beta(D_2)\}$. 
\end{Proposition}
\begin{IEEEproof}
	It follows from Theorem~\ref{theorem:caseIIE1} that $ \beta(D,G_o)\leq \text{max}\{ \beta(D_1),\beta(D_3)\}+\text{max}\{ \beta(D_2),\beta(D_3)\} $. So,	
	considering $ \text{min}\{ \beta(D_1),\beta(D_2)\} \leq \beta(D_3)\leq \text{max}\{ \beta(D_1),\beta(D_2)\}$, if $\beta(D_1)\geq \beta(D_2) $, then
	\begin{equation} \label{eq:B}
	\beta(D,G_o)\leq \beta(D_1)+\beta(D_3),  
	\end{equation} 
	and if $\beta(D_2)\geq \beta(D_1) $, then
	\begin{equation} \label{eq:C}
	\beta(D,G_o)\leq \beta(D_2)+\beta(D_3).  
	\end{equation} 
	From \eqref{eq:B} and \eqref{eq:C}, we get 
	\begin{equation} \label{eq:D}
	\beta(D,G_o)\leq \beta(D_3)+\text{max}\{ \beta(D_1),\beta(D_2)\}.
	\end{equation}
	Now from Theorem~\ref{theorem:betaD}, one can get $ \beta(D)= \beta(D_3)+\text{max}\{\beta(D_1),\beta(D_2) \}$. As $\beta(D)\leq  \beta(D,G_o) $ (Lemma~\ref{simplebound}), we get
	\begin{equation} \label{eq:check2a}
	\beta(D,G_o)\geq \beta(D_3)+\text{max}\{\beta(D_1),\beta(D_2) \}. 
	\end{equation}  
	From \eqref{eq:D} and \eqref{eq:check2a}, we get $ \beta(D,G_o)= \beta(D_3)+\text{max}\{\beta(D_1),\beta(D_2) \}$.   
\end{IEEEproof}

Now for UIC problems with more than two senders, we study some classes of interactions between the sub-digraphs of a digraph (representing the UIC problem) in the following sub-section. 

\section{Generalizing the results of some classes of TSUIC problems to multiple senders} \label{generalsetup}
In this section, we illustrate how the method proposed in this paper can be generalized to scenarios with more than two senders.

Let $ N'$ be the number of senders, each with at least one private message. Clearly, $ 1 \leq N' \leq N $. In this section, we consider a special case of multi-sender unicast-index coding (MSUIC), where the only common messages are present in all $ N' $ senders and the rest are all private messages. We call this MSUIC special MSUIC (SMSUIC). Following the earlier convention of notations used in TSUIC, the set of common messages and its corresponding sub-problem are denoted by $ \mathcal{P}_{N'+1} $ and $ D_{N'+1} $, respectively. Precisely, for $ D $, we have $ N'+1 $ sub-digraphs, where $ D_1, D_2,\dotsc,D_{N'}$ are the vertex-induced sub-digraphs of $ D $ associated with vertices of those requesting the private messages, and $ D_{N'+1} $ is associated with vertices of those requesting only common messages. In MSUIC, each vertex of $ \Gamma_t(D) $ is labeled as $ (\textbf{b}_{D_{1}}^{i_1},\textbf{b}_{D_{2}}^{i_2},\textbf{b}_{D_{3}}^{i_3},\dotsc,\textbf{b}_{D_{N'}}^{i_{N'}},\textbf{b}_{D_{N'+1}}^{k})$, where $ i_j \in \{1,2,\dotsc,2^{tn_{j}}\} $ for $ j\in \{1,2,\dotsc,N'\} $ and $ k\in \{ 1,2,\dotsc, 2^{tn_{N'+1}}\} $.

\begin{figure}[t]
	\centering
	\includegraphics[width=0.3\linewidth]{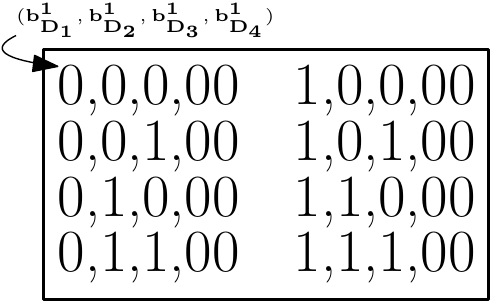} 
	\caption{Grouping of vertices in Block 1 of $ \Gamma_1(D)$ for a SMSUIC with three senders having $ \mathcal{M}_1=\{ x_1,x_4,x_5\} $, $ \mathcal{M}_2=\{ x_2,x_4,x_5\} $, $ \mathcal{M}_3=\{ x_3,x_4,x_5\} $, and $ t=1 $.}
	\label{fig:zeroblock}
\end{figure} 

\begin{figure}
	\centering
	\subfloat[]{\includegraphics[width=4.5cm, keepaspectratio]{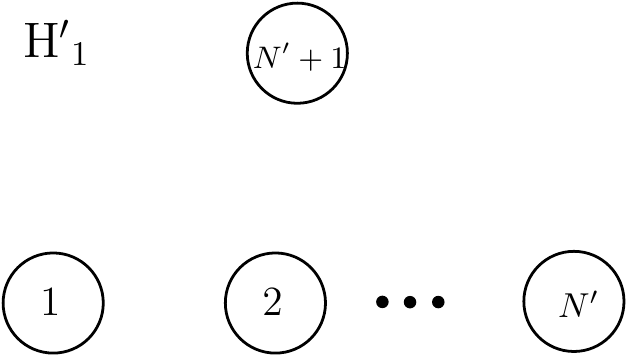}%
		\label{fig:examplenew1}}				
	\hfil
	\subfloat[]{\includegraphics[width=4.5cm, keepaspectratio]{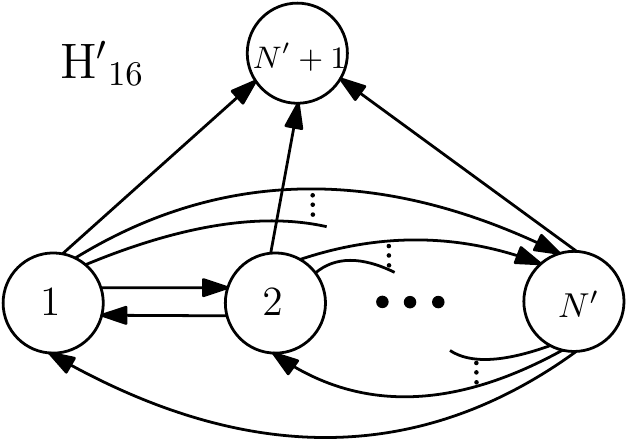}%
		\label{fig:examplenew2}}
	\caption{ Extension of (a) $ \mathrm{H}_1 $, and (b) $ \mathrm{H}_{16} $ to SMSUIC.}
	\label{examplenew} 
\end{figure}

Observe that the extensions of Definition~\ref{def:senderconstraint} (including the sender-constraint graph, $ G_o $), Definition~\ref{def:2scoloring}, and Lemmas~\ref{lemmaA}, \ref{lemmaB}, \ref{lemmaC} \& \ref{lemmaD} to SMSUIC are straightforward. Similar to the proof of Theorem~\ref{theorem:2}, one can prove the following in SMSUIC.
\begin{equation}
\beta_t(D,G_o)= \underset{J_1,J_2,\dotsc, J_{N'}}{\min}\ \frac{\lceil \log_2 |\mathcal{J}_1| \rceil+ \lceil \log_2 |\mathcal{J}_2|  \rceil+\dotsc+\lceil \log_2 |\mathcal{J}_{N'}| \rceil }{t}.
\end{equation} 

Now we extend the proposed grouping of the vertices of $ \Gamma_t(D) $ to MSUIC (including SMSUIC). Firstly, refer to Section~\ref{confusiongraphcoloring} and Appendix~\ref{append2a} for the notations, groupings and observations related to the vertices in any $ k $-th block of $\Gamma_t(D)$ in TSUIC. 
We follow a similar way of grouping of vertices in MSUIC, where any $ k $-th block of $\Gamma_t(D)$ has the following:

\begin{enumerate} 
	\item Vertices labeled by all $ ( \textbf{b}_{D_{1}}^{i_1}, \textbf{b}_{D_{2}}^{i_2},\textbf{b}_{D_{3}}^{i_3},\dotsc, \textbf{b}_{D_{N'}}^{i_{N'}}, \textbf{b}_{D_{N'+1}}^{k})$, $ i_j\in \{1,2,\dotsc,2^{tn_j}\} $, $ j\in \{1,2,\dotsc, N'\} $ and $k\in \{1,2,\dotsc, 2^{tn_{N'+1}}\}$, with the same $\textbf{b}_{D_{N'+1}}^{k}$ sub-label,
	\item any row sub-block consists of vertices labeled by all $ ( \textbf{b}_{D_{1}}^{i_1}, \textbf{b}_{D_{2}}^{i_2},\textbf{b}_{D_{3}}^{i_3},\dotsc, \textbf{b}_{D_{N'}}^{i_{N'}}, \textbf{b}_{D_{N'+1}}^{k})$, $ i_j\in \{1,2,\dotsc,2^{tn_j}\} $, $ j\in \{1,2,\dotsc, N'\} $, with the same $\textbf{b}_{D_{2}}^{i_2},\textbf{b}_{D_{3}}^{i_3},\dotsc, \textbf{b}_{D_{N'}}^{i_{N'}}, \textbf{b}_{D_{N'+1}}^{k}$ sub-labels, and
	\item any $ i_1$-th column sub-block consists of vertices labeled by all  $ ( \textbf{b}_{D_{1}}^{i_1}, \textbf{b}_{D_{2}}^{i_2},\textbf{b}_{D_{3}}^{i_3},\dotsc, \textbf{b}_{D_{N'}}^{i_{N'}}, \textbf{b}_{D_{N'+1}}^{k})$, $ i_j\in \{1,2,\dotsc,2^{tn_j}\} $, $ j\in \{1,2,\dotsc, N'\} $, with the same $\textbf{b}_{D_{1}}^{i_1}$ and $\textbf{b}_{D_{N'+1}}^{k}$ sub-labels. Moreover, in contrast to SSUIC, there are multiple sub-labels other than $ \textbf{b}_{D_{1}}^{i_1} $ and $ \textbf{b}_{D_{N'+1}}^{k} $ in MSUIC, so we arrange the vertices of any $ i_1$-th column sub-block as dictated by Figure~\ref{fig:msuicvertexarrangement} in Appendix~\ref{append1}. Clearly, a block has $ 2^{tn_1} $ column sub-blocks and $ 2^{t(\sum_{i=2}^{N'}n_i)} $ row sub-blocks. 
\end{enumerate}
Now we illustrate the grouping of the vertices with an example. Assume that we have three senders $ S_1 $, $ S_2 $ and $ S_3 $ with $ \mathcal{M}_1=\{ x_1,x_4,x_5\} $, $ \mathcal{M}_2=\{ x_2,x_4,x_5\} $ and $ \mathcal{M}_3=\{ x_3,x_4,x_5\} $. We get $ N'=3 $, and $ V(D_4)=\{4,5\} $. For $ x_i\in \{0,1\} $, $ i\in \{ 1,2,3,4,5\} $, we arrange the vertices of the first block as shown in Figure~\ref{fig:zeroblock}.

Based on our classification of interactions (referring to Figure~\ref{fig:allgraphscasepart1}), $\mathrm{H}_1$ has no arc, and $\mathrm{H}_{16}$ has the following: Vertices $ 1 $ and $ 2 $, each has an out-degree of two, whereas the vertex $ 3 $ has zero out-degree. Now considering the interactions between the sub-digraphs $ D_1, D_2,\dotsc,D_{N'},D_{N'+1} $, the extensions of $ \mathrm{H}_1 $ and $\mathrm{H}_{16}$ to SMSUIC are straightforward (refer to Figure~\ref{examplenew}). We labeled them by $\mathrm{H'}_1$ and $\mathrm{H'}_{16}$, respectively, in SMSUIC. Now we have the following proposition.
\begin{Proposition} \label{lemma:2extra}
	For any $ D $ having a fully-participated  interaction between its sub-digraphs $ D_1, D_2,\dotsc,D_{N'}, D_{N'+1} $, if $ f(D)\in \{\mathrm{H'}_1,\mathrm{H'}_{16}\} $, then $ \beta_t(D,G_o)=\sum_{i=1}^{N'+1}\beta_t(D_i)+\epsilon/t$ for some $\epsilon\in \{-N',-N'+1,\dotsc, 0\} $.  
\end{Proposition}
\begin{IEEEproof}
	Refer to Appendix~\ref{append8}.
\end{IEEEproof}
Similar to the proof of Theorem~\ref{theorem:caseIIA}, one can prove the following theorem using Proposition~\ref{lemma:2extra}.
\begin{Theorem}
	For any $ D $ having any interaction (i.e., either fully participated or partially participated) between its sub-digraphs $ D_1, D_2,\dotsc,D_{N'}, D_{N'+1} $, if $ f(D)$ (i.e., $ H' $) has some arcs among its vertices $ 1,2,\dotsc,N',N'+1 $ such that there is no out-going arc from $ N'+1 $ to any other vertex, then $ \beta_t(D,G_o)=\sum_{i=1}^{N'+1}\beta_t(D_i)+\epsilon/t$ for some $\epsilon\in \{-N',-N'+1,\dotsc, 0\} $, and $ \beta(D,G_o)=\sum_{i=1}^{N'+1}\beta(D_i)$.     
\end{Theorem}
\begin{Remark}
	Extending the other cases of \textnormal{TSUIC} to \textnormal{MSUIC} is a laborious task as it involves a construction of multi-dimensional blocks and sub-blocks in a confusion graph. 
\end{Remark}
\section{Discussions}
Consider any digraph $ D $ and its sub-digraphs $ D_1 $, $ D_2 $ and $ D_3 $. Let $ d^+_{D}(u) $ be the out-degree of a vertex $ u $ of $ D $. Now we make the following two observations in TSUIC:
\begin{itemize}  
	\item \emph{The role of side-information of the vertices in $ V(D_3) $ (vertices requesting the common messages) about the messages requested by vertices in $ V(D_1)\cup V(D_2) $ (vertices requesting the private messages) in \textnormal{TSUIC}:}
	It is proved in SSUIC that if the interaction between $ D_1 $, $ D_2 $ and $ D_3 $ is acyclic, i.e., $ f(D)$ belongs to one of the digraphs in CASE~I, then $ \beta(D)=\beta(D_1)+\beta(D_2)+\beta(D_3)$ (by using Theorem~\ref{theorem:betaD}). This means that the arcs contributing acyclic interactions between the sub-digraphs of $ D $ can be removed without affecting the optimal broadcast rate of $ D $; in other words, those are non-critical arcs. In this paper, we have proved that this result is also true in TSUIC (by Theorem~\ref{theorem:caseIIA}). Moreover, in TSUIC, we have proved that for $ D $, if the vertices in $ V(D_3) $ have no side-information about the messages requested by vertices in $ V(D_1)\cup V(D_2) $, i.e., $ d^{+}_{f(D)} (3)=\emptyset$, then by Theorem~\ref{theorem:caseIIA}, we have $ \beta(D,G_o)=\beta(D_1)+\beta(D_2)+\beta(D_3)$ (behaves like having acyclic interactions between $ D_1 $, $ D_2 $ and $ D_3 $). Under this condition, any arc that is contributing any interaction between $ D_1 $, $ D_2 $ and $ D_3 $ is non-critical. 
	\item \emph{Non-critical arcs in \textnormal{SSUIC} are not necessarily non-critical in \textnormal{TSUIC}: }
	We illustrate this with an example. Consider the TSUIC problem stated in Example~\ref{eg1} (whose $ f(D)=\mathrm{H}_{33}$). In SSUIC, we know that the optimal broadcast rate $ \beta(D)=2$. This problem has an arc $ (3,1) $ that is non-critical in SSUIC (its removal does not change the optimal broadcast rate) but it is critical in TSUIC. This can be understood from the following: In SSUIC, we can remove the arc $(3,1)\in A(D)$, and still form a valid index code $\{ x_1\oplus x_2,\ x_3\}$ that achieves $\beta(D)$. This infers that removing the arc $ (3,1) $ does not affect the optimal broadcast rate in SSUIC. However, in TSUIC, if we remove the arc $ (3,1)\in A(D)$, then the new problem, say $ D' $, has $ \beta(D',G_o)=3 $ (applying Theorem~\ref{theorem:caseIIA}), whereas we get a valid two-sender index code $ \{x_1\oplus x_3,\ x_2\oplus x_3\} $ of codelength two if we consider $ (3,1)\in A(D) $. Now it is evident that there exist cases in TSUIC where some side-information (e.g., $ (1,2)$ and $(2,1) $) cannot be exploited directly during encoding by senders because of the constraint due to the two senders. However, those side-information can be utilized during decoding process at receivers' end due to the presence of other helping side-information (e.g., $(3,1)$). So, these helping side-information can be critical in TSUIC. This observation was also made by Sadeghi et al.~\cite{parastoomultisender1} for MSUIC under a different performance metric (rate region with fixed capacity links).
\end{itemize}

\section{Concluding remarks and open problems}
In this paper, we studied two-sender unicast-index-coding problems and established their structural characteristics. Noting that SSUIC is a well-studied problem (though for any arbitrary instance, it is still an open problem), there have been many important contributions made in the literature. In this paper, we solved TSUIC instances by expressing the optimal broadcast rates in terms of that of SSUIC. To this end, we introduced a two-sender graph coloring of confusion graphs in TSUIC, and propose a way of grouping the vertices of a confusion graph for analysis. Using these techniques, we derived optimal broadcast rates of TSUIC problems, both in the asymptotic and non-asymptotic regime, as a function of the optimal broadcast rates of their sub-problems. We have also presented a class of TSUIC instances where the interactions between the sub-problems of the problem are not critical. We illustrated that our proposed approach to TSUIC can be extended to some cases with multiple senders. 

Some open problems for future works are the following:
\begin{itemize} 
	\item \textbf{Study of the critical edges in the TSUIC problems}: It is observed that the non-critical arcs in SSUIC can be critical arcs in TSUIC. This requires further study.
	\item \textbf{Study of a general distributed index coding}: As our study is a step towards understanding multi-sender index coding, it is left as a future work to extend the approaches implemented and the results obtained in this paper to more general setups.  
	\item \textbf{Finding the optimal broadcast rates of TSUIC problems with cyclic-partially-participated interactions:} The analysis of $ D $ with partially-participated interactions between its sub-digraphs $ D_1 $, $ D_2 $ and $ D_3 $ is left as a future work. \vspace{1ex}
\end{itemize} 
\appendices
\section{} \label{append1}

\begin{figure}[h!]
	\centering
	\subfloat[]{\includegraphics[height=9.5cm, keepaspectratio]{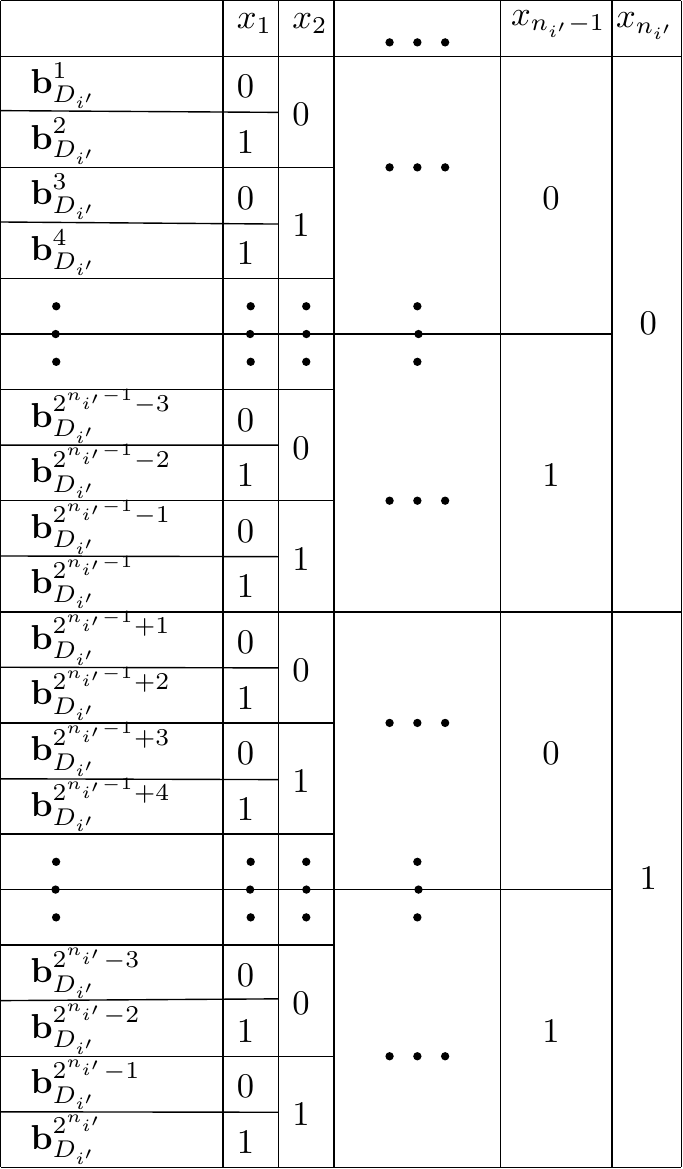}%
		\label{table:blockmap}}
	\hfil
	\subfloat[]{\includegraphics[height=9.5cm, keepaspectratio]{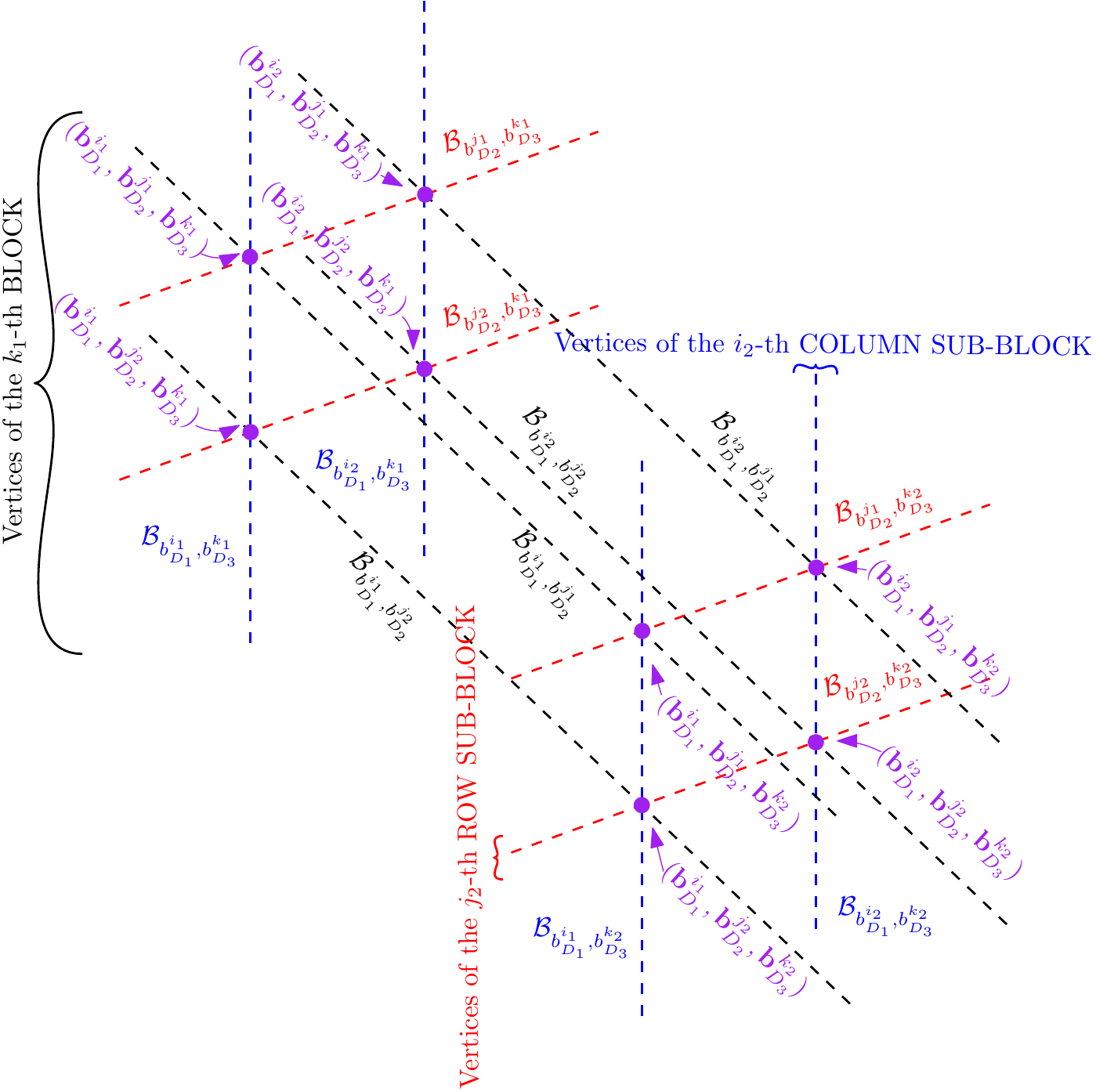}%
		\label{fig:coordinates}}
	\caption{ (a) Each $ n_{i'} $-bit tuple $\textbf{b}_{D_{i'}}^{j'}$ with its respective bits for $ t=1 $, where $ i'\in \{1,2,3\} $ and $ j'\in \{ 1,2,\dotsc, 2^{n_{i'}}\} $, and (b) representation of the vertices (e.g., $ (\textbf{b}_{D_{1}}^{i_1},\textbf{b}_{D_{2}}^{j_1},\textbf{b}_{D_{3}}^{k_1})$) and sets of vertices (e.g., $ \mathcal{B}_{\textbf{b}_{D_{2}}^{j_1},\textbf{b}_{D_{3}}^{k_1} } $), each represented by a dotted line, in a confusion graph.}
	\label{table}
\end{figure}

\begin{figure} [htbp]
	\centering
	\includegraphics[width=0.8\linewidth]{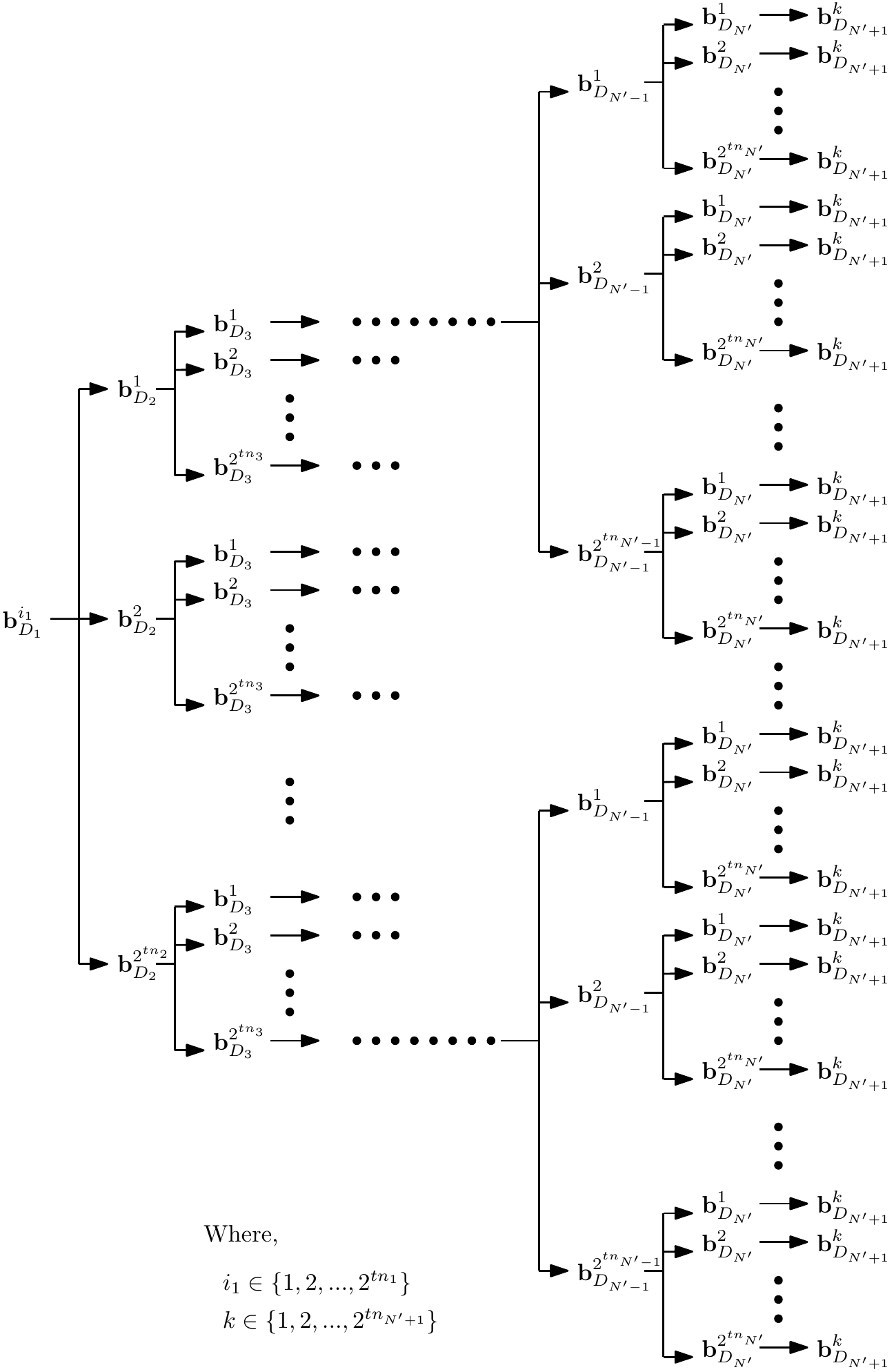}
	\caption{Arrangement of vertices in $ i_1 $-th column sub-block of a $ k $-th block of a $\Gamma_t(D)$, where, for example, the first, second and the last vertices are labeled $ ( \textbf{b}_{D_{1}}^{i_1}, \textbf{b}_{D_{2}}^{1},\textbf{b}_{D_{3}}^{1},\dotsc, \textbf{b}_{D_{N'-1}}^{1}, \textbf{b}_{D_{N'}}^{1}, \textbf{b}_{D_{N'+1}}^{k}) $, $ ( \textbf{b}_{D_{1}}^{i_1}, \textbf{b}_{D_{2}}^{1},\textbf{b}_{D_{3}}^{1},\dotsc, \textbf{b}_{D_{N'-1}}^{1}, \textbf{b}_{D_{N'}}^{2}, \textbf{b}_{D_{N'+1}}^{k}) $, and $ ( \textbf{b}_{D_{1}}^{i_1}, \textbf{b}_{D_{2}}^{2^{tn_2}},\textbf{b}_{D_{3}}^{2^{tn_3}},\dotsc, \textbf{b}_{D_{N'-1}}^{2^{tn_{N'-1}}}, \textbf{b}_{D_{N'}}^{2^{tN'}}, \textbf{b}_{D_{N'+1}}^{k}) $, respectively. Observe that this sub-block has $ 2^{t(\sum_{i=2}^{N'}n_i)} $ vertices in total.}
	\label{fig:msuicvertexarrangement}
\end{figure}

The two figures in Figure~\ref{table} outline the labels used to represent vertices of a confusion graph. The Figure~\ref{fig:msuicvertexarrangement} outlines the arrangement of vertices in any column sub-block of a block of a confusion graph for MSUIC. 

\section{Proposed grouping of the vertices of $ \Gamma_t(D) $, and its characteristics} \label{append2a}

\begin{figure}[t]
	\centering
	\includegraphics[height=12.5cm,keepaspectratio]{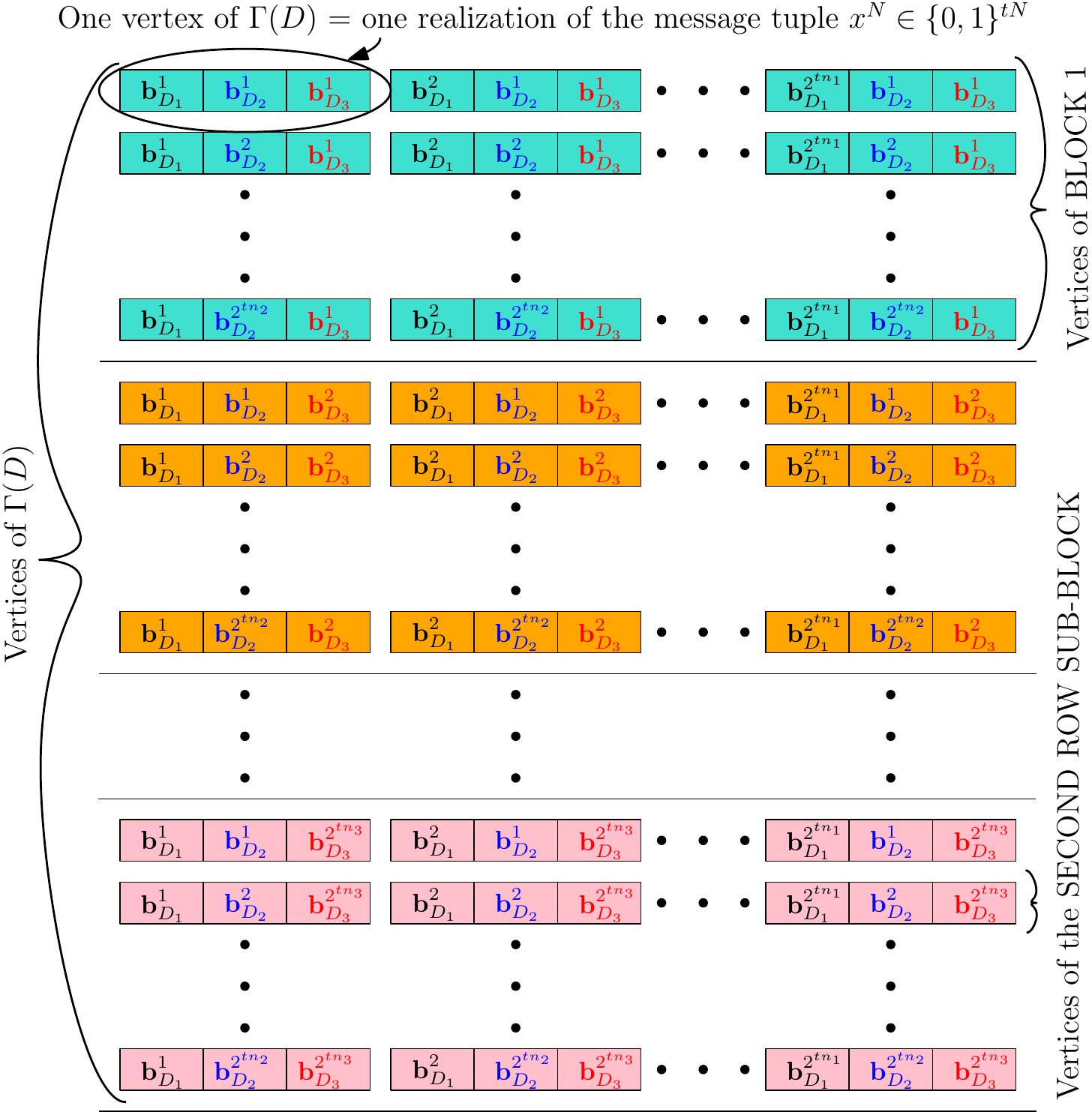}
	\caption{Functional block digraph of grouping the vertices of a confusion graph. The vertices of a confusion graph are all the possible realizations of words of $ tN $ bits.}
	\label{fig:functionalblock} \vspace{-2ex}
\end{figure}

A vertex of the confusion graph $ \Gamma_t(D) $ is represented by a tuple $ x^N $, where $ x^N=(x_1,x_2,\dotsc,x_N) $, and it is labeled by a unique $ (\textbf{b}_{D_{1}}^{i},\textbf{b}_{D_{2}}^{j},\textbf{b}_{D_{3}}^{k}) $ (see Figure~\ref{fig:functionalblock}). For the ease of analysis, by considering special groups of vertices, we divide a confusion graph to the following fundamental sub-graphs.
\begin{Definition} [Block] \label{block}
	The subgraph of $ \Gamma_t(D) $ induced by a vertex set that is formed by collecting all the vertices with the same $ \textbf{b}_{D_{3}}^{k} $ sub-label is called a $ k $-th block. 
\end{Definition}

Refer to Figure~\ref{fig:functionalblock} for a functional block diagram of a $ k $-th block. Moreover, this grouping provides $2^{t n_3} $ blocks in $ \Gamma_t(D) $. Clearly, all blocks in $ \Gamma_t(D) $ are \emph{isomorphic graphs}. This is because each block consists of all $ (\textbf{b}_{D_{1}}^{i},\textbf{b}_{D_{2}}^{j},\textbf{b}_{D_{3}}^{k})$ for the same $ \textbf{b}_{D_{3}}^{k} $ sub-label (the $ \textbf{b}_{D_{3}}^{k} $ sub-labels are different only for different blocks), and the edges in any block is due to the confusion at some receivers in $ V(D_1)\cup V(D_2) $. Moreover, the $ (tn_1+tn_2) $-bit tuples of messages requested by the vertices in $ V(D_1)\cup V(D_2) $ are labeled by $ (\textbf{b}_{D_{1}}^{i},\textbf{b}_{D_{2}}^{j}) $. For convenience, we further group the vertices of a block with the same $ \textbf{b}_{D_{3}}^{k} $ sub-label in two ways, but before this we introduce the following sets of vertices.

For the indices $ i,j,k $, let $\mathcal{B}_{\textbf{b}_{D_{1}}^{i},\textbf{b}_{D_{2}}^{j} }\triangleq \{(\textbf{b}_{D_{1}}^{i},\textbf{b}_{D_{2}}^{j},\textbf{b}_{D_{3}}^{k}):\text{for some fixed}\ (\textbf{b}_{D_{1}}^{i},\textbf{b}_{D_{2}}^{j})\}$ with cardinality $ 2^{tn_3} $, $ \mathcal{B}_{\textbf{b}_{D_{1}}^{i},\textbf{b}_{D_{3}}^{k} }\triangleq \{(\textbf{b}_{D_{1}}^{i},\textbf{b}_{D_{2}}^{j},\textbf{b}_{D_{3}}^{k}):\text{for some fixed}\ (\textbf{b}_{D_{1}}^{i},\textbf{b}_{D_{3}}^{k})\}$ with cardinality $ 2^{tn_2} $, and $ \mathcal{B}_{\textbf{b}_{D_{2}}^{j},\textbf{b}_{D_{3}}^{k} }$ is similarly defined.

Each block is further divided into two smaller sub-graphs, which are defined as follows:
\begin{Definition} [Column sub-block] \label{columnsubblock}
	The sub-graph of $ \Gamma_t(D) $ induced by a vertex set that is formed by collecting all the vertices with the same $ \textbf{b}_{D_{1}}^{i} $ and $ \textbf{b}_{D_{3}}^{k} $ sub-labels is called an $ i $-th column sub-block. 
\end{Definition} 

\begin{Definition} [Row sub-block] \label{rowsubblock}
	The sub-graph of $ \Gamma_t(D) $ induced by a vertex set that is formed by collecting all the vertices with the same $ \textbf{b}_{D_{2}}^{j} $ and $ \textbf{b}_{D_{3}}^{k} $ sub-labels is called a $ j $-th row sub-block. 
\end{Definition} 

The sub-grouping of vertices provides $2^{t n_1} $ and $2^{t n_2} $  column and row sub-blocks, respectively, within each block. In addition, the vertex sets of a column and a row sub-blocks are represented by $ \mathcal{B}_{\textbf{b}_{D_{1}}^{i},\textbf{b}_{D_{3}}^{k}}$ and $ \mathcal{B}_{\textbf{b}_{D_{2}}^{j},\textbf{b}_{D_{3}}^{k}}$, respectively. Clearly, all $ i $-th column sub-blocks of a $ \Gamma_t(D) $ are isomorphic graphs. This is because each column sub-block consists of all $ (\textbf{b}_{D_{1}}^{i},\textbf{b}_{D_{2}}^{j},\textbf{b}_{D_{3}}^{k})$ for the same $(\textbf{b}_{D_{1}}^{i},\textbf{b}_{D_{3}}^{k}) $ sub-labels (the $ (\textbf{b}_{D_{1}}^{i},\textbf{b}_{D_{3}}^{k}) $ sub-labels are different only for different column sub-blocks), and the edges within any column sub-block is due to the confusion only at some receivers in $ V(D_2) $. Moreover, the $ tn_2 $-bit tuples of messages requested by the vertices in $ V(D_2) $ are labeled by $ \textbf{b}_{D_{2}}^{j}$. From a similar reasoning as presented above for the case of column sub-blocks, it is not difficult to see that all the row sub-blocks are also isomorphic graphs, and the edges within any row sub-block is due to the confusion only at some receivers in $ V(D_1) $.

Now we illustrate the grouping of the vertices by an example. 
\begin{Example}
	Consider the \textnormal{TSUIC} problem stated in Example~\ref{eg1}. The confusion graph of the problem, $ \Gamma_1(D) $, has the following: $ \textbf{b}_{D_1}^1=0,\textbf{b}_{D_1}^2=1 $, $ \textbf{b}_{D_2}^1=0,\textbf{b}_{D_2}^2=1 $, $ \textbf{b}_{D_3}^1=0,\textbf{b}_{D_3}^2=1 $, $ \mathcal{B}_{\textbf{b}_{D_{2}}^{1},\textbf{b}_{D_{3}}^{1} }=\{(0,0,0),(1,0,0) \}$, $ \mathcal{B}_{\textbf{b}_{D_{2}}^{2},\textbf{b}_{D_{3}}^{1} }=\{(0,1,0),(1,1,0) \}$, $ \mathcal{B}_{\textbf{b}_{D_{2}}^{1},\textbf{b}_{D_{3}}^{2} }=\{(0,0,1),(1,0,1) \}$, $ \mathcal{B}_{\textbf{b}_{D_{2}}^{2},\textbf{b}_{D_{3}}^{2} }=\{(0,1,1),(1,1,1) \}$, $ \mathcal{B}_{\textbf{b}_{D_{1}}^{1},\textbf{b}_{D_{3}}^{1} }=\{(0,0,0),(0,1,0) \}$, $ \mathcal{B}_{\textbf{b}_{D_{1}}^{2},\textbf{b}_{D_{3}}^{1} }=\{(1,0,0),(1,1,0) \}$, $ \mathcal{B}_{\textbf{b}_{D_{1}}^{1},\textbf{b}_{D_{3}}^{2} }=\{(0,0,1),(0,1,1) \}$, $ \mathcal{B}_{\textbf{b}_{D_{1}}^{2},\textbf{b}_{D_{3}}^{2} }=\{(1,0,1),(1,1,1) \}$, $ \mathcal{B}_{\textbf{b}_{D_{1}}^{1},\textbf{b}_{D_{2}}^{1} }=\{(0,0,0),(0,0,1) \}$, $ \mathcal{B}_{\textbf{b}_{D_{1}}^{2},\textbf{b}_{D_{2}}^{1} }=\{(1,0,0),(1,0,1) \}$, $ \mathcal{B}_{\textbf{b}_{D_{1}}^{1},\textbf{b}_{D_{2}}^{2} }=\{(0,1,0),(0,1,1)\}$, and $ \mathcal{B}_{\textbf{b}_{D_{1}}^{2},\textbf{b}_{D_{2}}^{2} }\hskip-3pt=\hskip-2pt \{(1,1,0),(1,1,1) \}$. Furthermore, the sub-graph of $ \Gamma_1(D) $ induced by the vertices in the sets $ \{(0,0,0), (1,0,0), (0,1,0),\\ (1,1,0) \} $ and $ \{(0,0,1), (1,0,1), (0,1,1), (1,1,1) \} $ form $ k=1 $ and $ k=2 $ blocks, respectively (see Figure~\ref{fig:example2}). In addition, we have four different row sub-blocks, each formed by the vertices in one of the following sets: $ \mathcal{B}_{\textbf{b}_{D_{2}}^{1},\textbf{b}_{D_{3}}^{1} } $, $ \mathcal{B}_{\textbf{b}_{D_{2}}^{2},\textbf{b}_{D_{3}}^{1} }$, $ \mathcal{B}_{\textbf{b}_{D_{2}}^{1},\textbf{b}_{D_{3}}^{2} } $, and $ \mathcal{B}_{\textbf{b}_{D_{2}}^{2},\textbf{b}_{D_{3}}^{2} }$ (for a general outline refer to Figure~\ref{fig:coordinates} in Appendix~\ref{append1}).
\end{Example}

\subsection{Some lemmas}
The following lemmas (\ref{prop:isomorphicd1}, \ref{prop:isomorphicd2}, \ref{prop:coloringd1}, \ref{prop:coloringd2} and \ref{prop:coloringd3}) state the characteristics of our proposed grouping of vertices of a confusion graph $\Gamma_t(D) $ in TSUIC. These are helpful to understand the construction of a confusion graph, and are used in the proofs of our results. 

\begin{Lemma} \label{prop:isomorphicd1}
	$ \Gamma_t(D_1) $ and any $ j $-th row sub-block are isomorphic graphs.
\end{Lemma}
\begin{IEEEproof} 
	Let $ G $ be any $ j $-th row sub-block. We know that any vertex in $ G $ has the same $ (\textbf{b}_{D_{2}}^{j},\textbf{b}_{D_{3}}^{k}) $ sub-labels, and all vertices of $ \Gamma_t(D) $ with the same $ (\textbf{b}_{D_{2}}^{j},\textbf{b}_{D_{3}}^{k}) $ sub-label are included in the sub-block. Thus in $ G $, any edge between its vertices is only due to the confusion at some receiver belonging to $ V(D_1) $ (corresponding to the change in bits of $ \textbf{b}_{D_{1}}^{i} $ sub-label of the vertices). We know that $\Gamma_t(D_1) $ has vertices $ V(\Gamma_t(D_1))=\{(\textbf{b}_{D_{1}}^{i})\} $, and any edge between its vertices is due to the confusion at some receiver belonging to $ V(D_1) $. Observe that $ | V(\Gamma_t(D_1))|=2^{tn_1}=|\mathcal{B}_{\textbf{b}_{D_{2}}^{j},\textbf{b}_{D_{3}}^{k} }|=|V(G)|$. Now $ (\textbf{b}_{D_{1}}^{i_1},\textbf{b}_{D_{1}}^{i_2})\in E(\Gamma_t(D_1)) $ if $ ((\textbf{b}_{D_{1}}^{i_1},\textbf{b}_{D_{2}}^{j},\textbf{b}_{D_{3}}^{k}),(\textbf{b}_{D_{1}}^{i_2},\textbf{b}_{D_{2}}^{j},\textbf{b}_{D_{3}}^{k})) \in E(G)$ and vice-versa. This is because the edges are due to the confusion of the tuples, representing those vertices, at some receiver belonging to $ V(D_1) $. Consequently, $ \Gamma_t(D_1) $ and $ G $ are isomorphic graphs.
\end{IEEEproof}

We illustrate Lemma~\ref{prop:isomorphicd1} by an example. Consider the TSUIC problem stated in Example~\ref{eg1}. The confusion graph of $ D_1 $, $ \Gamma_1(D_1) $, has two vertices $ 0 $ and $ 1 $ connected by an edge as they are confused at receiver $ 1 $. Observe that $ j\in \{1,2 \} $. Now any $ j$-th row sub-block of $ \Gamma_1(D) $ has two vertices $(0,\textbf{b}_{D_{2}}^{j},\textbf{b}_{D_{3}}^{k}) $ and $(1,\textbf{b}_{D_{2}}^{j},\textbf{b}_{D_{3}}^{k}) $ connected by an edge as they are confused at receiver $ 1 $, and clearly, this vertex-induced sub-graph is isomorphic to $ \Gamma_1(D_1) $.

In a similar way to the proof of Lemma~\ref{prop:isomorphicd1}, one can prove the following Lemma.
\begin{Lemma} \label{prop:isomorphicd2}
	Each pair of the following graphs are isomorphic: (i) $ \Gamma_t(D_2) $ and any $ i $-th column sub-block, (ii) $ \Gamma_t(D_3) $ and the sub-graph of $ \Gamma_t(D) $ induced by the vertices in $  \mathcal{B}_{\textbf{b}_{D_{1}}^{i},\textbf{b}_{D_{2}}^{j} } $ for any $ (\textbf{b}_{D_{1}}^{i},\textbf{b}_{D_{2}}^{j}) $, and (iii) any $ k $-th block of a $ \Gamma_t(D) $ and $ D' $, where $ D' $ is an induced graph of $ D $ by the vertex set $ V(D)\setminus V(D_3)$.
\end{Lemma}

\begin{Lemma} \label{prop:coloringd1}
	In two-sender graph coloring of any row sub-block, $\goodchi(\Gamma_t(D_1))$ is the minimum number of total ordered pairs of colors required to color the sub-block, and the minimum number of colors associated with $ S_1 $ and $ S_2 $ are $\goodchi(\Gamma_t(D_1))$ and one, respectively. 
\end{Lemma}
\begin{IEEEproof} 
	Let $ G $ be any $ j $-th row sub-block. Observe that $ G $ includes all vertices of $ \Gamma_t(D) $ with the same $ (\textbf{b}_{D_{2}}^{j},\textbf{b}_{D_{3}}^{k}) $ sub-labels (due to our proposed method of grouping the vertices of $ \Gamma_t(D) $). So, the edges between the vertices of $G$ are only due to the confusion at some receivers belonging to $ V(D_1) $. From Lemma~\ref{lemmaA}, any pair of vertices of $ G $ connected by an edge must have different colors associated with $ S_1 $ and the same color associated with $ S_2 $. Thus the minimum number of colors associated with $ S_2 $ is one. From Lemma~\ref{prop:isomorphicd1}, $ G $ is isomorphic to $ \Gamma_t(D_1) $, so the minimum number of colors associated with $ S_1 $ must be $\goodchi(\Gamma_t(D_1))$.   
\end{IEEEproof}
In a similar way to the proof of Lemma~\ref{prop:coloringd1}, one can prove the following Lemma.
\begin{Lemma} \label{prop:coloringd2}
	In two-sender graph coloring of any column sub-block, $\goodchi(\Gamma_t(D_2))$ is the minimum number of total ordered pairs of colors required to color the sub-block, and the minimum number of colors associated with $ S_1 $ and $ S_2 $ are one and $\goodchi(\Gamma_t(D_2))$, respectively.
\end{Lemma}

\begin{Lemma}  \label{prop:coloringd3}
	In two-sender graph coloring of the sub-graph of $ \Gamma_t(D) $ induced by the vertices in $  \mathcal{B}_{\textbf{b}_{D_{1}}^{i},\textbf{b}_{D_{2}}^{j} } $ for any $ (\textbf{b}_{D_{1}}^{i},\textbf{b}_{D_{2}}^{j}) $, the minimum number of total ordered pairs of colors required to color the vertex-induced sub-graph is $\goodchi(\Gamma_t(D_3))$.
\end{Lemma}
\begin{IEEEproof} 
	We know that $ \Gamma_t(D_3) $ requires the minimum of $\goodchi(\Gamma_t(D_3))$ different colors in its coloring in SSUIC. Thus it has the minimum of $\goodchi(\Gamma_t(D_3))$ independent vertex sets. From Lemma~\ref{prop:isomorphicd2}, the sub-graph of $ \Gamma_t(D) $ induced by the vertices in $  \mathcal{B}_{\textbf{b}_{D_{1}}^{i},\textbf{b}_{D_{2}}^{j} } $ is isomorphic to $ \Gamma_t(D_3) $, so it has the minimum of $\goodchi(\Gamma_t(D_3))$ independent vertex sets. In two-sender graph coloring, we assign each independent vertex set a unique ordered pair of colors. Thus the vertex-induced sub-graph requires the minimum of $\goodchi(\Gamma_t(D_3))$ ordered pairs of colors. 
\end{IEEEproof}

\section{Proof of Proposition~\ref{lemma:2}} \label{append2}

\subsection{An example}
\begin{figure}[t!]
	\centering
	\subfloat[]{\includegraphics[width=3cm, keepaspectratio]{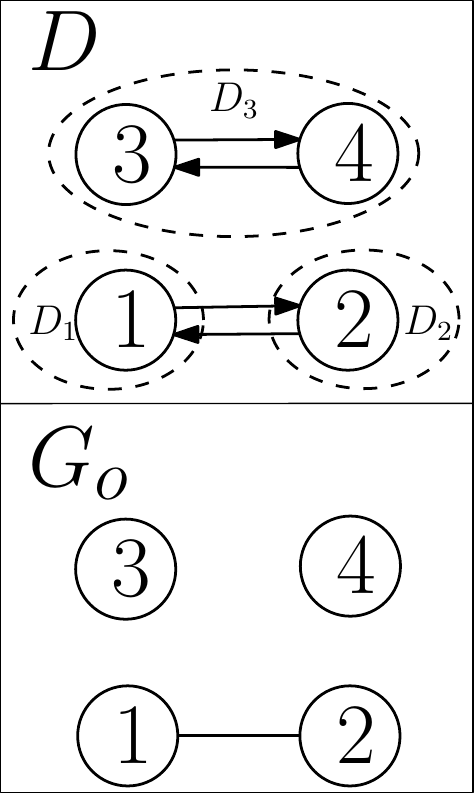}%
		\label{fig:example2a}}				
	\hfil
	\subfloat[]{\includegraphics[width=12cm, keepaspectratio]{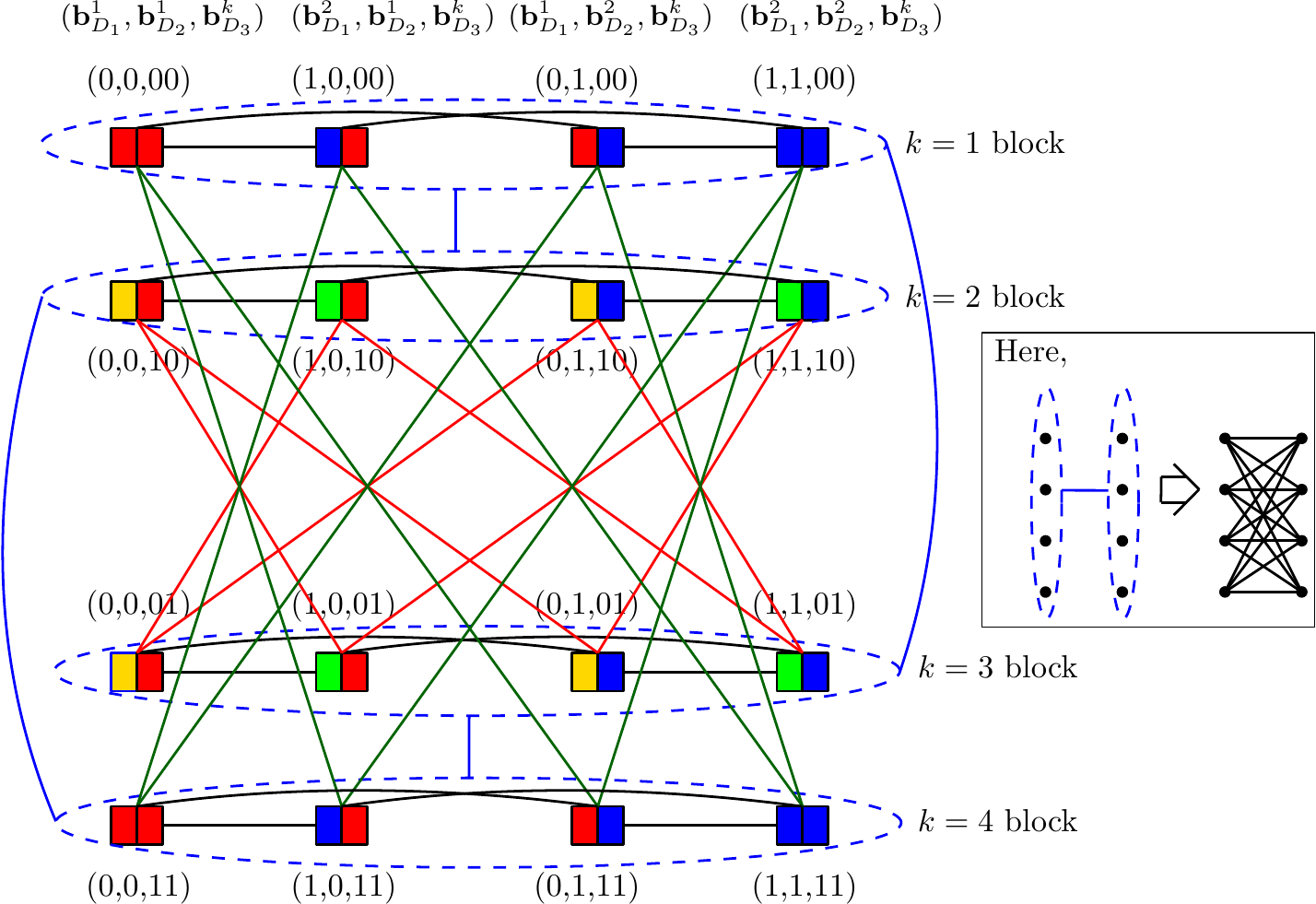}%
		\label{fig:example2b}}
	\caption{(a) A given side-information digraph $ D $ such that $f(D)=\mathrm{H}_{15}$, and a source-constraint graph $ G_o $, (b) the confusion graph $\Gamma_{1}(D)$, and its two-sender graph coloring (where each vertex is assigned with an ordered pair of colors such that the first color is always associated with $ S_1 $ and the second color is always associated with $ S_2 $). The edges are shown in color only for an illustration purpose, it is not an edge coloring.}
	\label{fig:exampletheorem2}
\end{figure}

Before proving proposition~\ref{lemma:2}, with the help of the following example, we provide an overview of the construction of the confusion graph and its two-sender graph coloring, which after generalization leads to the proof of Proposition~\ref{lemma:2}.
\begin{Example}
	Consider a \textnormal{TSUIC} problem $ (D,G_o)$ of the following: $ (1|2), (2|1), (3|4), (4|3) $, and $ \mathcal{M}_1=\{1,3,4 \} $, $ \mathcal{M}_2=\{2,3,4\} $ with $ t=1 $.
\end{Example}
For $ D $, we have $ f(D)=\mathrm{H}_{15} $. Refer to Figure~\ref{fig:exampletheorem2} for its confusion graph $\Gamma_1(D)$ whose vertices are grouped according to our proposed method. Now in details, we illustrate the construction of $\Gamma_1(D)$ and its two-sender graph coloring. These are used to derive $\beta_{t=1}(D,G_o)$. For the problem, we have $ V(D_1)=\{1\} $, $ V(D_2)=\{2\} $, and $ V(D_3)=\{3,4 \} $. $\Gamma_1(D)$ has $ 2^4=16 $ vertices labeled by all possible realizations of a word with four bits, each one is represented by a unique  $(\textbf{b}_{D_{1}}^{i},\textbf{b}_{D_{2}}^{j},\textbf{b}_{D_{3}}^{k})$ label, where $ i,j\in \{1,2 \} $, $ k\in \{ 1,2,3,4\} $, sub-labels $ \textbf{b}_{D_{1}}^{i},\textbf{b}_{D_{2}}^{j}\in \{0,1\} $ and $ \textbf{b}_{D_{3}}^{k}\in \{00,01,10,11\} $. 

Before analyzing $\Gamma_1(D)$, for convenience, we define the following types of edges of $ \Gamma_t(D) $, for any $ t\geq 1 $: 
\begin{Definition} [Inter-block edge and Intra-block edge] \label{def:edge}
	An edge between two vertices each belonging to a different block\footnote{Refer to Definition~\ref{block} in Appendix~\ref{append2a}.} of $ \Gamma_t(D) $ (e.g., an edge between a vertex of the $ k_1 $-th block and a vertex of the $ k_2 $-th block of $ \Gamma_t(D) $) is called an \emph{inter-block edge}, and an edge within the vertices of a block of $ \Gamma_t(D) $ (e.g., an edge between any two vertices of the $ k $-th block of $ \Gamma_t(D) $) is called an \emph{intra-block edge}. 
\end{Definition}     

\subsubsection{Intra-block coloring}
Now consider the block with $ k=1 $ (that is, the block with all vertices with the same $ \textbf{b}_{D_{3}}^{1} $ label). It has four vertices labeled by $ (0,0,00), (1,0,00), (0,1,00)$ and $(1,1,00) $. One can find the intra-block edges (due to confusion at some receivers in $ V(D_1\cup D_2) $), and inter-block edges (due to confusion at some receivers in $ V(D) $) as shown in Figure~\ref{fig:exampletheorem2}. We observe that all the blocks are isomorphic to each other. Now we color $\Gamma_1(D)$ starting from the block with $ k=1 $. We color similarly for any other individual block. Consider its $ (j=1) $-th row sub-block (refer to Definition~\ref{rowsubblock} in Appendix~\ref{append2a} for its definition). It has two vertices $ (0,0,00)$ and $(1,0,00)$. As these two tuples ($ (0,0,00)$ and $(1,0,00)$) are confused at receiver $ 1 $, so $ S_1 $ must assign different colors, and $ S_2 $ must assign the same color (by Lemma~\ref{lemmaA}). Say we assign $ (0,0,00)\rightarrow (\text{RED}, \text{RED})$ and $(1,0,00)\rightarrow (\text{BLUE}, \text{RED})$. We color similarly for each individual row sub-block (that is, the sub-block with all vertices with the same $\textbf{b}_{D_{2}}^{j}, \textbf{b}_{D_{3}}^{k} $ labels). Now consider its $ (i=1)$-th column sub-block\footnote{Refer to Definition~\ref{columnsubblock} in Appendix~\ref{append2a}.}. It has two vertices $ (0,0,00)$ and $(0,1,00)$. As these two tuples ($ (0,0,00)$ and $(0,1,00)$) are confused at receiver $ 2 $, so $ S_2$ must assign different colors, and $ S_1 $ must assign the same color (by Lemma~\ref{lemmaB}). Say we assign $ (0,0,00)\rightarrow (\text{RED}, \text{RED})$ and $(0,1,00)\rightarrow (\text{RED}, \text{BLUE})$. We color similarly for each individual column sub-block. By carrying this way of coloring (as of the sub-blocks) to all the vertices of the block with $ k=1 $, altogether, we have the following: $ (0,0,00)\rightarrow (\text{RED}, \text{RED})$, $ (1,0,00)\rightarrow (\text{BLUE}, \text{RED})$, $ (0,1,00)\rightarrow (\text{RED}, \text{BLUE})$, and $ (1,1,00)\rightarrow (\text{BLUE}, \text{BLUE})$. We say a two-sender graph coloring is the \emph{best possible coloring} if it corresponds to the minimum sum of the bits, which is required to uniquely index the colors associated with each sender, in TSUIC. As $ |V(D_1)|=1$, $\Gamma_{1}(D_1)$ is a graph with two vertices (labeled by $ 0 $ and $ 1 $) connected by an edge. Thus $ \goodchi(\Gamma_{1}(D_1))=2$. Similarly, we get $ \goodchi(\Gamma_{1}(D_2))=2$. Now considering Lemmas~\ref{prop:isomorphicd1}, \ref{prop:isomorphicd2}, \ref{prop:coloringd1}, \ref{prop:coloringd2}, and our proposed grouping of the vertices of a confusion graph, one can get $ \goodchi(\Gamma_{1}(D_1))\times \goodchi(\Gamma_{1}(D_2))$ ordered pairs of colors in the best possible coloring of a block of $ \Gamma_1(D) $ in two-sender graph coloring. Thus, in any block, the four ordered pairs of colors (two colors associated per sender), that we have assigned to the vertices of $ \Gamma_{1}(D) $, is the best possible coloring.  

\subsubsection{Inter-block coloring}
We address the inter-block edges whist coloring. We consider any two blocks, and perform two-sender graph coloring. Firstly, consider the two blocks with $ k=1 $ (all tuples having $ \textbf{b}_{D_{3}}^{1}=00$) and $ k=2 $ (all tuples having $ \textbf{b}_{D_{3}}^{2}=10$). Observe that these two tuples are confused at receiver $ 3 $. As receivers $ 3 $ does not have $ \{x_1, x_2\} $ in its side-information, any tuple of the block with $ k=1 $ and any tuple of the block with $ k=2 $ are confused at receiver $ 3 $. Thus every vertex of the block with $ k=1 $ are connected to each vertex of the block with $ k=2 $ by an inter-block edge. Consequently, we do not need to consider other inter-block edges due to confusions at some other receivers for this case. Now during two-sender graph coloring of these two blocks, we need to have two different ordered pairs of colors (one for each block). Furthermore, as $ x_3 $ is a common message to both senders, it suffices to have two completely different color sets (each for one block) associated with one of the senders. In other words, one of the senders can contribute additional colors to resolve these confusions (indicated by the inter-block edges between the blocks with $ k$ equal to $ 1 $ and $ 2 $). For example, assume that $ S_1 $ contributes the additional colors to resolve the confusions (inter-block edges) between the vertices of these blocks. Now we have the following coloring for the vertices of the block with $ k=2 $: $ (0,0,10)\rightarrow (\text{YELLOW}, \text{RED})$, $ (1,0,10)\rightarrow (\text{GREEN}, \text{RED})$, $ (0,1,10)\rightarrow (\text{YELLOW}, \text{BLUE})$, and $ (1,1,10)\rightarrow (\text{GREEN}, \text{BLUE})$. 

Next, we consider the two blocks with $ k=1 $ (all tuples having $ \textbf{b}_{D_{3}}^{1}=00$) and $ k=4 $ (all tuples having $ \textbf{b}_{D_{3}}^{4}=11$). Clearly, these blocks have no inter-block edges due to the confusion at receivers $ 3 $ and $ 4 $. The inter-block edges are due to the confusion at receivers $ 1 $ and $ 2 $, and one can see them as shown in Figure~\ref{fig:exampletheorem2}. It is not difficult to verify that if we color the vertices of the block with $ k=4 $ by the same coloring function done for the vertices of the block with $ k=1 $, which is a function of $(\textbf{b}_{D_{1}}^{i},\textbf{b}_{D_{2}}^{j})$ sub-labels of the vertices, then the coloring is still valid. Thus in order to color the vertices of the block with $ k=4 $, we do not need any additional colors for senders than that assigned to the vertices of the block with $ k=1 $. Finally, we color similarly as above for the blocks with $k$ equal to $ 2 $ and $ 3 $. 

Now observe $\Gamma_1(D)$ by assuming each block as one super-vertex, the edges that connects all the vertices of one block to every vertex of another block, and vice-versa (edges due to confusion at some receivers in $ V(D_3) $) as a single super-edge connecting those two super-vertices, and neglect all the inter-block edges due to the confusion at some receivers except receivers in $ V(D_3)$, we see that the resulting graph is $\Gamma_{1}(D_3)$. Clearly, in two-sender graph coloring, we require $ \goodchi(\Gamma_{1}(D_3))$ times of $ \goodchi(\Gamma_{1}(D_1))\times \goodchi(\Gamma_{1}(D_2))$ (which is required for each block) ordered pairs of colors in total, i.e., $ 2\times 4=8 $ ordered pairs of colors, which means $\beta_{1}(D,G_o)\geq 3$. The lower bound $ 3 $ is achievable by the coloring scheme in Figure~\ref{fig:example2b}. 
\subsection{Ingredients for the proof}
We prove the following five lemmas that are used to prove Proposition~\ref{lemma:2}.
\begin{Lemma} \label{lemma:1}
	For any $ k $-th block of the confusion graph $\Gamma_t(D)$ of a digraph $ D $, $ \goodchi(\Gamma_t(D_1))\times \goodchi(\Gamma_t(D_2)) $ is the minimum ordered pairs of colors in two-sender graph coloring, where the number of colors associated with $ S_1 $ and $ S_2 $ are $ \goodchi(\Gamma_t(D_1))$ and $ \goodchi(\Gamma_t(D_2))$, respectively.    
\end{Lemma}
\begin{IEEEproof}
	For any $ D $, based on our proposed way of grouping the vertices of $ \Gamma_t(D) $ (see Appendix~\ref{append2a}), we write all the vertices of any $ k $-th block of $ \Gamma_t(D) $ in the following matrix form:
	\[  
	\mathbb{B}_k=      
	\begin{bmatrix}
	(\textbf{b}_{D_{1}}^{1},\textbf{b}_{D_{2}}^{1},\textbf{b}_{D_{3}}^{k})   & (\textbf{b}_{D_{1}}^{2},\textbf{b}_{D_{2}}^{1},\textbf{b}_{D_{3}}^{k}) &\dots  & (\textbf{b}_{D_{1}}^{2^{tn_1}},\textbf{b}_{D_{2}}^{1},\textbf{b}_{D_{3}}^{k}) \\
	(\textbf{b}_{D_{1}}^{1},\textbf{b}_{D_{2}}^{2},\textbf{b}_{D_{3}}^{k})   & (\textbf{b}_{D_{1}}^{2},\textbf{b}_{D_{2}}^{2},\textbf{b}_{D_{3}}^{k}) &\dots  & (\textbf{b}_{D_{1}}^{2^{tn_1}},\textbf{b}_{D_{2}}^{2},\textbf{b}_{D_{3}}^{k}) \\
	\vdots    & \vdots 	 &\ddots & \vdots \\
	(\textbf{b}_{D_{1}}^{1},\textbf{b}_{D_{2}}^{2^{tn_2}},\textbf{b}_{D_{3}}^{k})   & (\textbf{b}_{D_{1}}^{2},\textbf{b}_{D_{2}}^{2^{tn_2}},\textbf{b}_{D_{3}}^{k}) &\dots  & (\textbf{b}_{D_{1}}^{2^{tn_1}},\textbf{b}_{D_{2}}^{2^{tn_2}},\textbf{b}_{D_{3}}^{k}) 
	\end{bmatrix}.
	\]    
	$ \mathbb{B}_k $ provides a visualization of the arrangement of vertices in the $ k $-th block. 
	The coloring of any row sub-block (one row of $ \mathbb{B}_k $ provides the arrangement of its vertices) requires the minimum of $ \goodchi(\Gamma_t(D_1)) $ different colors associated with $ S_1 $ and exactly one color associated with $ S_2 $ (due to Lemma~\ref{prop:coloringd1}). Now considering the coloring function of $ S_1 $, i.e., $ J_1(\textbf{b}_{D_{1}}^{i},\textbf{b}_{D_{3}}^{k}) $, the same coloring function must be applied to all row sub-blocks of the block. 
	Now the coloring of any column sub-block (one column of $ \mathbb{B}_k $ provides the arrangement of its vertices) requires the minimum of $ \goodchi(\Gamma_t(D_2)) $ different colors associated with $ S_2 $ and exactly one color associated with $ S_1 $ (due to Lemma~\ref{prop:coloringd2}). Similarly, considering the coloring function of $ S_2 $, i.e., $ J_2(\textbf{b}_{D_{2}}^{j},\textbf{b}_{D_{3}}^{k}) $, the same coloring function must be applied to all column sub-blocks of the block.
	Altogether, we get the minimum of $ \goodchi(\Gamma_t(D_1))\times \goodchi(\Gamma_t(D_2)) $ ordered pairs of colors to color a block of the confusion graph $\Gamma_t(D)$ in two-sender graph coloring.      
\end{IEEEproof}

\begin{Lemma} \label{lemma:nointeredgeD3}
	Consider a two-sender graph coloring function $ J_o $ that properly colors the confusion graph $ \Gamma_t(D) $. If there is no inter-block edge due to the confusion at some receiver in $ V(D_3) $ between any blocks of $ \Gamma_t(D) $, then $ J_o(\textbf{b}_{D_{1}}^{i},\textbf{b}_{D_{2}}^{j},\textbf{b}_{D_{3}}^{1})=J_o(\textbf{b}_{D_{1}}^{i},\textbf{b}_{D_{2}}^{j},\textbf{b}_{D_{3}}^{2})=\cdots=J_o(\textbf{b}_{D_{1}}^{i},\textbf{b}_{D_{2}}^{j},\textbf{b}_{D_{3}}^{2^{tn_3}})$, for all $ i$ and $j$, is a valid two-sender graph coloring.  	  
\end{Lemma}
\begin{IEEEproof}
	We first prove the lemma considering any two blocks, say $ k_1 $-th block and $ k_2 $-th block. If there is no inter-block edge due to the confusion at any receiver in $ V(D_3) $ between the $ k_1 $-th and $ k_2 $-th blocks of $ \Gamma_t(D)$, then we have two cases: (i) no inter-block edge and (ii) some inter-block edges due to the confusion at some receivers in $ V(D_1)\cup V(D_2) $. In case~(i), since $ k_1 $-block and $ k_2 $-block are isomorphic, we can color a block by two-sender graph coloring, and keep the same copy of coloring in another block (i.e., $ J_o(\textbf{b}_{D_{1}}^{i_1},\textbf{b}_{D_{2}}^{j_1},\textbf{b}_{D_{3}}^{k_1})=J_o(\textbf{b}_{D_{1}}^{i_1},\textbf{b}_{D_{2}}^{j_1},\textbf{b}_{D_{3}}^{k_2})$, $ \forall i_1,j_1$). Now for case (ii), suppose that there exists an edge $ ((\textbf{b}_{D_{1}}^{i_1},\textbf{b}_{D_{2}}^{j_1},\textbf{b}_{D_{3}}^{k_1}),(\textbf{b}_{D_{1}}^{i_2},\textbf{b}_{D_{2}}^{j_2},\textbf{b}_{D_{3}}^{k_2} ))\in E(\Gamma_t(D))$. Observe that $ k_1\neq k_2 $ because $ k_1 $ and $ k_2 $ are two different blocks. Moreover, since the edge is due to the confusion at some receivers in $ V(D_1)\cup V(D_2) $, we must have $ (i_1,j_1)\neq (i_2,j_2)$. Now if there exists the edge $( (\textbf{b}_{D_{1}}^{i_1},\textbf{b}_{D_{2}}^{j_1},\textbf{b}_{D_{3}}^{k_1}),(\textbf{b}_{D_{1}}^{i_2},\textbf{b}_{D_{2}}^{j_2},\textbf{b}_{D_{3}}^{k_2})) $, then there must exist an edge $( (\textbf{b}_{D_{1}}^{i_1},\textbf{b}_{D_{2}}^{j_1},\textbf{b}_{D_{3}}^{k_1}),(\textbf{b}_{D_{1}}^{i_2},\textbf{b}_{D_{2}}^{j_2},\textbf{b}_{D_{3}}^{k_1})) $. This edge is between the vertices of the same block, and the confusion must have already resolved by the coloring $ J_o $. Thereby, $ J_o(\textbf{b}_{D_{1}}^{i},\textbf{b}_{D_{2}}^{j},\textbf{b}_{D_{3}}^{k_1})=J_o(\textbf{b}_{D_{1}}^{i},\textbf{b}_{D_{2}}^{j},\textbf{b}_{D_{3}}^{k_2})$, $ \forall i,j$, is a valid coloring.     
	
	Since the choice of $ k_1 $ and $ k_2 $ is arbitrary, a two-sender graph coloring of $ \Gamma_t(D) $ with $ J_o(\textbf{b}_{D_{1}}^{i},\textbf{b}_{D_{2}}^{j},\textbf{b}_{D_{3}}^{1})=J_o(\textbf{b}_{D_{1}}^{i},\textbf{b}_{D_{2}}^{j},\textbf{b}_{D_{3}}^{2})=\cdots=J_o(\textbf{b}_{D_{1}}^{i},\textbf{b}_{D_{2}}^{j},\textbf{b}_{D_{3}}^{2^{tn_3}})$, for all $ i$ and $j$, is a valid two-sender graph coloring.	
\end{IEEEproof}
\begin{Lemma} \label{lemma:interblockedge}
	For a digraph $ D $ having fully-participated interactions between its sub-digraphs $ D_1 $, $ D_2 $ and $ D_3 $, the confusion at some receivers in $V(D_1)$ does not contribute any inter-block edges in $ \Gamma_t(D) $ if $D_1\rightarrow D_3$ in $D$ (equivalently, $ (1,3)\in A(f(D))$), and the confusion at some receivers in $V(D_2)$ does not contribute any inter-block edges in $ \Gamma_t(D) $ if $D_2\rightarrow D_3$ in $D$ (equivalently, $ (2, 3)\in A(f(D))$).
\end{Lemma}
\begin{IEEEproof}
	There is no edge due to the confusion at some receivers in $ V(D_1) $ between any pair of vertices $ ( (\textbf{b}_{D_{1}}^{i_1},\textbf{b}_{D_{2}}^{j_1},\textbf{b}_{D_{3}}^{k_1}), (\textbf{b}_{D_{1}}^{i_2},\textbf{b}_{D_{2}}^{j_2},\textbf{b}_{D_{3}}^{k_2})) $, $ k_1\neq k_2 $ (for inter-block edges), 
	because any vertex in $V(D_1) $ has $ \{ x_u:u\in V(D_3) \}$ in its side-information and the corresponding $ \textbf{b}_{D_{3}}^{k_1} $ and $ \textbf{b}_{D_{3}}^{k_2} $ labels of the two vertices are different. This proves the first assertion. Repeating the same argument for $ D_2 $, we get the second assertion (for the case $D_2\rightarrow D_3$ in $D$).	
\end{IEEEproof}

\begin{Lemma} \label{floorceiling}
	For any real numbers $ A $ and $ B $, $\lceil A+B \rceil = \lceil A \rceil+\lceil B \rceil+\epsilon' $, where $ \epsilon'\in \{-1,0\} $. 
\end{Lemma}
\begin{IEEEproof}
	As we know that for any real number $ A $, we have $ A\leq \lceil A \rceil$, and $ \lceil A \rceil-A < 1$ (this implies $ \lceil A \rceil< A+1  $ or $\lceil A \rceil-1< A$). So we get $ A\leq \lceil A \rceil < A+1 $. This is true for any other real number $ A+B $, so $ A+B \leq \lceil A+B \rceil< A+B+1 $. Altogether, we get 
	\begin{align} \label{eq:one}
	\lceil A \rceil+ \lceil B \rceil-2 &<A+B\leq  \lceil A+B \rceil < A+B+1\leq \lceil A \rceil+ \lceil B \rceil+1 \nonumber \\
	\lceil A \rceil+ \lceil B \rceil-2 &< \lceil A+B \rceil < \lceil A \rceil+ \lceil B \rceil+1.
	\end{align}
	There are only two integers in $ (\lceil A \rceil+ \lceil B \rceil-2,\lceil A \rceil+ \lceil B \rceil+1) $, and they are $ \lceil A \rceil+ \lceil B \rceil-1 $ and $ \lceil A \rceil+ \lceil B \rceil$, so $\lceil A+B \rceil = \lceil A \rceil+\lceil B \rceil+\epsilon' $, where $ \epsilon'\in \{-1,0\} $.	
\end{IEEEproof}

\begin{Lemma} \label{sumofbetta}
	For any $ \Gamma_t(D) $, if a minimum of $\goodchi(\Gamma_t(D_1))\times \goodchi(\Gamma_t(D_2))\times \goodchi(\Gamma_t(D_3)) $ ordered pairs of colors are required in its two-sender graph coloring, then $ \beta_{t}(D,G_o)= \beta_{t}(D_1)+ \beta_{t}(D_2)+\beta_{t}(D_3)+\epsilon /t $, $\epsilon\in \{-2,-1,0\} $. 
\end{Lemma}

\begin{IEEEproof}
	Let $ \goodchi'(\Gamma_t(D_3)) $ and $ \goodchi''(\Gamma_t(D_3)) $ be the non-negative non-zero integer factors of $ \goodchi(\Gamma_t(D_3)) $, and $ \goodchi^1(\Gamma_t(D_3)) $ and $ \goodchi^2(\Gamma_t(D_3)) $ be the best choice over all $ \goodchi'(\Gamma_t(D_3)) $ and $ \goodchi''(\Gamma_t(D_3)) $, respectively, such that the term $ \lceil \log_2 (\goodchi(\Gamma_t(D_1))\times \goodchi'(\Gamma_t(D_3))) \rceil + \lceil \log_2 (\goodchi(\Gamma_t(D_2))\times \goodchi''(\Gamma_t(D_3))) \rceil$ is minimized. 
	The colors associated with $ S_i,\ i \in \{1,2\} $, is always an integer multiple of $\goodchi(\Gamma_t(D_i))$ whilst coloring $\Gamma_t(D)$ due to its symmetry. Thus along with Lemmas~\ref{lemmaD} and \ref{lemma:1}, one can find that  $\goodchi(\Gamma_t(D_1))\times \goodchi^1(\Gamma_t(D_3))$ and $\goodchi(\Gamma_t(D_2))\times \goodchi^2(\Gamma_t(D_3))$ are the colors associated with $ S_1 $ and $ S_2 $, respectively, in order to produce a minimum of $\goodchi(\Gamma_t(D_1))\times \goodchi(\Gamma_t(D_2))\times \goodchi(\Gamma_t(D_3)) $ ordered pairs of colors in the two-sender graph coloring of $ \Gamma_t(D) $.
	Now from Theorem~\ref{theorem:2}, we get
	\begin{align}
	&t\times \beta_{t}(D,G_o) \nonumber \\
	&=\lceil \log_2 (\goodchi(\Gamma_t(D_1))\times \goodchi^1(\Gamma_t(D_3))) \rceil + \lceil \log_2 (\goodchi(\Gamma_t(D_2))\times \goodchi^2(\Gamma_t(D_3))) \rceil \nonumber \\
	&= \lceil \log_2 (\goodchi(\Gamma_t(D_1)))+\log_2 (\goodchi^1(\Gamma_t(D_3))) \rceil+ \lceil \log_2 (\goodchi(\Gamma_t(D_2)))+\log_2(\goodchi^2(\Gamma_t(D_3))) \rceil \nonumber \\
	&=\lceil \log_2 (\goodchi(\Gamma_t(D_1)))\rceil +\lceil \log_2 (\goodchi^1(\Gamma_t(D_3)))\rceil+\lceil \log_2 (\goodchi(\Gamma_t(D_2)))\rceil+\lceil \log_2 (\goodchi^2(\Gamma_t(D_3)))\rceil+\epsilon_1  \nonumber \\
	&= \lceil \log_2 (\goodchi(\Gamma_t(D_1)))\rceil+ \lceil \log_2 (\goodchi(\Gamma_t(D_2)))\rceil+\lceil \log_2 (\goodchi^1(\Gamma_t(D_3))\times \goodchi^2(\Gamma_t(D_3)))\rceil+\epsilon_1+\epsilon_2 \nonumber \\
	&\beta_{t}(D,G_o)= \beta_{t}(D_1)+ \beta_{t}(D_2)+\beta_{t}(D_3)+\epsilon /t,  \label{eq:beta27a}
	\end{align}
	where $\epsilon_1\in \{ -2,-1,0\}$ and $\epsilon_2\in \{0,1\}$ are obtained by using Lemma~\ref{floorceiling}, and $ \epsilon=(\epsilon_2+\epsilon_1) \in \{-2,-1,0,1\} $, $ \beta_t(D_m)=\frac{\lceil \log_2 (\goodchi(\Gamma_t(D_m)))\rceil }{t}$, for $ m\in \{1,2,3\} $. 
	As we know that $ \beta_{t}(D,G_o)\leq \beta_{t}(D_1)+ \beta_{t}(D_2)+\beta_{t}(D_3)$ (a simpler upper bound in TSUIC), the value of $ \epsilon $ in \eqref{eq:beta27a} cannot be greater than zero. Thus $\epsilon\in \{-2,-1,0\} $.
\end{IEEEproof}

\subsection{Proof of Proposition~\ref{lemma:2}}
Whilst constructing a confusion graph, we follow our proposed grouping of vertices of the confusion graph described in Appendix~\ref{append2a}. For convenience, let $ D $ be denoted by $ D^{i} $ if $ f(D)=\mathrm{H}_{i} $, $ i\in \{1,16\} $ (see Figure~\ref{fig:allgraphscasepart1}). 

\subsubsection{Construction and coloring of $ \Gamma_t(D^{16}) $}
We present the construction and two-sender graph coloring of $ \Gamma_t(D^{16}) $, where $ D^{16} $ has the fully-participated interactions between the sub-digraphs $D_1$, $D_2 $ and $D_3$.

\emph{(A) Construction of $ \Gamma_t(D^{16}) $:}
All edges of $ \Gamma_t(D^{16}) $ are listed in the following:
\begin{enumerate} 
	\item [(i)] Edges in $ E(\Gamma_t(D^{16})) $ due to the confusion at some vertices in $ V(D_1) $: The confusion at any vertex in $ V(D_1) $ contributes to only intra-edges due to Lemma~\ref{lemma:interblockedge}.
	\item [(ii)] Edges in $ E(\Gamma_t(D^{16})) $ due to the confusion at some vertices in $ V(D_2) $: The confusion at any vertex in $ V(D_2) $ contributes to only intra-edges due to Lemma~\ref{lemma:interblockedge}.
	\item [(iii)] Edges in $ E(\Gamma_t(D^{16})) $ due to the confusion at some vertices in $ V(D_3) $: If there exists an edge due to the confusion at some vertices in $ V(D_3) $ between any vertex pair $  ( (\textbf{b}_{D_{1}}^{i_1},\textbf{b}_{D_{2}}^{j_1},\textbf{b}_{D_{3}}^{k_1}),\\ (\textbf{b}_{D_{1}}^{i_2},\textbf{b}_{D_{2}}^{j_2},\textbf{b}_{D_{3}}^{k_2}))$, then each of the vertices in the $ k_1 $-th block has edges with all the vertices in the $ k_2 $-th block. This is because any vertex in $ V(D_3) $ has no message requested by any vertex in $ V(D_1)\cup V(D_2) $ as its side-information. This results no effect due to a change in bits of $ \textbf{b}_{D_{1}}^{i} $ or $\textbf{b}_{D_{2}}^{j} $ sub-label once we have an edge due to confusion at some receivers in $ V(D_3) $, which corresponds to the change in bits of the $ \textbf{b}_{D_{3}}^{k}$ sub-label. 
\end{enumerate}

\emph{(B) Coloring of $ \Gamma_t(D^{16}) $:}
In SSUIC, we know that the minimum numbers of colors required to color $ D_1 $, $ D_2 $ and $ D_3 $ separately are $ \goodchi(\Gamma_t(D_1)) $, $ \goodchi(\Gamma_t(D_2)) $ and $ \goodchi(\Gamma_t(D_3)) $, respectively. From Lemma~\ref{lemma:1}, in two-sender graph coloring, vertices in any $ k $-th block of $ \Gamma_t(D^{16}) $ are colored properly with the minimum of $\goodchi(\Gamma_t(D_1)) \times \goodchi(\Gamma_t(D_2)) $ ordered pairs of colors, where the minimum number of colors associated with $ S_1 $ and $ S_2 $ are $ \goodchi(\Gamma_t(D_1)) $ and $ \goodchi(\Gamma_t(D_2)) $, respectively. Referring to the construction of $ \Gamma_t(D^{16}) $, the inter-block edges are solely due to the confusion at some vertices in $ V(D_3) $ (from (i), (ii) and (iii) of the construction), and if there exists an inter-block edge between any two vertices, the first one belonging to $ k_1 $-th block and the second one belonging to $ k_2 $-th block, then we have edges from every vertex of the $ k_1 $-th block to all vertices of the $ k_2 $-th block. This states that it is necessary to have two different sets of ordered pairs of $\goodchi(\Gamma_t(D_1)) \times \goodchi(\Gamma_t(D_2)) $ colors, one for each block if there is an edge between these blocks. Furthermore, it is sufficient to consider the different color sets associated with one of the senders for those blocks in order to obtain the different sets of $\goodchi(\Gamma_t(D_1)) \times \goodchi(\Gamma_t(D_2)) $ ordered pairs of colors. As we require the minimum of $\goodchi(\Gamma_t(D_3))$ ordered pairs of colors to color vertices in any $ \mathcal{B}_{\textbf{b}_{D_{1}}^{i},\textbf{b}_{D_{2}}^{j} } $ (refer to Lemma~\ref{prop:coloringd3}), so the total number of minimum ordered pairs of colors required to color $ \Gamma_t(D^{16}) $ in two-sender graph coloring is $\goodchi(\Gamma_t(D_1))\times \goodchi(\Gamma_t(D_2))\times \goodchi(\Gamma_t(D_3)) $. Now from Lemma~\ref{sumofbetta}, we get $ \beta_{t}(D^{16},G_o)= \beta_{t}(D_1)+ \beta_{t}(D_2)+\beta_{t}(D_3)+\epsilon /t$, where $\epsilon\in \{-2,-1,0\} $. 

\subsubsection{Construction and coloring of $ \Gamma_t(D^{1}) $}
We present the construction and two-sender graph coloring of $ \Gamma_t(D^{1}) $. In contrast to $ D^{16}$ above, $D^{1} $ has no interaction between $ D_1$, $D_2$ and $ D_3 $. This results in extra edges, both intra-block and inter-block edges, in $ \Gamma_t( D^{1}) $ with respect to $ \Gamma_t(D^{16}) $. We observe that one can build $ \Gamma_t( D^{1})$ on the top of  $ \Gamma_t(D^{16}) $ by adding these extra edges. 	

\emph{(A) Construction of $ \Gamma_t(D^{1}) $:}
The extra edges of $ \Gamma_t( D^{1}) $ with respect to $ \Gamma_t(D^{16}) $ are both intra-block and inter-block edges. 

\emph{(B) Coloring of $ \Gamma_t(D^{1}) $:}
The extra intra-block edges do not change the requirements of ordered pairs of colors in two-sender graph coloring of a block of $ \Gamma_t(D^1) $ due to Lemma~\ref{lemma:1}. Now we address the extra inter-block edges.   
For the extra inter-block edges in $ E(\Gamma_t( D^{1})) $ due to the confusion at some vertices in $ V(D_1) $ and $ V(D_2) $ in two-sender graph coloring, we have the following: If there is no inter-block edge due to the confusion at some vertices in $ V(D_3) $, then we can do two-sender graph coloring of these blocks as stated by Lemma~\ref{lemma:nointeredgeD3}. This implies that we can do two-sender graph coloring of all these blocks by $\goodchi(\Gamma_t(D_1))\times \goodchi(\Gamma_t(D_2))$ ordered pairs of colors, where the minimum colors associated with $ S_1 $ and $ S_2 $ are $\goodchi(\Gamma_t(D_1))$ and $\goodchi(\Gamma_t(D_2))$, respectively. As the vertex $ 3 $ has no out-going arcs in both $ \mathrm{H}_{1} $ and $ \mathrm{H}_{16} $, the edges in $ E(\Gamma_t(D^{1})) $ due to the confusion at some vertices in $ V(D_3) $ are the same as of (iii) of the construction of $ \Gamma_t(D^{16} )$. Thus similar to the case of $ \Gamma_t(D^{16}) $, referring to the edges in $ E(\Gamma_t( D^{1})) $ due to the confusion at some vertices in $ V(D_3) $, if any $k_1 $-th and $ k_2 $-th blocks have the inter-block edges (including all inter-block edges due to the confusion at some vertices in $ V(D_1)\cup V(D) $), then it is necessary to have two different sets of ordered pairs of $\goodchi(\Gamma_t(D_1)) \times \goodchi(\Gamma_t(D_2)) $ colors, one for each block. Furthermore, it is sufficient to consider different color sets associated with one of the senders for these blocks in order to achieve the necessary ordered pairs of colors. Altogether, a TSUIC coloring of $ \Gamma_t( D^{1}) $ can be done similar to $ D^{16} $ with the minimum of $\goodchi(\Gamma_t(D_1))\times \goodchi(\Gamma_t(D_2))\times \goodchi(\Gamma_t(D_3)) $ ordered pairs of colors.  Now from Lemma~\ref{sumofbetta}, we get $ \beta_{t}( D^{1},G_o)=\beta_{t}(D_1)+ \beta_{t}(D_2)+\beta_{t}(D_3)+\epsilon /t $, $\epsilon\in \{-2,-1,0\} $. \hskip10pt $\square$

\section{Proof of Theorem~\ref{theorem:caseIIB1}} \label{append3}

For the problems in TSUIC, we prove this theorem by constructing a valid index code based on single-sender index codes. 
Before starting proof, unless stated otherwise, we assume the following for any vertex-induced sub-digraph $ D_{i'} $, for some index $ i'\in \{1,2,3\}$, in SSUIC:
\begin{enumerate} 
	\item Let $ \mathcal{C}(D_{i'}) $ be an index code (linear or non-linear) having a codeword length of $ |\mathcal{C}(D_{i'})| $ bits, for a given $ t $ (message bits), that achieves $ \beta_t(D_{i'}) $. For convenience, we represent $ |\mathcal{C}(D_{i'})|$ by $\ell^*(\mathcal{C}(D_{i'}))$ such that $\ell^*(\mathcal{C}(D_{i'}))=\beta_t(D_{i'})$. 
	\item Let the sequence of bits in $ \mathcal{C}(D_{i'}) $ be $ (w^{i'}_1, w^{i'}_2, \dotsc, w^{i'}_{\ell^*(\mathcal{C}(D_{i'}))})$, where $ w^{i'}_m\in \{0,1\} $, $ m\in \{ 1,2,\dotsc, \ell^*(\mathcal{C}(D_{i'}))\} $. 
	\item Let $ \mathcal{C}^1(D_{i'}) = (w^{i'}_1, w^{i'}_2, \dotsc, w^{i'}_{\ell^*_1(\mathcal{C} (D_{i'}))})$ and $ \mathcal{C}^2(D_{i'}) = (w^{i'}_{\ell^*_1(\mathcal{C} (D_{i'}))+1},w^{i'}_{\ell^*_1(\mathcal{C} (D_{i'}))+2},\dotsc,\\ w^{i'}_{\ell^*(\mathcal{C} (D_{i'}))})$ with $ |\mathcal{C}^2(D_{i'})|=\ell^*_2(\mathcal{C}(D_{i'})) $ be two parts of the sequence of bits of a codeword of $ \mathcal{C}(D_{i'}) $ such that $\mathcal{C}(D_{i'})=(\mathcal{C}^1(D_{i'}), \mathcal{C}^2(D_{i'}))$ with $ \ell^*(\mathcal{C}(D_{i'}))=\ell^*_1(\mathcal{C}(D_{i'}))+ \ell^*_2(\mathcal{C}(D_{i'})) $. 
	\item For any two codes $ \mathcal{C}(D_{i'}) $ and $\mathcal{C}(D_{j'}) $ with codeword lengths of $ \ell^*(\mathcal{C}(D_{i'}))$ and $\ell^*(\mathcal{C}(D_{j'})) $ bits, respectively, $ \mathcal{C}(D_{i'})\oplus \mathcal{C}(D_{j'})$ refers to the bit-wise XOR of bits of $ \mathcal{C}(D_{i'})$ and $\mathcal{C}(D_{j'})$ with zero padding if $ \ell^*(\mathcal{C}(D_{i'}))\neq \ell^*(\mathcal{C}(D_{j'})) $. This means $ \mathcal{C}(D_{i'})\oplus \mathcal{C}(D_{j'})$ contains $ \text{max}\{ \ell^*(\mathcal{C}(D_{i'})), \ell^*(\mathcal{C}(D_{j'})) \} $ bits. For example, if $ \mathcal{C}(D_{i'})=(101) $ and $ \mathcal{C}(D_{j'})=(001101) $, then $ \mathcal{C}(D_{i'})\oplus \mathcal{C}(D_{j'})= (101\textcolor{blue}{000})\oplus (001101)=(100101)$.
\end{enumerate}


(First case: $\beta_t(D,G_o)=\beta_t(D_3) $ if $ \beta_{t}(D_3) \geq \beta_{t}(D_1)+\beta_{t}(D_2)$) The given condition $ \beta_{t}(D_3) \geq \beta_{t}(D_1)+\beta_{t}(D_2)$ implies that $ |\mathcal{C}(D_3)|\geq |\mathcal{C}(D_1)|+|\mathcal{C}(D_2)|$ (i.e., $  \ell^*(\mathcal{C}(D_3))\geq  \ell^*(\mathcal{C}(D_1))+ \ell^*(\mathcal{C}(D_2)) $) for a finite $ t $. Now in TSUIC, we propose that $ S_1 $ transmits $ \mathcal{C}_1=\mathcal{C}^1(D_3) \oplus \mathcal{C}(D_1) $ of $ \ell^*_1(\mathcal{C}(D_3))= \ell^*(\mathcal{C}(D_1))$ bits, and $ S_2 $ transmits $ \mathcal{C}_2=\mathcal{C}^2(D_3) \oplus \mathcal{C}(D_2) $ of $  \ell^*_2(\mathcal{C}(D_3))$ bits because $  \ell^*_2(\mathcal{C}(D_3))\geq  \ell^*(\mathcal{C}(D_2))$ as we have $  \ell^*(\mathcal{C}(D_3))\geq  \ell^*(\mathcal{C}(D_1))+ \ell^*(\mathcal{C}(D_2)) $ and $ \ell^*_1(\mathcal{C}(D_3))= \ell^*(\mathcal{C}(D_1))$. Each receiver receives $ ( \ell^*_1(\mathcal{C}(D_3))+ \ell^*_2(\mathcal{C}(D_3)))$-bit $ (\mathcal{C}_1,\mathcal{C}_2)$. Now the decoding is done in the following way: 
(i) All the vertices in $ V(D_1) $ will decode their requested messages from $\mathcal{C}_1$ and its side-information that also includes $ \{x_i:i\in V(D_3) \} $ (as there is a fully-participated $ D_1\rightarrow D_3 $ in $D$),
(ii) all the vertices in $ V(D_2) $ will decode their requested messages from $ \mathcal{C}_2 $ and its side-information that also includes $ \{x_i:i\in V(D_3) \} $ (as there is a fully-participated $ D_2\rightarrow D_3 $ in $D$), and
(iii) all the vertices in $ V(D_3) $ will decode their requested messages from $ (\mathcal{C}_1,\mathcal{C}_2) $ and its side-information that also includes $ \{x_i:i\in V(D_1)\cup V(D_2) \} $ (as there is fully-participated $ D_3\rightarrow (D_1\cup D_2) $ in $D$). Thus $ ( \ell^*_1(\mathcal{C}(D_3))+ \ell^*_2(\mathcal{C}(D_3)))$-bit $ (\mathcal{C}_1,\mathcal{C}_2) $ is a valid index code in TSUIC for this case, and 
\begin{equation} \label{p:1}
\beta_t(D,G_o)\leq ( \ell^*_1(\mathcal{C}(D_3))+ \ell^*_2(\mathcal{C}(D_3)))/t= \ell^*(\mathcal{C}(D_3))/t=\beta_t(D_3).
\end{equation}
In SSUIC, $ \beta_{t}(D)\geq \beta_{t}(D_3) $ because $ D_3 $ is a sub-graph of $ D $. Now in TSUIC,
\begin{equation} \label{p:2}
\beta_t(D,G_o)\geq \beta_{t}(D)\geq \beta_{t}(D_3).
\end{equation} 
From \eqref{p:1} and \eqref{p:2}, we have $ \beta_t(D,G_o)= \beta_t(D_3)$. 

(Second case: $\beta_t(D,G_o)=\beta_{t}(D_1)+\beta_{t}(D_2) $ if $ \beta_{t}(D_3) \leq \beta_{t}(D_1)+\beta_{t}(D_2)$) The given condition $ \beta_{t}(D_3) \leq \beta_{t}(D_1)+\beta_{t}(D_2)$ implies that $ |\mathcal{C}(D_3)|\leq |\mathcal{C}(D_1)|+|\mathcal{C}(D_2)|$. Now we have the following three sub-cases: (i) $ |\mathcal{C}(D_3)|\geq \text{max}\{ |\mathcal{C}(D_1)|,|\mathcal{C}(D_2)| \}$, (ii) $ |\mathcal{C}(D_3)|\leq |\mathcal{C}(D_1)| $, and (iii) $ |\mathcal{C}(D_3)|\leq |\mathcal{C}(D_2)| $. For these sub-cases, we propose the following:

(Sub-case (i): $ |\mathcal{C}(D_3)|\geq \text{max}\{ |\mathcal{C}(D_1)|,|\mathcal{C}(D_2)| \}$) $ S_1 $ transmits $ \mathcal{C}_1=\mathcal{C}^1(D_3) \oplus \mathcal{C}(D_1) $ of $ \ell^*_1(\mathcal{C}(D_3))= \ell^*(\mathcal{C}(D_1))$ bits, and $ S_2 $ transmits $ \mathcal{C}_2=\mathcal{C}^{2}(D_3) \oplus \mathcal{C}(D_2) $ of $ \ell^*(\mathcal{C}(D_2))$ bits because $ \ell^*(\mathcal{C}(D_2))\geq  \ell^*_2(\mathcal{C}(D_3))$ as we have $ \ell^*(\mathcal{C}(D_3))\leq  \ell^*(\mathcal{C}(D_1))+ \ell^*(\mathcal{C}(D_2))$ and $ \ell^*_1(\mathcal{C}(D_3))= \ell^*(\mathcal{C}(D_1))$. Each receiver receives $ ( \ell^*(\mathcal{C}(D_1))+ \ell^*(\mathcal{C}(D_2)))$-bit $ (\mathcal{C}_1,\mathcal{C}_2)$. Now one can verify that the decoding is done in the same way as stated in the first case. Thus $ ( \ell^*(\mathcal{C}(D_1))+ \ell^*(\mathcal{C}(D_2)))$-bit $ (\mathcal{C}_1,\mathcal{C}_2) $ is a valid index code in TSUIC for this sub-case. 

(Sub-case (ii): $ |\mathcal{C}(D_3)|\leq |\mathcal{C}(D_1)| $) $ S_1 $ transmits $ \mathcal{C}_1=\mathcal{C}(D_3) \oplus \mathcal{C}(D_1) $ of $  \ell^*(\mathcal{C}(D_1))$ bits (because $ \ell^*(\mathcal{C}(D_3))\leq  \ell^*(\mathcal{C}(D_1))$), and $ S_2 $ transmits $ \mathcal{C}_2=\mathcal{C}(D_2) $ of $  \ell^*(\mathcal{C}(D_2))$ bits. Now the decoding is done in the following way: 
(i) All the vertices in $ V(D_1) $ will decode their requested messages from $ \mathcal{C}_1 $ and its side-information that also includes $ \{x_i:i\in V(D_3) \} $ (as there is a fully-participated $ D_1\rightarrow D_3 $ in $ D$),
(ii) all the vertices in $ V(D_2) $ will decode their requested messages from $ \mathcal{C}_2 $ and its side-information, and
(iii) all the vertices in $ V(D_3) $ will decode their requested messages from $ \mathcal{C}_1$ and its side-information that also includes $ \{x_i:i\in V(D_1)\} $ (as there is a fully-participated $ D_3\rightarrow D_1 $ in $ D $). Thus $ ( \ell^*(\mathcal{C}(D_1))+ \ell^*(\mathcal{C}(D_2)))$-bit $ (\mathcal{C}_1,\mathcal{C}_2) $ is a valid encoding in TSUIC for this sub-case. 

(Sub-case (iii): $ |\mathcal{C}(D_3)|\leq |\mathcal{C}(D_2)| $) As we have a fully-participated $ D_3\rightarrow (D_1\cup D_2) $ in $ D$, so by swapping $ D_1 $ and $ D_2 $ (meaning we swap the two senders) in the sub-case (ii), one can prove that $ ( \ell^*(\mathcal{C}(D_1))+ \ell^*(\mathcal{C}(D_2)))$-bit $ (\mathcal{C}_1,\mathcal{C}_2) $ is a valid index code in TSUIC for this sub-case.

Altogether for the second case, 
\begin{equation} \label{p:1a}
\beta_t(D,G_o)\leq ( \ell^*(\mathcal{C}(D_1))+ \ell^*(\mathcal{C}(D_2)))/t=\beta_t(D_1)+\beta_t(D_2).
\end{equation}
By considering $ \beta_t(D,G_o)\geq \beta_t(D_1)+\beta_t(D_2) $ (by Lemma~\ref{lemma:1a}) and \eqref{p:1a}, we get, $\beta_t(D,G_o)=\beta_t(D_1)+\beta_t(D_2) $.

Now combining these two cases (First and Second cases), we get
\begin{equation} \label{eq:A}
\beta_t(D,G_o)=\text{max} \{\beta_t(D_3),\beta_{t}(D_1)+\beta_{t}(D_2) \}.
\end{equation}
Now taking a limit $ t\rightarrow \infty $ on both sides of \eqref{eq:A}, we get 
\begin{align} \label{eq:x}
\underset{t\rightarrow \infty}{\text{lim}}\ \beta_t(D,G_o)&=\underset{t\rightarrow \infty}{\text{lim}}\ \text{max} \{\beta_t(D_3),\beta_{t}(D_1)+\beta_{t}(D_2) \}\nonumber \\
\underset{t\rightarrow \infty}{\text{lim}}\ \beta_t(D,G_o)&=\text{max} \{\underset{t\rightarrow \infty}{\text{lim}}\ \beta_t(D_3),\ \underset{t\rightarrow \infty}{\text{lim}}\ (\beta_{t}(D_1)+\beta_{t}(D_2)) \}. 
\end{align}
We know that $\beta=\underset{t}{\mathrm{inf}}\ \beta_t=\underset{t\rightarrow \infty}{\lim}\ \beta_t$ (by Definition~\ref{def:boradcastrate}), and ``\emph{a limit of a finite sum of functions equals the sum of the limit of each functions, if the limit of each function exists},'' so we get $\beta(D,G_o)=\text{max} \{ \beta(D_3),\beta(D_1)+\beta(D_2) \}$ from \eqref{eq:x}. \hskip10pt $\square$	
\section{Proof of Theorem~\ref{theorem:caseIIC1}} \label{append5}

For the problems in TSUIC, we prove this theorem by constructing a valid index code based on single-sender index codes. Refer to the first paragraph of Appendix~\ref{append3} for notations. 

In TSUIC, we propose that $ S_1 $ transmits $ \mathcal{C}_1=\mathcal{C}(D_{1}) \oplus \mathcal{C}(D_{3}) $ consisting of $\text{max} \{ \ell^*(\mathcal{C}(D_1)),\\ \ell^*(\mathcal{C}(D_3)) \}$ bits, and $ S_2 $ transmits $ \mathcal{C}_2=\mathcal{C}(D_2)$ of $  \ell^*(\mathcal{C}(D_2)) $ bits. Each receiver receives $ ( \ell^*(\mathcal{C}(D_2))+\text{max} \{ \ell^*(\mathcal{C}(D_1)), \ell^*(\mathcal{C}(D_3)) \})$-bit $ (\mathcal{C}_1,\mathcal{C}_2) $. Now the decoding is done in the following way: 
(i) All the vertices in $ V(D_1) $ will decode their requested messages from $ \mathcal{C}_1 $ and its side-information that also includes $ \{x_i:i\in V(D_3) \} $ (as there is a fully-participated $ D_1\rightarrow D_3 $ in $ D$),
(ii) all the vertices in $ V(D_2) $ will decode their requested messages from $ \mathcal{C}_2 $ and its side-information, and
(iii) all the vertices in $ V(D_3) $ will decode their requested messages from $\mathcal{C}_1$ and its side-information that also includes $ \{x_i:i\in V(D_1) \} $ (as there is a fully-participated $ D_3\rightarrow D_1 $ in $ D$). Thus $ ( \ell^*(\mathcal{C}(D_2))+ \text{max} \{ \ell^*(\mathcal{C}(D_1)), \ell^*(\mathcal{C}(D_3))\} )$-bit $ (\mathcal{C}_1,\mathcal{C}_2) $ is a valid index code in TSUIC for this case, and
\begin{equation} \label{p:2a}
\beta_t(D,G_o)\leq \frac{1}{t}( \ell^*(\mathcal{C}(D_2))+\text{max} \{ \ell^*(\mathcal{C}(D_1)), \ell^*(\mathcal{C}(D_3)) \})=\beta_t(D_2)+\text{max}\{\beta_t(D_1),\beta_t(D_3) \}.
\end{equation}
Now by Lemma~\ref{lemma:1a}, we have $ \beta_t(D,G_o)\geq \beta_t(D_1)+\beta_t(D_2) $, and from \eqref{p:2a}, if $ \beta_t(D_1)\geq \beta_t(D_3)$, then $ \beta_t(D,G_o)\leq  \beta_t(D_1)+\beta_t(D_2) $. Altogether, we get $ \beta_t(D,G_o)= \beta_t(D_1)+\beta_t(D_2) $. 

For the sub-digraph $ D_{j'} $, $ j'\in \{1,2,3 \} $, we know that there exists an index code $ \mathcal{C}(D_{j'}) $ of $  \ell^*(\mathcal{C}(D_{j'})) $ bits such that $  \ell^*(\mathcal{C}(D_{j'}))/t $ tends to $\beta(D_{j'})$ if $ t\rightarrow \infty $, and for any $ t\geq 1 $, $ \beta(D_{j'})\leq  \ell^*(\mathcal{C}(D_{j'}))/t$. So, we write $  \ell^*(\mathcal{C}(D_{j'}))/t=\beta(D_{j'})+\epsilon_t(D_{j'}) $, for some $ \epsilon_t(D_{j'})\geq  0$ such that $ \epsilon_t(D_{j'})$ tends to zero if message length $ t $ tends to infinity. For $ D$, considering the same code formation, which is a valid two-sender index code, as stated for the cases considering the finite message length, we get $ S_1 $ and $ S_2 $ transmitting sub-codewords of $ \text{max}\{  \ell^*(\mathcal{C}(D_1)),  \ell^*(\mathcal{C}(D_3))\} $ and $  \ell^*(\mathcal{C}(D_2)) $ bits, respectively. For any $ t\geq 1 $, there exists a two-sender index code of the following bit length: 
\begin{align*}
p_1+p_2= \text{max}\{  \ell^*(\mathcal{C}(D_1)),  \ell^*(\mathcal{C}(D_3))\}+  \ell^*(\mathcal{C}(D_2)).
\end{align*}
Now dividing both sides by $ t $ in the above equation, we get	
\begin{align}
\frac{p_1+p_2}{t} &= \text{max}\{  \ell^*(\mathcal{C}(D_1))/t,  \ell^*(\mathcal{C}(D_3))/t\}+  \ell^*(\mathcal{C}(D_2))/t \nonumber \\
&=  \text{max}\{\beta(D_{1})+\epsilon_t(D_{1}), \beta(D_{3})+\epsilon_t(D_{3}) \}+ \beta(D_{2})+\epsilon_t(D_{2}), 
\end{align}
where $\epsilon_t(D_{j'})\geq 0$ for $ j'\in \{1,2,3 \} $. For any $ j'\in \{1,2,3 \} $, as $\epsilon_t(D_{j'})\rightarrow 0$ for $ t\rightarrow \infty $, we get
\begin{equation} \label{eq:f}
\beta(D,G_o) \leq \underset{t\rightarrow \infty}{\lim}\ (p_1+p_2)/t= \text{max}\{\beta(D_{1}), \beta(D_{3}) \}+ \beta(D_{2}). 
\end{equation}
Alternatively, we can get \eqref{eq:f} by taking a limit $ t\rightarrow \infty $ on both sides of \eqref{p:2a} because $\beta=\underset{t\rightarrow \infty}{\lim}\ \beta_t$ (by Definition~\ref{def:boradcastrate}).

Clearly, if $ \beta(D_1)\geq \beta(D_3) $, then from \eqref{eq:f}, we get
\begin{equation} \label{eq:g}
\beta(D,G_o)\leq \beta(D_1)+\beta(D_2).
\end{equation}
Now from Lemma~\ref{lemma:1a} and \eqref{eq:g}, we get $\beta(D,G_o)=\beta(D_1)+\beta(D_2) $ if $ \beta(D_1)\geq \beta(D_3) $. \hskip10pt $\square$
\section{Proof of Theorem~\ref{theorem:caseIIE1}} \label{append7}
\begin{IEEEproof}
	For the problems in TSUIC, we prove this theorem by constructing a valid index code based on single-sender index codes. Refer to the first paragraph of Appendix~\ref{append3} for notations.
	
	In TSUIC, we propose that $ S_1 $ transmits $ \mathcal{C}_1=\mathcal{C}(D_1) \oplus \mathcal{C}(D_3) $ of $ \text{max}\{ \ell^*(\mathcal{C}(D_1)), \ell^*(\mathcal{C}(D_3))\}$ bits, and $ S_2 $ transmits $ \mathcal{C}_2=\mathcal{C}(D_2) \oplus \mathcal{C}(D_3) $ of $ \text{max}\{ \ell^*(\mathcal{C}(D_2)), \ell^*(\mathcal{C}(D_3))\}$ bits. Each receiver receives $ (\text{max}\{ \ell^*(\mathcal{C}(D_1)), \ell^*(\mathcal{C}(D_3))\}+\text{max}\{ \ell^*(\mathcal{C}(D_2)), \ell^*(\mathcal{C}(D_3))\})$-bit $ (\mathcal{C}_1,\mathcal{C}_2)$. 
	Now the decoding is done in the following way: (i) All the vertices in $ V(D_1) $ will decode their requested messages from $ \mathcal{C}_1\oplus \mathcal{C}_2$ and its side-information that also includes $ \{x_i:i\in V(D_2) \} $ (as there is a fully-participated $ D_1\rightarrow D_2 $ in $ D$), 	
	(ii) if $ f(D)\in \{\mathrm{H}_{i'}: i'\in \{33,34,35\} \}$, all the vertices in $ V(D_2) $ will decode their requested messages from $ \mathcal{C}_1\oplus \mathcal{C}_2$ and its side-information that also includes $ \{x_i:i\in V(D_1) \} $ (as there is a fully-participated $ D_2\rightarrow D_1 $ in $ D$), and if $f(D)=\mathrm{H}_{36}$, all the vertices in $ V(D_2) $ will decode their requested messages from $ \mathcal{C}_2$ and its side-information that also includes $ \{x_i:i\in V(D_3) \} $ (as there is a fully-participated $ D_2\rightarrow D_3 $ in $ D$),  	
	(iii) all the vertices in $ V(D_3) $ will decode their requested messages from $ \mathcal{C}_1$  and its side-information that also includes $ \{x_i:i\in V(D_1)\} $ (as there is a fully-participated $ D_3\rightarrow D_1$ in $ D$).
	Thus $ (\text{max}\{ \ell^*(\mathcal{C}(D_1)), \ell^*(\mathcal{C}(D_3))\}+\text{max}\{ \ell^*(\mathcal{C}(D_2)), \ell^*(\mathcal{C}(D_3))\})$-bit $ (\mathcal{C}_1,\mathcal{C}_2) $ is a valid index code in TSUIC for this case, and 
	\begin{align} \label{p:3a}
	\beta_t(D,G_o) &\leq (\text{max}\{ \ell^*(\mathcal{C}(D_1)), \ell^*(\mathcal{C}(D_3))\}+\text{max}\{ \ell^*(\mathcal{C}(D_2)), \ell^*(\mathcal{C}(D_3))\})/t \nonumber \\
	& \leq \text{max}\{\beta_t(D_1),\beta_t(D_3)\}+\text{max}\{\beta_t(D_2),\beta_t(D_3)\}. 
	\end{align}
	Now if $ \beta_t(D_3)\leq \text{min}\{\beta_t(D_1),\beta_t(D_2)\}$, from \eqref{p:3a}, we get $\beta_t(D,G_o)\leq \beta_t(D_1)+\beta_t(D_2)$. From Lemma~\ref{lemma:1a}, we have $ \beta_t(D,G_o)\geq \beta_t(D_1)+\beta_t(D_2) $. Thus $\beta_t(D,G_o)=\beta_t(D_1)+\beta_t(D_2) $ if $ \beta_t(D_3)\leq \text{min}\{\beta_t(D_1),\beta_t(D_2)\}$. 
	
	Now by taking a limit $ t\rightarrow \infty $ on both sides of \eqref{p:3a}, we get 
	\begin{align}  \label{eq:i}
	\underset{t\rightarrow \infty}{\lim}\ \beta_t(D,G_o) &\leq \underset{t\rightarrow \infty}{\lim}\ \text{max}\{\beta_t(D_1),\beta_t(D_3)\}+\underset{t\rightarrow \infty}{\lim}\ \text{max}\{\beta_t(D_2),\beta_t(D_3)\} \nonumber \\
	\beta(D,G_o) &\leq \text{max}\{\beta(D_1),\beta(D_3)\}+\text{max}\{\beta(D_2),\beta(D_3)\}.	
	\end{align}	
	This is because $\beta=\underset{t\rightarrow \infty}{\lim}\ \beta_t$ (by Definition~\ref{def:boradcastrate}). Now from \eqref{eq:i}, we get $\beta(D,G_o)\leq \beta(D_1)+\beta(D_2) $ if $ \beta(D_3)\leq \text{min}\{\beta(D_{1}), \beta(D_{2})\} $, and $  \beta(D,G_o)\geq \beta(D_1)+\beta(D_2)$ from Lemma~\ref{lemma:1a}. So, $\beta(D,G_o)= \beta(D_1)+\beta(D_2) $ if $ \beta(D_3)\leq \text{min}\{\beta(D_{1}), \beta(D_{2})\} $.	
\end{IEEEproof}

\section{Proof of Proposition~\ref{lemma:2extra}} \label{append8}
Before proving Proposition~\ref{lemma:2extra}, we have the following lemmas related to row and column sub-blocks of any $ k $-th block, $k\in \{1,2,\dotsc, 2^{tn_{N'+1}}\}$.
\begin{Lemma} \label{lemma:extra1}
	For any row sub-block of a $ k $-th block, the minimum number of $ N' $-tuples of colors required to color it in \textnormal{SMSUIC} is $ \goodchi(\Gamma_t(D_1))\times 1 \times \dotsc \times 1 $, where the minimum colors associated with $ S_1 $ and remainder senders are $ \goodchi(\Gamma_t(D_1)) $ and one each, respectively. 
\end{Lemma}
\begin{IEEEproof}
	Based on our proposed grouping of vertices (also see Figure~\ref{fig:msuicvertexarrangement}), any row sub-block of a $ k $-th block consists of vertices labeled by all $ ( \textbf{b}_{D_{1}}^{i_1}, \textbf{b}_{D_{2}}^{i_2},\textbf{b}_{D_{3}}^{i_3},\dotsc, \textbf{b}_{D_{N'}}^{i_{N'}}, \textbf{b}_{D_{N'+1}}^{k})$, $ i_j\in \{1,2,\dotsc,2^{tn_j}\} $ for $ j\in \{1,2,\dotsc, N'\} $, with the same $(\textbf{b}_{D_{2}}^{i_2},\textbf{b}_{D_{3}}^{i_3},\dotsc, \textbf{b}_{D_{N'}}^{i_{N'}}, \textbf{b}_{D_{N'+1}}^{k})$ sub-labels. Clearly, the edges in the row sub-block are only due to the confusion at vertices in $V(D_1) $, so $ S_1 $ colors differently if there is any confusion, and any sender $ S_j $, $ j\neq 1 $, provides the same color to all vertices of the row sub-block. Moreover, observe that any row sub-block and $ \Gamma_t(D_1) $ are isomorphic graphs (one can extend the proof of Lemma~\ref{prop:isomorphicd1} to get this result). The proof completes by noting that any $ S_j $, $ j\in \{1,2,\dotsc, N'\} $, requires a minimum of $  \goodchi(\Gamma_t(D_j)) $ colors to color a confusion graph $ \Gamma_t(D_j) $ whose vertices are labeled by $ \textbf{b}_{D_{j}}^{i_j} $, $ i_j\in \{1,2,\dotsc,2^{tn_j} \} $. 	
\end{IEEEproof}	

\begin{Lemma} \label{lemma:extra2}
	For any column sub-block of a $ k $-th block, the minimum number of $ N' $-tuples of colors required to color it in \textnormal{SMSUIC} is $ 1\times \goodchi(\Gamma_t(D_2))\times \goodchi(\Gamma_t(D_3))\times\dotsc \times \goodchi(\Gamma_t(D_{N'})) $, where the minimum colors associated with $ S_1 $ and $ S_j $ are one and $ \goodchi(\Gamma_t(D_j)) $, respectively, for all $ j\in \{2,3,\dotsc,N'\} $.   
\end{Lemma}
\begin{IEEEproof}	
	Based on our proposed grouping of vertices, any column sub-block of a $ k $-th block consists of vertices labeled by all $ ( \textbf{b}_{D_{1}}^{i_1}, \textbf{b}_{D_{2}}^{i_2},\textbf{b}_{D_{3}}^{i_3},\dotsc, \textbf{b}_{D_{N'}}^{i_{N'}}, \textbf{b}_{D_{N'+1}}^{k})$, $ i_j\in \{1,2,\dotsc,2^{tn_j}\} $ for $ j\in \{1,2,\dotsc, N'\} $, with the same $\textbf{b}_{D_{1}}^{i_1}$ and $ \textbf{b}_{D_{N'+1}}^{k} $ sub-labels. As a result, the edges in the column sub-block are only due to the confusion at some vertex in $\bigcup_{i=2}^{N'} V(D_i)$, and $ S_1 $ assigns the same color to all vertices in the sub-block. Now for any sender $ S_j $, $ j\in \{2,3,\dotsc,N'\} $, the message tuples associated to all sub-labels ($ \textbf{b}_{D_{1}}^{i_1}, \textbf{b}_{D_{2}}^{i_2},\textbf{b}_{D_{3}}^{i_3},\dotsc, \textbf{b}_{D_{N'}}^{i_{N'}}, \textbf{b}_{D_{N'+1}}^{k}$) except $ \textbf{b}_{D_{j}}^{i_j} $ are ``DON'T CARE'' since all $ \textbf{b}_{D_{1}}^{i_1}, \textbf{b}_{D_{2}}^{i_2},\textbf{b}_{D_{3}}^{i_3},\dotsc, \textbf{b}_{D_{N'}}^{i_{N'}} $ sub-labels are associated only with private messages, and $ \textbf{b}_{D_{N'+1}}^{k} $, which is associated with common messages, is same for all vertices of the $ k $-th block. Thus $ S_j $ colors any two or more vertices with the same $ \textbf{b}_{D_{j}}^{i_j} $ but with different other sub-labels in the sub-block with the same color. For example, $ S_2 $ assigns the same color to the vertices labeled by $ ( \textbf{b}_{D_{1}}^{1}, \textbf{b}_{D_{2}}^{i_2},\textbf{b}_{D_{3}}^{1},\dotsc, \textbf{b}_{D_{N'}}^{1}, \textbf{b}_{D_{N'+1}}^{k} ) $, $ ( \textbf{b}_{D_{1}}^{2}, \textbf{b}_{D_{2}}^{i_2},\textbf{b}_{D_{3}}^{2},\dotsc, \textbf{b}_{D_{N'}}^{2}, \textbf{b}_{D_{N'+1}}^{k} ) $ and $ (  \textbf{b}_{D_{1}}^{3}, \textbf{b}_{D_{2}}^{i_2},\textbf{b}_{D_{3}}^{3},\dotsc, \textbf{b}_{D_{N'}}^{3}, \textbf{b}_{D_{N'+1}}^{k}) $ for some $ i_2 $. Furthermore, it is the only one sender which can resolve the confusion by assigning different colors if there is confusion of the tuples associated with $ \textbf{b}_{D_{j}}^{i_j} $, $ j\in \{1,2,\dotsc,N' \} $. Clearly, for any column sub-block, the coloring function of $ S_j $ depends on the message tuples associated only to the sub-label $ \textbf{b}_{D_{j}}^{i_j} $, $ i_j\in \{1,2,\dotsc,2^{tn_j}\} $. 
	Note that any $ S_j $, $ j\in \{1,2,\dotsc, N'\} $, requires a minimum of $  \goodchi(\Gamma_t(D_j)) $ colors to color a confusion graph $ \Gamma_t(D_j) $ whose vertices are labeled by $ \textbf{b}_{D_{j}}^{i_j} $, $ i_j\in \{1,2,\dotsc,2^{tn_j} \} $. Along with the proposed grouping of vertices (see Figure~\ref{fig:msuicvertexarrangement}), altogether, for any column sub-block, each $ S_j $ requires a minimum of $  \goodchi(\Gamma_t(D_j)) $ colors. In SMSUIC, for any column sub-block, if any two vertices are connected by an edge due to the confusion of the message tuples associated with the following: 
	(i) Only with sub-label $ \textbf{b}_{D_{j}}^{i_j} $, then the sender $ S_j $ assigns different colors in the color tuples of those vertices, and
	(ii) more than two sub-labels (for example $ \textbf{b}_{D_{2}}^{i_2}$ and $\textbf{b}_{D_{3}}^{i_3} $), then all the associated senders (for example both $ S_2 $ and $ S_3 $) assign different colors in the color tuples of those vertices. This is because the confusion of the message tuples associated with the sub-labels of those vertices occurred at different sub-labels related only to private messages. Overall, considering the symmetry of the column sub-block, the minimum of $ 1\times \goodchi(\Gamma_t(D_2))\times \goodchi(\Gamma_t(D_3))\times\dotsc \times \goodchi(\Gamma_t(D_{N'})) $ $ N' $-tuples of colors is required to color it in MSUIC, where the minimum colors associated with $ S_1 $ and $ S_j $ are one and $ \goodchi(\Gamma_t(D_j)) $, respectively, for all $ j\in \{2,3,\dotsc,N'\} $.     	
\end{IEEEproof}	

\begin{Lemma} \label{lemma:extra3}
	For any $ k $-th block, the minimum number of $ N' $-tuples of colors required to color it in \textnormal{MSUIC} is $ \goodchi(\Gamma_t(D_1))\times \goodchi(\Gamma_t(D_2))\times \goodchi(\Gamma_t(D_3))\times\dotsc \times \goodchi(\Gamma_t(D_{N'})) $.
\end{Lemma}
\begin{IEEEproof}	
	Consider any two different $ i_1 $-th and $ i'_1 $-th column sub-blocks of a $ k $-th block, where $ i_1,i'_1\in \{1,2,\dotsc,2^{tn_1} \} $. Observe that these two column sub-blocks are isomorphic, and the vertices of these two sub-blocks have different $ \textbf{b}_{D_{1}}^{i_1} $ sub-label such that any $ i_1 $-th sub-block has vertices labeled by all $ ( \textbf{b}_{D_{1}}^{i_1}, \textbf{b}_{D_{2}}^{i_2},\textbf{b}_{D_{3}}^{i_3},\dotsc, \textbf{b}_{D_{N'}}^{i_{N'}}, \textbf{b}_{D_{N'+1}}^{k})$, $ i_j\in \{1,2,\dotsc,2^{tn_j}\} $ for $ j\in \{2,\dotsc, N'\} $, with the same $ \textbf{b}_{D_{1}}^{i_1} $ sub-label. Assume that we properly color $ i_1 $-th column sub-block with a minimum of $ 1\times \goodchi(\Gamma_t(D_2))\times \goodchi(\Gamma_t(D_3))\times\dotsc \times \goodchi(\Gamma_t(D_{N'})) $  $ N' $-tuples of colors in multi-sender graph coloring (by Lemma~\ref{lemma:extra2}). Now we keep the same coloring functions of all $ S_j $, $ j\in \{2,\dotsc, N'\} $, that operate in $ i_1 $-th column sub-block to $ i'_1$-th column sub-block. Considering our proposed grouping of the vertices, its symmetry, and the above observation, this is a valid multi-sender graph coloring of all $ S_j $ for those two column sub-blocks. Moreover, $ S_1 $ is the only one sender that provides different colors if there is confusion at some vertices in $ V(D_1) $ (where the confusion is associated with the sub-labels $ \textbf{b}_{D_{1}}^{i_1} $ and $ \textbf{b}_{D_{1}}^{i'_1} $), and once any coloring function of $ S_j $, $ j\in \{ 2,3,\dotsc,N'\} $ resolves the confusion at vertices in $ V(D_j) $ within one column sub-block, then it also resolves the confusion at vertices in $ V(D_j) $ belonging to the different column sub-blocks. For example, if a vertex $ u $ in $ i_1 $-th column sub-block is connected by an edge to a vertex $ v' $ in $ i'_1 $-th column sub-block due to the confusion at a vertex in $ V(D_j) $, then $ u $ is also connected to another vertex $ v$ present in $ i_1 $-th column with the same sub-label $\textbf{b}_{D_{j}}^{i_j}$ as of $ v' $. The confusion is resolved by our assumption in any $ i_1 $-th column, so it must be true for the two column sub-blocks as we repeat the same coloring function of all $ S_j $, $ j\in \{ 2,3,\dotsc,N'\} $ to both the column sub-blocks.
	
	Now for sender $ S_1 $, whilst coloring in MSUIC, it assigns the same color to all vertices of a column sub-block as it has the same $\textbf{b}_{D_{1}}^{i_1}$ sub-label. If we consider all column sub-blocks of the block, then $ S_1 $ properly colors with a minimum of $ \goodchi(\Gamma_t(D_1)) $ colors (from Lemma~\ref{lemma:extra1}). Altogether, $ \goodchi(\Gamma_t(D_1)) $ times  the $ N' $-tuples of colors required to color one column sub-block, i.e., $ \goodchi(\Gamma_t(D_1))\times \goodchi(\Gamma_t(D_2))\times \goodchi(\Gamma_t(D_3))\times\dotsc \times \goodchi(\Gamma_t(D_{N'})) $ is the minimum number of $ N' $-tuples of colors required to color a block in SMSUIC.		  
\end{IEEEproof}	

Let $ D $ with $ f(D)=\mathrm{H}'_1 $ and $ f(D)=\mathrm{H}'_{16} $ be denoted by $ D^1_{\mathrm{M}} $ and $D_\mathrm{M}^{16}$, respectively. Based on our proposed grouping of vertices of $ \Gamma_t(D_\mathrm{M}^{16}) $, observe that any block has all vertices with the same label $ \textbf{b}_{D_{N'+1}}^{k} $, and this label only changes when block changes. This is exactly the same case as in TSUIC (where $ \textbf{b}_{D_{3}}^{k} $ changes when block changes). Now along with the consideration of the grouping of vertices, their symmetry, and Lemma~\ref{lemma:extra3}, one can construct and color $ \Gamma_t(D_\mathrm{M}^{16})  $ similar to $ \Gamma_t(D^{16}) $, and $ \Gamma_t(D_\mathrm{M}^{1})  $ similar to $ \Gamma_t(D^{1}) $ stated in the proof of Proposition~\ref{lemma:2} in Appendix~\ref{append2}. This results the following lemma.
\begin{Lemma} \label{lemma:extra4}
	For any $ \Gamma_t(D_\mathrm{M}^{16}) $ or $ \Gamma_t(D_\mathrm{M}^{1}) $, the minimum number of $ N' $-tuples of colors required to color it in \textnormal{SMSUIC} is $ \goodchi(\Gamma_t(D_1))\times \goodchi(\Gamma_t(D_2))\times \goodchi(\Gamma_t(D_3))\times\dotsc \times \goodchi(\Gamma_t(D_{N'}))\times \goodchi(\Gamma_t(D_{N'+1})) $.
\end{Lemma}

Similar to the proof of Lemma~\ref{sumofbetta}, by utilizing Lemma~\ref{lemma:extra4}, it is not difficult to prove the following lemma:

\begin{Lemma} \label{lemma:extra5}
	For any $ \Gamma_t(D)$, if $ \goodchi(\Gamma_t(D_1))\times \goodchi(\Gamma_t(D_2))\times \goodchi(\Gamma_t(D_3))\times\dotsc \times \goodchi(\Gamma_t(D_{N'}))\times \goodchi(\Gamma_t(D_{N'+1})) $ $ N' $-tuples of colors required in \textnormal{SMSUIC}, then $ \beta_t(D,G_o)=\sum_{i=1}^{N'+1}\beta_t(D_i)+\epsilon/t$ for some $\epsilon\in \{-N',{-N'+1},\dotsc, 0\} $.   	
\end{Lemma} 

Now the proof of Proposition~\ref{lemma:2extra} follows from Lemmas~\ref{lemma:extra4} and \ref{lemma:extra5}.

\bibliographystyle{IEEEtran}
\bibliography{Bibliography}


\end{document}